\documentclass[fleqn,usenatbib]{mnras}

\usepackage{newtxtext,newtxmath}

\usepackage[T1]{fontenc}

\DeclareRobustCommand{\VAN}[3]{#2}
\let\VANthebibliography\thebibliography
\def\thebibliography{\DeclareRobustCommand{\VAN}[3]{##3}\VANthebibliography}

\usepackage{graphicx}	
\usepackage{amsmath}	
\usepackage{natbib}
\usepackage{graphics}
\usepackage{longtable}[1]
\usepackage{lscape}
\usepackage{color}
\usepackage{rotating}
\usepackage{cite}
\usepackage[svgnames,x11names,table]{xcolor}

\usepackage{epstopdf}
\usepackage{multicol,float}

\usepackage{textcomp}
\usepackage{hyperref}
\usepackage{ulem}

\newcommand{\m}{$-$}

\defcitealias{2011MNRAS.417..835F}{F11}
\defcitealias{2010ApJ...721.1348R}{R10}

\title[Star clusters and clumps in Arp 147]{Massive star clusters and clumps in the collisional ring galaxy Arp 147}

\author[Z. Randriamanakoto et al.]{
Z. Randriamanakoto,$^{1,\,2}$\thanks{E-mail: zara@saao.ac.za},  M. Rakototafika$^{2}$, B. Mongwane$^{3}$, P. V\"ais\"anen$^{4,\,1}$ and M. Rakotomanga$^{5}$\\
$^{1}$\,South African Astronomical Observatory, PO Box  9, Observatory 7935, South Africa \\
$^{2}$\,Department of Physics, University of Antananarivo, PO Box 906, Antananarivo 101, Madagascar\\
$^{3}$\,Department of Mathematics \& Applied Mathematics, University of Cape Town, Rondebosch 7701, South Africa \\
$^{4}$\,{Finnish Centre for Astronomy with ESO (FINCA), Vesilinnantie 5, FI-20014 University of Turku, Finland}\\
$^{5}$\,School of Mathematics, University of Leeds, Leeds LS2 9JT, United Kingdom
}

\date{Accepted 2026 April 12. Received 2026 April 09; in original form 2025 August 29}

\pubyear{\the\year{}}

\begin{document}
\label{firstpage}
\pagerange{\pageref{firstpage}--\pageref{lastpage}}
\maketitle

\begin{abstract}
We conduct a photometric study of star clusters (or knots) in the collisional ring galaxy (CRG) Arp 147 to trace the star formation history across its empty ring. Using HST \textcolor{black}{F450W, F606W and F814W} images, we find that Arp 147 hosts 211 knots and six kpc-size clumps, nearly 60 per cent of which have ages below 10 Myr, and two thirds have masses above  $\rm 10^{5}\,M_{\odot}$. The cluster mass function (CMF) of knots with ages between $10 - 200$ Myr deviates from a power-law and follows a Schechter function with a characteristic truncation mass of \textcolor{black}{${\rm M}_{c} = 6.2 \times 10^{5} \, {\rm M}_{\odot}$}. This shape of the CMF is more prominent for a subsample of knots in the eastern region of the ring. Over the same age interval, we derive a low rate of disruption \textcolor{black}{($\delta \sim 0.25$)} from the cluster age function and a cluster formation efficiency (CFE) of $\sim$ 3 per cent. In contrast, the CFE in the $1 - 10$ Myr age range is nearly 40 per cent. \textcolor{black}{We note the lack of high-resolution UV and H$\alpha$ observations to help break age-extinction degeneracy which affects the derived ages for dusty young clusters and old ones with low reddening.
Nevertheless, this study has shown, at least to a first-order approximation, that collision-triggered starburst events happening across the CRG} offer an ideal environment for a second generation of young blue knots to form in abundance. It also \textcolor{black}{suggests} that the drop-through collision between the two galaxies can fuel at least mild cluster disruption over time. 

\end{abstract}




\section{Introduction}

Collisional ring galaxies (CRGs) are peculiar objects that represent "a special class" of ring galaxies (RiGs). They are rare in the local universe (only accounting $\approx$\,0.01\,per cent of the observed galaxies, see e.g. \citealt{1985ARA&A..23..147A,2009ApJS..181..572M})
but are expected to be ubiquitous at higher redshifts given the predominance of galaxy collisions. However, \citet{2020NatAs...4..957Y} have reported a scarcity of the most distant CRGs ($z >$\,0.1) as opposed to predictions from previous works \citep[e.g.][]{2004ApJ...612..679L,2008MNRAS.389.1275D}. 
The discrepancy may be due to the challenges of identifying high-{\it z} CRGs that are usually not prone to fragmentation. They are not bright enough to be detected in contrast to the nearby ones hosting blue clumps with ages between $10 - 100$ Myr which make them easily identifiable. This inconsistency in the frequency 
of CRGs remains an open question and is worth to be addressed to better constrain galaxy merger rates. 
 It is also relevant to study RiGs given that local CRGs have striking resemblances with high-{\it z} clumpy galaxies \citep{2005ApJ...631...85E,2006ApJ...651..676E}.  Although the latter have arc-like appearance such as in partial RiGs, they do not have visible disks and with their clumpy structure bluer at all redshifts than those of RiGs. Nearby CRGs are thus of great interest for characterising the supposedly rare high-{\it z} CRGs and in an effort to better understand the origin of star formation (SF) in clumpy systems \citep{2021MNRAS.507.6140I}.

CRGs are formed through the head-on collision between two galaxies, in which a small galaxy companion passes through the larger disk galaxy in its central region \citep[e.g.][]{1976ApJ...209..382L,1977ApJ...212..616T,1996FCPh...16..111A,2018MNRAS.481.2951E}.
As the intruder flies away after the impact, a bright structured ring forms from a radially expanding density wave. With a diameter that can reach up to a hundred kpc \citep{2008MNRAS.386L..38G}, 
the high-density gas in the ring provides ideal sites of enhanced SF, since gas and dust that migrated to form the ring are prone to collapse and produce starburst activities \citep{1976ApJ...209..382L,
1987ApJ...312..566A,2018MNRAS.473..585R}. 
These new episodes of SF, which last at least 100 Myr, are so intense that they favour the formation of young massive clusters (YMCs) and star-forming clumps (the largest units of SF in a galaxy) along the ring. 

YMCs are bright compact objects with a typical age range of $\tau \sim 10 - 100$ Myr and total stellar masses ${\rm M}_{\rm cl} \gtrsim 10^4 \, \text{M}_{\odot}$ (see \citealt{2003dhst.symp..153W}, \citealt{2010ARA&A..48..431P}, \citealt{ 2020SSRv..216...69A} for comprehensive reviews). Although they are a common occurrence in various types of galaxies, 
collisions and mergers between galaxies are known to increase the number of these young blue knots \citep[e.g.][]
{1999AJ....118.1551W,2005ApJ...628..231B,2010AJ....139.1369P,2018MNRAS.473..585R,2013ApJ...775L..38R,2019MNRAS.482.2530R}.
In fact, previous works by \citet{2010MNRAS.405..857G}, \citet{2012MNRAS.426.3008K}, \citet{ 2020SSRv..216...69A} and references therein have reported that the cluster formation efficiency ($\Gamma$ or CFE), which refers to the fraction of SF occurring in bound stellar clusters, increases with an increasing star formation rate (SFR) density ($\Sigma_{\rm SFR}$) of the host galaxy. Hosts with extreme environments are found to be $\sim 3 - 5$ times more efficient in producing YMCs as compared to the cases of gas-poor spiral galaxies. 
Other works by e.g. \citet{2017ApJ...849..128C,2023ApJ...949..116C} and \citet{2023MNRAS.519.3749C}, however, have reported that $\Gamma$ remains constant and independent from the intensity of the host $\Sigma_{\rm SFR}$. Any correlation, in this case, between the two parameters would thus simply be due to a subtle bias from age-extinction degeneracy and/or a mixed-age comparison of observational data. 

In spite of CRGs known to host sites of strong SF with a SFR range of $\approx 0.1 - 20\,{\rm M_{\odot}\,yr^{-1}}$ (e.g. \citealt{1997AJ....113..201A,2006MNRAS.370.1607W,2010MNRAS.408..234M}), 
their HII regions and blue star-forming knots are 
still poorly investigated. Only a handful of observational works can be found in the literature: a sample of 11 CRGs by \citet{1997AJ....113..201A}, a sample of 8 northern CRGs by \citet{1998AJ....116.2757B}, VII Zw 466 by \citet{1978ApJ...224..796T}, Arp 143 by \citet{2009ApJ...693.1650B}, NGC 922 by \citet{2010AJ....139.1369P}, AM 0644$-$741 by \citet{2011ApJ...739...97H} and \citet{2024MNRAS.529.4369G}, FM 287-14 by \citet{2013A&A...559A...8F}, and AM 0035$-$335  a.k.a. the Cartwheel by e.g. \citet{2022MNRAS.514.1689Z} and \citet{2023MNRAS.519.5492M,2024MNRAS.527.2816M}.
The Cartwheel galaxy is one of the most spectacular objects that has been studied extensively at several wavelengths. 
Its peculiar morphology, i.e. prominent inner and outer rings linked together by faint spokes, emerged from the collision of a spiral target with a much smaller intruder $\sim$ 300 Myr ago \citep[e.g.][]{1996ApJ...467..241H}. With a SFR of $\approx 20\,{\rm M_{\odot}\,yr^{-1}}$ \citep{1997AJ....113..201A,2010MNRAS.408..234M}, 
the high concentration of bright HII regions, H$\alpha$ and radio continuum emission in this CRG archetype reveals that SF activity predominantly happens across its rings, especially in the outer ring hosting more than 200 H$\alpha$-emitting knots \citep{1996ApJ...467..241H,2022MNRAS.514.1689Z}. 
Unlike the relatively distant Cartwheel ($D_L \sim$ 133 Mpc), NGC 922 is one of the nearest CRGs known to date ($D_L \sim$ 43 Mpc). 
The encounter with a dwarf compact companion $\sim$ 330 Myr ago resulted in the C-shape morphology of the galaxy along with an off-nuclear ring and a SFR of $\approx 8\,{\rm M_{\odot}\,yr^{-1}}$  \citep{2006MNRAS.370.1607W}.    
\citet{2010AJ....139.1369P} found that the nuclear region of NGC 922 harbours the older massive clusters ($\tau > $\,50 Myr and M$_{\rm cl} > 10^5\,{\rm M}_\odot$), while almost 70 per cent of the population with ages younger than 7 Myr resides in the ring. Based on simulations, the authors predict that the blue knots are prone to cluster disruption and dissolution. 

Cluster analyses of both the Cartwheel and NGC 922 indicate that 
the ring environment likely plays a role in the clump formation and disruption mechanisms of CRGs. Theoretical works by e.g. \citet{2012MNRAS.420.1158M} and \citet{2018MNRAS.473..585R} also reported evidence for enhanced SFRs in rings, which make CRGs potentially excellent birth sites for a large number of blue YMCs.  
How much influence do the expanding ring and the subsequent fragmentation process have in defining the lifetime of these objects? Which types of clusters in terms of their mass, age, and spatial distribution can survive the ring fragmentation (occurring $\sim$ 100 Myr after the impact, \citealt{1977ApJ...212..616T}) 
combined with internal factors (e.g., supernova explosion phase, cluster evolutionary fading)?
Answers to such questions provide valuable insights into the field of star clusters and SF in general. They can be addressed by studying CRGs with an empty ring in particular. Denoted as "RE" by \citet{1976ApJ...208..650T}, 
these atypical and rare CRGs have no distinct component that can be unambiguously identified as a central nucleus.

\begin{table}
\caption{Physical properties of Arp 147.}
\begin{tabular}{l l c}
 \hline \hline
  Parameter & Value & References  \\
  \hline 
  IRAS name & 03087$+$0107 & (1) \\
  Other names & IC 298, CGCG 390-016 & (1) \\
  (Sub)class & RING, RE & (2),\,(3) \\\\
  
  RA (J2000)  & ~~03~11~18.9 & (1) \\
  DEC (J2000) & $+$01~18~53 &  (1) \\\\
  $D_{L}$ & \textcolor{black}{132 Mpc} & \textcolor{black}{(2),\,(4)}\\
  Physical scale & \textcolor{black}{640 pc/arcsec} & \textcolor{black}{(2),\,(4)}\\

 ${\rm M_{ring}/M_{intruder}}$ & 0.57 & \textcolor{black}{(5)} \\
 On sky separation$^a$ & 10 -- 13 kpc & \textcolor{black}{(4)},\,(6)  \\
  
  \hline
  \noalign{\smallskip}  
  {\bf The Ring} \\
  \noalign{\smallskip} 
    Diameter & \textcolor{black}{$\sim$\,12 kpc} & \textcolor{black}{(4)}\\
  Inclination angle & 25$^\circ$ & \textcolor{black}{(4)}\\
  Photometric \& Kinematic PA & 33$^\circ$, 70.5$^\circ$ & \textcolor{black}{(4)} \\
  Expansion velocity & 225 km s$^{-1}$ & \textcolor{black}{(4)} \\\\
  $B-$mag & $\sim$ 15.9 mag & (1) \\
  Stellar mass$^{b}$ & $\sim 6-10 \times 10^{10}\,{\rm M}_{\odot}$ & \textcolor{black}{(4)} \\
      SFR$^a$ & $\sim 4 - 12\,{\rm M_{\odot}~yr^{-1}}$ & (4),\,(5),\,(6)\\ 
        Age$^a$ & \textcolor{black}{$\sim 40 -80$\,Myr} & \textcolor{black}{(4)},\,(6),\,(7)\\
    {\it Z}$^{b}$ & 0.19 -- 0.40 $Z_{\odot}$ & \textcolor{black}{(4)} \\

  \hline
\noalign{\smallskip}
\multicolumn{3}{@{} p{8.5cm} @{}}{\footnotesize{\textcolor{black}{{\it Notes. } $^a$The listed value range of the parameter is drawn by combining the works from multiple references. $^b$The parameter spans a certain range of values as derived by the work from a single reference.} 
{\it Reference list.} (1) NED Database; (2) \citet{1991S&T....82Q.621D}; (3) \citet{1976ApJ...208..650T}; \textcolor{black}{(4)} 
\citetalias{2011MNRAS.417..835F}; \textcolor{black}{(5)} \citet{2008AJ....136.1259R}; (6) \citetalias{2010ApJ...721.1348R}; (7) 
\citet{2012MNRAS.420.1158M}.}}
\end{tabular}
\label{tab:Arp147-prop}
\end{table}

Arp 147 is a well-known prototype of RE galaxies  located at roughly \textcolor{black}{132 Mpc away ($z \sim 0.03$, 1 arcsec corresponds to 640 pc)} (\citealt{1991S&T....82Q.621D,2011MNRAS.417..835F}, hereafter \citetalias{2011MNRAS.417..835F}). Its ring has a stellar mass of $\sim 6-10 \times 10^{10}\,{\rm M}_{\odot}$ and a metallicity of $0.19 - 0.40\,{\rm Z_{\odot}}$
\citepalias{2011MNRAS.417..835F}. The impact of the head-on collision between a spiral galaxy (the target) and its minor companion (the intruder) that occurred \textcolor{black}{ $\approx\,40 - 80$ Myr} ago is believed to have triggered the creation of the ring of gas and stars (\citealt{1992ApJ...399L..51G}; \citetalias{2011MNRAS.417..835F}; \citealt{2012MNRAS.420.1158M}). 
\citetalias{2011MNRAS.417..835F} have conducted spatially resolved kinematic study of Arp 147 to investigate the properties of the ring including its SF activity. They found that the ring is almost circular with a diameter of $\sim$\,12 kpc,  
i.e. relatively larger than most local CRGs  \citep{2009ApJS..181..572M}. They also reported that the ring hosts two (young and old) stellar population components based on H$\alpha$ equivalent width (EW) and a blue colour map. It is believed that the collision triggered young population to form and gets mixed up with a native old stellar population of the parent spiral disk galaxy. 
Finally, the authors derived starburst ages younger than 70 Myr across the ring, which explain the existence of strongly star-forming HII regions, regardless of the age gradient in the recent SF, i.e. the W-SW quadrant is bluer compared to the others with an increasingly redder colour counterclockwise.

\begin{figure*}
\begin{center}
 \resizebox{0.7\hsize}{!}{\includegraphics[trim= 10cm 0cm 10cm 0cm]{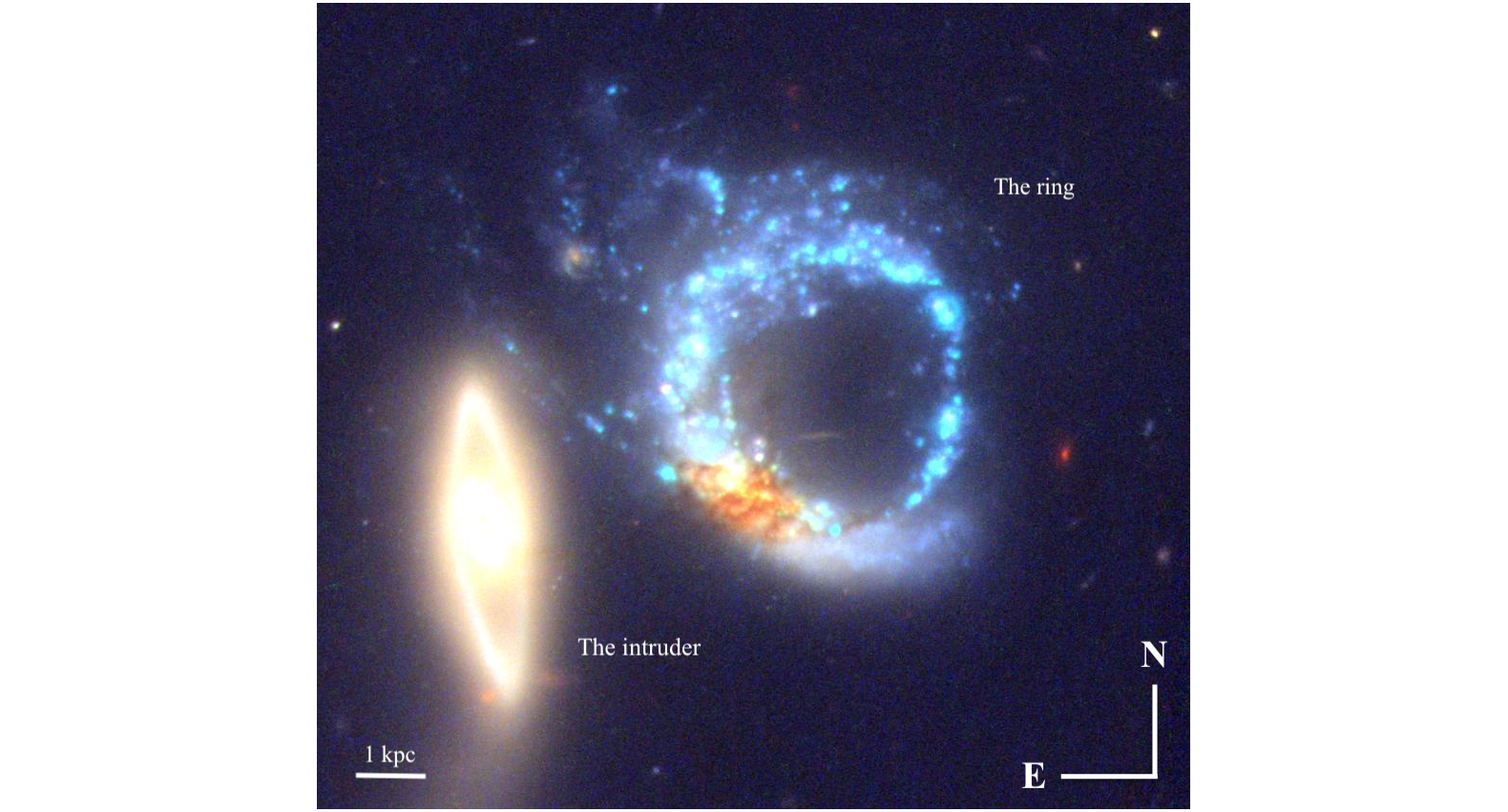}}   
\end{center}
\caption{False colour optical image of Arp 147 taken with the HST/WFPC2 \textcolor{black}{F450W, F606W, and F814W} broadband filters. The intrusion of the red elliptical galaxy (east) into the field of the spiral galaxy tidally stretched the latter into a blue ring-like structure with bright star-forming regions. A plume extending from the N-NE of the ring also hosts trails of blue knots. The reddish area on the S-SE side of the ring is believed to host the nucleus of the original spiral galaxy before its collision with the intruder. The $\sim$ 0.7 by 0.7 arcmin image is oriented with North pointing up and East to the left. The horizontal line marks a scale bar of 1 kpc.} 
\label{fig:3col-HST}
\end{figure*}

\citet{2008AJ....136.1259R} derived a SFR of 4.1 and 8.6 \,${\rm M_{\odot} \,yr^{-1}}$ from H$\alpha$ and far-infrared (FIR) emission, respectively. Other SFR values in a similar range, based on observations in the X-ray  ($\sim\,9 -12 \,{\rm M_{\odot} \,yr^{-1}}$ by \citealt{2010ApJ...721.1348R}, hereafter \citetalias{2010ApJ...721.1348R}),  H$\alpha$ ($4.7\,{\rm M_{\odot} \,yr^{-1}}$ by  \citetalias{2011MNRAS.417..835F}), and mid-IR ($4.1\,{\rm M_{\odot} \,yr^{-1}}$ by \citetalias{2010ApJ...721.1348R}), are also found in the literature. Table\,\ref{tab:Arp147-prop} summarises the general physical properties of Arp 147. Hereafter, we will use the name Arp 147 mainly
referring to the ring of the system.  
Based on the derived values of SFR, which are moderately $\sim 2 -3$ times higher than those of local spirals,
it is not surprising that Arp 147 is a nursery for massive blue star-forming knots \textcolor{black}{and ultraluminous X-ray sources (ULXs).}  The latter sources are off-center accreting black holes with luminosities $L_X \geq \times 10^{39}\,{\rm erg\,s^{-1}}$ \citep[e.g.][]{1989ARA&A..27...87F}. \citetalias{2010ApJ...721.1348R} reported that the most intense starburst activity in Arp 147 may have ended some 15 Myr ago given that none of its 8 ULXs 
is more luminous than  $10^{40}\,{\rm erg\,s^{-1}}$. \textcolor{black}{They also found that the positions of ULXs and star-forming knots across the ring are strongly related in a way that regions with a substantial number of ULXs host as well a large population of blue knots. In particular, the X-ray sources are either near or in a blue knot. However, they did not explore whether the ages and masses of the knots correlate with their ULX counterparts, assuming the apparent proximity between these two SF tracers are real.} 

\textcolor{black}{Based on these previous works,} Arp 147 is a perfect testbed for assessing the role of the intense collision in shaping the key characteristics of these knots. \textcolor{black}{In particular, it is worth checking whether their physical properties are in agreement with the observational studies \citepalias[e.g.][]{2010ApJ...721.1348R,2011MNRAS.417..835F} reporting the SF activity across the ring.
However, there are no detailed photometric analysis of the CRG's individual knots and clumpy regions published in the literature. It is thus necessary to derive diagnostic tools such as cluster mass function (CMF) and cluster age distribution function (CAF) to trace past and ongoing mechanisms triggering their birth and disruption.} 

\textcolor{black}{This paper builds on our preliminary results in \citet{2023afas.confE..83R,2025afas.confE.136R}. Herewith, we provide a substantial expansion of the work which} characterises the properties of massive clusters and star-forming clumps of Arp 147 using HST/WFPC2 data archive from the Mikulski Archive for Space Telescopes (MAST). 
The cluster analysis is complemented with ULX catalogues \citepalias{2010ApJ...721.1348R} and archival {\it Spitzer}/IRAC infrared data (ID 20369) to further investigate the star formation history (SFH) of the CRG \textcolor{black}{on sub-galactic scales}. 
The paper is outlined as follows: in Section \ref{sec:obs}, we describe the HST data and object detection. Sections \ref{sec:select} and \ref{sec:source-props} outline the photometric catalogue and the source properties, respectively. 
We present our results and their implications in Section \ref{sec:more-res}, \textcolor{black}{compare the main findings with other CRG works in Section \ref{sec:compare}}, and provide a summary of this work in Section \ref{sec:conclusion}. \textcolor{black}{Throughout this work, we assume the following cosmological parameters: 
$H_0 = 73\,{\rm km~s^{-1}~ Mpc^{-1}}$, $\Omega_{\rm m} = 0.24$, and $\Omega_{\rm \Lambda} = 1 - \Omega_{\rm m}$ \citep{2007ApJS..170..377S}.} 

\section{DATA AND SOURCE EXTRACTION}\label{sec:obs}

\begin{table*}
\caption{HST/WFPC2 observation log of Arp 147 obtained in 2008 under GO Program 11902 (PI M. Livio).}
\begin{center}
\begin{tabular}{l c c c c c c c}
 \hline \hline
  Filter & Cent.Wavelength & Exp.time & PSF/FWHM & $A^G_{\rm \lambda}$ & $m_0, a_c$ & mag.lim & N \\
  & (\AA) & (s) & (pix) & (mag) & (mag, mag) & (mag) &($\sigma_m \leq 0.20$ mag)  \\ 
  \hline 
  F450W  & 4556 & 8800 & 3.13 & 0.322  & 21.987, $-0.906$ & 26.62 & 253 \\
  F606W  & 5997 & 6600 & 3.26 & 0.228 & 22.887, $-1.017$ & \textcolor{black}{26.14} & 267 \\
  
  F814W & 8012 & 6600 & 3.24 & 0.146 & 21.639, $-1.017$ & \textcolor{black}{25.64} & 236
  \\

  \hline
\end{tabular}
\label{tab:data}
\end{center} 
\end{table*}

\subsection{HST observations}
We retrieved F450W, F606W, and F814W  
 science images of Arp\,147 from MAST (GO Program: 11902, PI M. Livio).  With a spatial sampling of 0.046 arcsec \textcolor{black}{(equivalent to $\sim 29$\,pc at the distance $D_L = 132$\,Mpc)}, these images observed with the HST/WFPC2 Planetary Camera (PC) were mainly used to demonstrate science capabilities of the instrument after a brief hiatus in science operation. They are already processed and calibrated by the HST pipeline reduction software. Figure\,\ref{fig:3col-HST} shows a RGB composite of Arp 147 with \textcolor{black}{F450W, F606W, and F814W filters} in the blue, green, and red channels, respectively. Star-forming regions can be identified as bright blue stellar knots populating the ring and the plume extending from the N$-$NE region of the galaxy. The reddish area on the S$-$SE side of the ring is believed to host the displaced remnant nucleus of the original spiral galaxy before the elliptical intruder on the east region of the field has tidally stretched it into the observed ring-like structure 
\citep{1992ApJ...399L..51G,2012MNRAS.420.1158M}.  Table\,\ref{tab:data} provides a summary of the HST observations used in this work. 
The average PSF/FWHM resolution of a point source is $\sim 0.15$ arcsec in all three filters.

\subsection{Object extraction and ring identification}\label{sec:phot}
 We run {\tt SExtractor} \citep{1996A&AS..117..393B} on the unsharp-masked\footnote{Unsharp-masking is a common technique used to enhance the contrast of an image so that diffuse and extended sources become sharper and stand a chance to be detected.} version of \textcolor{black}{an image combining F450W and F814W observations} to optimise source detection. We co-added images 
from these two filters given that \textcolor{black}{F450W} has the deepest exposure (8800\,s) while \textcolor{black}{F814W}  observations are less affected by dust and extinction. Besides activating standard deblending parameters to extract sources from crowded regions of the ring, we also applied a minimum detection limit of 3.5$\sigma$ above the background (bg) rms noise as well as a minimum of 9 contiguous pixels.
These best fit parameters led to the detection of 258 objects that are mainly located in the ring and the plume of the system, i.e.\,within a circle radius of $\sim 17$ arcsec from the centre of the ring\footnote{We estimate the centre of the ring by eye where RA = 03:11:18.3 and DEC = $+$01:18:56 in J2000 coordinates.}. 
Had we increased the detection threshold to 5$\sigma$, we would have missed $\sim$ 25 per cent of potential source candidates. Further visual inspection of the composite image 
was done to ensure that all potential cluster candidates were included in the output catalogue. 

 We adopt the following steps to define the ring and subsequently help separate the selected objects associated to the ring and other regions such as the plume (see Section \ref{sec:cat}): {\it i)} we first perform a visual identification of the ring from the multicolour image to serve as a starting point 
when defining the ring in a more robust way; {\it ii)} we then derive a background map of the CRG using {\tt SExtractor} and draw different contours to highlight its morphology; {\it iii)} we associate any region with a bg value above (below) 2$\sigma$ 
as part of the ring (outer region). We define the bg value of an object as the median value of a 5 by 5 pixel background region centered at its coordinates.

\subsection{Aperture photometry}\label{sec:box}
We use {\tt IRAF/PHOT} task
to perform aperture photometry on each of the original \textcolor{black}{multiband} images. The source instrumental magnitudes were computed using a fixed aperture radius of 2 pixels (0.1 arcsec) and sky background annuli between 5 and 8 pixels (0.25 and 0.4 arcsec, respectively). The small aperture radius is chosen to avoid sampling of nearby sources in the crowded regions of the ring.
To correct for the missing flux from the fixed aperture radius, we calculate aperture correction, $a_c$, for each band based on the curve-of-growth of 4 bright and isolated sources in the field. The value of $a_c$ is derived by taking the median of the difference in magnitudes between 2 pixels to infinity (equivalent to $\sim$ 10 pixels or 0.5 arcsec in this work).

The Vega mag zero point, $m_0$, and the Galactic foreground extinction, $A^G_{\rm \lambda}$, were retrieved online and also applied to the magnitudes of each source. 
The values of the aperture correction, the zero point, and the Galactic extinction for the \textcolor{black}{F450W, F606W, and F814W filters}  are listed in Table \ref{tab:data}. Uncertainties in the absolute magnitudes result from the {\tt IRAF/PHOT} photometric errors added in quadrature with the uncertainties in $m_0$ and the estimated $a_c$. 

\section{Source selection and age modelling }\label{sec:select}

\subsection{Terminology and source catalogue}\label{sec:cat}
At the distance of Arp 147, most of the detected sources are YMC complexes  
rather than individual YMCs \citep{2013MNRAS.431..554R}. 
Hereafter, we refer to both objects as blue knots. Among the star-forming regions spread across the ring are 6 large-scale star-forming complexes which we refer to as clumps in this work. In nearby galaxies, blue kpc-sized clumps, which incorporate many smaller complexes and individual YMCs, host the largest units of SF with masses generally above $10^6\,{\rm M}_{\odot}$ \citep[e.g.][]
{2006ApJ...651..676E,2010AJ....139.1369P}. They are worth studying since they represent the smallest structures that can be resolved in high-{\it z} clumpy galaxies \citep[e.g.][]{2005ApJ...631...85E,2006MNRAS.370.1607W}. 

To identify the knots in Arp 147, we follow the selection process developed by \citet{2013ApJ...775L..38R,2013MNRAS.431..554R}. The main difference arises from cross-matching the resulting coordinates of the detected objects in all three filters instead of relying on a criterion based on the concentration index CI vs. FWHM plot. The value of CI is no longer reliable at the distance of the CRG \textcolor{black}{($D_L \sim 130$ Mpc)} where clusters mostly resemble point-like sources \citep{2022MNRAS.513.4232R}. Prior to cross-identification, selection steps include rejection of objects detected outside the frame in Fig. \ref{fig:3col-HST} and removal of foreground Milky Way stars and background galaxies.

We only retain sources with SNR $ \geq$ 5 ($\sigma_m \leq 0.20$ mag)
in all three filters to construct a robust catalogue of high confidence photometric candidates. We end our source selection by visually inspecting the output catalogue to remove any remaining false detections and sources associated with the intruder. 
Out of the 211 selected objects, 81 per cent (171) belong to the ring, and the rest (40) are found elsewhere, including the plume on the N-NE side: they are, respectively, flagged as 0 and 1 in the online photometric catalogue as sampled in Table \ref{tab:knots-cat}. We note that the classification is based on the corresponding bg value of the source as derived in Section \ref{sec:phot}. 

After identifying the knots, we identify a \textit{clump}  as any kpc-scale star-forming region fulfilling the following conditions: {\it i)} firstly, a star-forming unit area larger than 0.2 kpc$^2$, i.e., a box size larger than \textcolor{black}{$\sim 440$ pc} ($\sim$\,0.7 arcsec). We chose this threshold using flux contour cuts on the \textcolor{black}{F606W} image and 
for each clump to include at least two bright knots with an \textcolor{black}{F606W} magnitude brighter than \textcolor{black}{\m12.3} mag located within 0.75 arcsec from each other. It is important to detect the clumps in the the \textcolor{black}{F606W image}  to avoid undersampling at redder wavelengths due to the clumps' lower contrast with the galaxy's disk profile \citep{2012ApJ...753..114W}. {\it ii)}. Secondly, the closest neighbouring clump is located at least 2.5 arcsec away. 
Because of a Jeans instability, star-forming clumps are found to be regularly spaced in circumnuclear rings, which represent an environment similar to the CRG ring \citep[e.g.][]{1993AJ....105.1344B,1997ApJ...480..235E}. {\it iii)} Thirdly, the corresponding initial extinction estimate $A_V^0$ of the region is below 1.3 mag to omit any clump located in the reddish area of the ring. This area not only has a highly variable background but it is also prone to large age and mass uncertainties due to age-extinction degeneracy, \textcolor{black}{which is a fundamental bias in cluster analysis and stellar population modelling. For instance, because of the degeneracy, the colour and derived age of a dusty young cluster resemble those of a genuinely older one without extinction.} Refer to Section \ref{sec:fit} on how to derive $A_V^0$ \textcolor{black}{-- a parameter used to restrict the allowed extinction range in order to minimise age-extinction degeneracy during cluster age fitting}. 
We end our identification with a visual inspection of the shortlisted candidates for quality control. 

\begin{figure}
\begin{center} 
  \resizebox{1.0\hsize}{!}{\includegraphics[trim= 0cm 0.7cm 0cm 0cm]{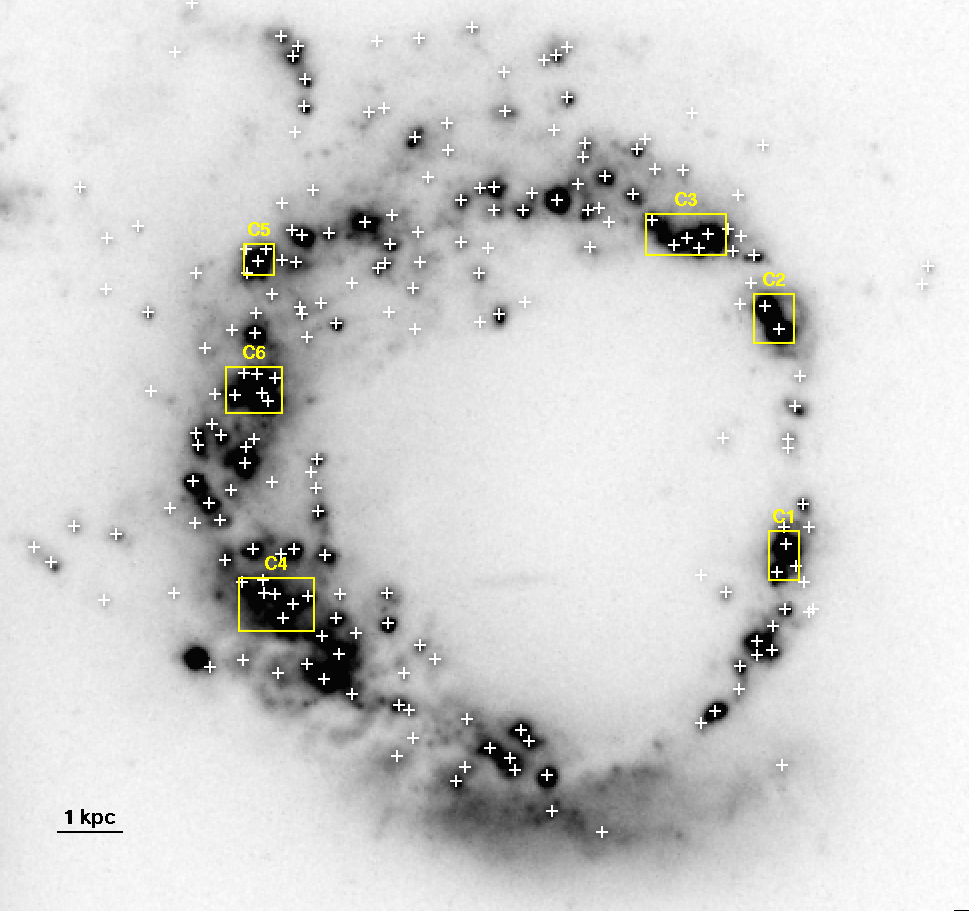}}  
\end{center}
\caption{Spatial distributions of the knots (crosses) and the kpc-sized clumps (open squares) on the \textcolor{black}{F606W image} of Arp 147. The horizontal line marks a scale bar of 1 kpc.}
\label{fig:clumps}
\end{figure}

The open squares in Fig. \ref{fig:clumps} represent the positions of the selected clumps across the ring, while the crosses denote the spatial distribution of the knots. 
Finally, we perform box photometry on each clump to derive the corresponding \textcolor{black}{multiband} magnitudes. Their ages and masses are computed in Section \ref{sec:fit} using least-square spectral energy distribution (SED) fitting.  Table \ref{tab:clumps-cat} summarises the properties of the selected clumps which are reported in Section \ref{sec:clumps-res}. \textcolor{black}{We note that some of these objects could be giant star-forming HII regions, i.e. ionised regions of the most recent SF from the intense UV radiation of newly formed massive stars,  rather than stellar clumps. In fact, using SWIFT integral field spectrograph (IFS), \citetalias{2011MNRAS.417..835F} have reported in their work  that both H$\alpha$ luminosities and electron densities 
in the four quadrants of the ring have typical values for star-forming HII regions, especially the ones in the eastern part. The optical IFS data, however, are inadequate to measure e.g. EW\,(H$\alpha$) in smaller regions like our clumps due to the poor seeing conditions and the coarse spaxel scale of the instrument. Follow-up observations with a high-resolution spectrograph such as VLT/MUSE (to analyse the abundance of nebular emission lines like H$\alpha$ and H$\beta$, see e.g. \citealt{2022MNRAS.514.1689Z}) or JWST/MIRI (to trace the IR dust emission from PAH molecules, see e.g. \citealt{2023ApJ...944L..16E}) should be conducted to confirm the nature of these bright kpc-sized objects.} 

\begin{table*}
\caption{General properties of the blue knots in Arp 147 listed in order of increasing RA. The full table is available online in a machine-readable format.}
\begin{center}
\begin{tabular}{l c c c c c c c c c}
 \hline \hline
  ID & RA & DEC  & F450W & F606W & F814W & log\,($\tau$) & log\,(${\rm M_{cl}}$) & $A_V$ & Flag \\
  & (deg) & (deg) & (mag) & (mag) & (mag) & (yr) &(M$_{\odot}$) & (mag) & (0 or 1) \\ 
  \hline 

     1 & 47.823622 & 1.316880 & 24.93 $\pm$ 0.09 & 24.60 $\pm$ 0.06 & 24.59 $\pm$ 0.11 & 6.85 & 4.90 & 0.73 & 1 \\
     2 & 47.823664 & 1.316752 & 25.39 $\pm$ 0.11 & 25.28 $\pm$ 0.09 & 24.91 $\pm$ 0.13 & 7.30 & 4.98 & \textcolor{black}{0.44} & 1 \\
     3 & 47.824436 & 1.314504 & 25.60 $\pm$ 0.13 & 25.56 $\pm$ 0.18 & 25.13 $\pm$ 0.18 & 7.30 & 4.87 & 0.43 & 0 \\
     4 & 47.824459 & 1.315070 & 24.39 $\pm$ 0.11 & 23.98 $\pm$ 0.15 & 24.15 $\pm$ 0.17 & 6.90 & 4.79 & 0.25 & 0 \\
     5 & 47.824462 & 1.314486 & 25.56 $\pm$ 0.15 & 25.27 $\pm$ 0.13 & 24.89 $\pm$ 0.15 & 7.48 & 5.28 & 0.67 & 0 \\
     6 & 47.824496 & 1.314693 & 25.20 $\pm$ 0.13 & 24.80 $\pm$ 0.13 & 24.40 $\pm$ 0.11 & 8.48 & 5.78 & \textcolor{black}{0.17} & 0 \\
     7 & 47.824503 & 1.315232 & 23.19 $\pm$ 0.05 & 22.78 $\pm$ 0.03 & 23.60 $\pm$ 0.07 & 6.30 & 4.93 & 0.00 & 0 \\
     8 & 47.824517 & 1.316116 & 24.96 $\pm$ 0.09 & 24.70 $\pm$ 0.08 & 24.48 $\pm$ 0.11 & 6.85 & \textcolor{black}{4.33} & \textcolor{black}{0.02} & 0 \\
     9 & 47.824551 & 1.315903 & 23.85 $\pm$ 0.06 & 23.71 $\pm$ 0.05 & 24.00 $\pm$ 0.09 & 6.60 & 4.87 & 0.25 & 0 \\
     10 & 47.824555 & 1.314801 & 24.06 $\pm$ 0.11 & 23.53 $\pm$ 0.06 & 23.45 $\pm$ 0.09 & 6.85 & \textcolor{black}{4.84} & \textcolor{black}{0.14} & 0 \\
      \hline

  \multicolumn{10}{@{} p{14.5cm} @{}}{\footnotesize{{\it Notes. }  Column 1: source identification; columns 2 \& 3: J2000 equatorial coordinates; columns $4-6$: Vega-based apparent magnitudes with their uncertainties in \textcolor{black}{F450W, F606W, and F814W} bands, respectively;  columns 7 \& 8: the source age and mass in logarithmic base; column 9: the estimated visual extinction;  column 10: additional information about the source position in the field of Arp 147 where 0 refers to any source inside the ring and 1 elsewhere including the plume.}}

\end{tabular}
\label{tab:knots-cat}
\end{center} 
\end{table*}

\begin{table*}
\caption{Physical parameters of the clumpy regions.}
\begin{center}
\begin{tabular}{l c c c c  c c c c c c}
 \hline \hline
  ID & RA & DEC  & \textcolor{black}{Area} & \textcolor{black}{$M_{\rm F606W}^{\rm brightest}$} & F450W & F606W & F814W & log\,($\tau$) & log\,(${\rm M_{CL}}$) & $A_V$  \\ 
  & (deg) & (deg) & (kpc$^2$) & (mag) & (mag) & (mag) & (mag) & (yr) & (M$_{\odot}$) & (mag) \\ 
  \hline 

    C1 & 47.824708	& 1.314946 & \textcolor{black}{0.33} &  \textcolor{black}{\m13.60} & 20.44 & 20.18 & 20.28 & 6.90 & 6.28 &  0.25\\
    
    C2 & 47.824777	& 1.316591 & \textcolor{black}{0.42} & \textcolor{black}{\m14.66} &19.80 & 19.67 & 19.94 &  6.60 &  6.40 &  0.25\\
    
    C3 & 47.825391	& 1.317175 & \textcolor{black}{0.72} & \textcolor{black}{\m14.17} &19.61 & 19.38 & 19.60 & 6.60 &  6.48 &  0.25\\
    
    C4 & 47.828234	& 1.314605 & \textcolor{black}{0.88} & \textcolor{black}{\m12.63} &20.11 & 19.96 & 19.83 & 6.85 &  6.83 &  0.83\\
    
    C5 & 47.828356	& 1.316999 & \textcolor{black}{0.20} & \textcolor{black}{\m13.35} &20.88 & 20.74 & 20.44 & 6.85 &  5.95 &  0.06\\
    
    C6 & 47.828392	& 1.316096 & \textcolor{black}{ 0.56} & \textcolor{black}{\m13.98} &19.88 & 19.60 & 19.39 & 6.85 &  6.37 &  0.09\\

  \hline
  \multicolumn{11}{@{} p{14.5cm} @{}}{\footnotesize{{\it Notes. }  Column 1: source identification; columns 2 \& 3: J2000 equatorial coordinates; column 4: area of the clump; column 5: Vega-based \textcolor{black}{F606W} absolute magnitude of the brightest knot within the clump; columns $6-8$: Vega-based apparent magnitudes of the clump in \textcolor{black}{F450W, F606W, and F814W} bands, respectively; columns 9 \& 10: the fitted age and mass of the clump in logarithmic base; column 11: the estimated visual extinction.}} 
\end{tabular}
\label{tab:clumps-cat}
\end{center} 
\end{table*}

\subsection{Completeness limits}
\label{sec:comp}
\begin{figure}
\centering
\includegraphics[width=0.45\textwidth]{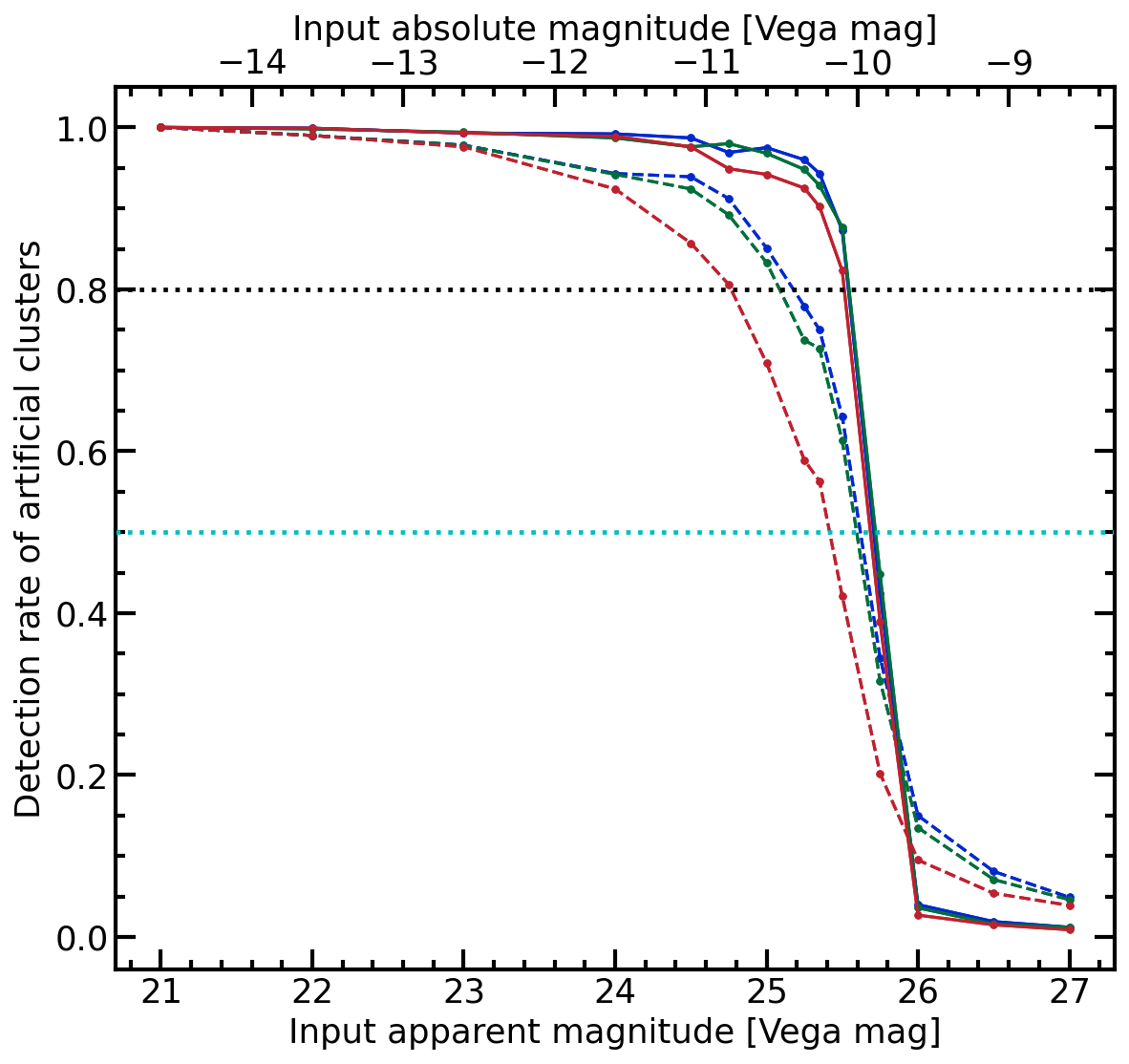}
\caption{Recovered completeness fractions after applying a $\sigma_m$ cut to the detected catalogues in the \textcolor{black}{F450W} (blue lines), \textcolor{black}{F606W} (green), and \textcolor{black}{F814W} (red) images.
Solid and dashed lines show the recovery rates outside and inside the ring, respectively. 
The horizontal lines denote the 50 (cyan) and 80 (black) per cent completeness limits.}
\label{fig:comp}
\end{figure}

We estimate the detection limits and test the reliability of our source identification and selection algorithm by running custom-made Monte Carlo completeness simulations. The full description of the simulation code can be found in \citet{2013MNRAS.431..554R,2019MNRAS.482.2530R}. Briefly, the code is composed of five consecutive steps: creation of artificial clusters from point-source PSF models, object detection, aperture photometry and calibration, source selection, and final outputs. We use {\tt IRAF/PSF} task to construct the PSF models from a catalogue of bright and isolated stars in the field. By adopting the same procedures, we compute the completeness limits of our data in the \textcolor{black}{F450W, F606W, and F814W} images between 21 and 27 magnitude range. For each filter, we run the pipeline twice by considering regions inside and outside (except the companion) the ring.
Fig. \ref{fig:comp} shows the predicted completeness fractions of the source catalogue in the \textcolor{black}{F450W} (blue lines), \textcolor{black}{F606W} (green), and \textcolor{black}{F814W} (red) images. The solid and dashed lines correspond to the recovered rates for the regions outside and then inside the ring, respectively. 
We find that the photometric catalogue in Section \ref{sec:cat} is 80 per cent complete down to \textcolor{black}{$m_{\rm F450W} =  25.2$, $m_{\rm F606W} = 25.1$ and $m_{\rm F814W} = 24.8$ mag. If we were to consider a blue knot at 10 Myr of age, these magnitude limits respectively translate to lower mass limits of $2.6 \times 10^4 \ {\rm M}_{\odot}$, $2.9 \times 10^4 \ {\rm M}_{\odot}$ and $4.1 \times 10^4 \ {\rm M}_{\odot}$, assuming the mass-to-light ratio at that age and {\tt Yggdrasil} models \citep{2011ApJ...740...13Z}. The set of parameters used to retrieve the models are presented in Section \ref{sec:fit}.} 

\subsection{Age and mass least-square fitting}\label{sec:fit}

 We derive the ages and masses of the star-forming knots and the kpc-sized clumps using the least-square fitting technique fully described in \citet{2019MNRAS.482.2530R} following \citet{2003A&A...397..473B} and \citet{2004MNRAS.347..196A}. The code is based on a one-to-one comparison between a set of observed magnitudes (from \textcolor{black}{F450W, F606W, and F814W} filters in this work) and a grid of single-age stellar population (SSP) models. We use {\tt Yggdrasil} evolutionary tracks with ages between 1 Myr and 3 Gyr to determine the best fit parameters \citep{2011ApJ...740...13Z}. 
We assume a Kroupa initial mass function (IMF), an instantaneous burst, and PADOVA AGB isochrones.
We also adopt a starburst attenuation law of $R_V$ = 4.05 \citep{2000ApJ...533..682C}, a metallicity of $Z = 0.004$ and then $Z = 0.008$,  
and a gas covering factor $f_{\rm cov}$\footnote{This input parameter regulates the relative contribution of 
photoionised gas from nebular emission and continuum to the predicted SED.} 
= 0, 0.5 and 1. \textcolor{black}{Unlike $f_{\rm cov}$ where we explore all options available, we only retrieve models assuming these two metallicities to align with the derived values of $Z = 0.19 - 0.40\,Z_{\odot}$ (i.e. $Z \sim 0.004 - 0.008$ for $Z_{\odot} = 0.02$) for Arp 147 \citepalias{2011MNRAS.417..835F}. Nonetheless, we note that previous works by e.g. \citet{2014ApJ...787..142G} and \citet{2017ApJ...847..112D} have reported a higher metallicity of $Z \sim 0.02$ (i.e. solar) in their YMC studies. Such values are indeed expected if red supergiants (RSGs) dominate the NIR light of a $\sim$ 7 Myr old cluster. 
The resulting output parameters depending on the values of $Z$ and $f_{\rm cov}$ are investigated in Section \ref{sec:res-age}.}

In an effort to better estimate the extinction $A_V$ for each source, \textcolor{black}{we account for the uneven distribution of the extinction across the ring. This is done by} constraining the input extinction range as a function of an initial extinction parameter $A_V^0$. The value of $A_V^0$ is computed from the extinction map in Fig. \ref{fig:Av-map} that we built based on a broadband \textcolor{black}{$\rm F450W - F814W$} colour map with a smoothed background. Further details of the method can be found in \citet{2019MNRAS.482.2530R}. Based on the \textcolor{black}{derived} extinction map, we expect star-forming knots in the S-SE region (potential host of the nuclear remnant) to be highly extinguished ($A_V^0 >$ 1.5 mag) while the rest of the objects mostly have $A_V^0 <$ 0.5 mag. \textcolor{black}{We note that our approach of assigning an initial extinction parameter to each knot only minimises but does not entirely resolve inaccurate ages due to reddening.}

\section{Global properties of the blue knots and kpc-sized clumps}\label{sec:source-props}

\subsection{Colours}\label{sec:color}
\begin{figure*}
\centering
\begin{tabular}{c}
  \resizebox{.9\hsize}{!}{
\includegraphics{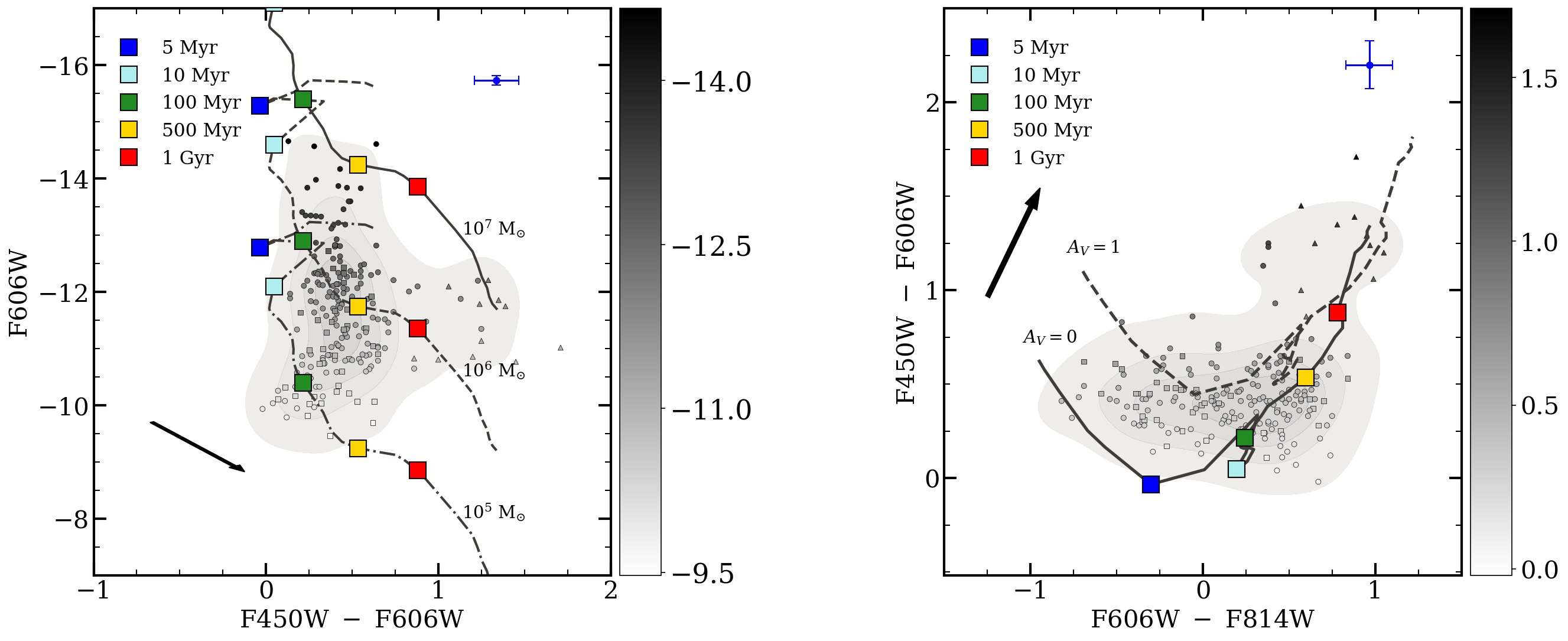}}
\end{tabular}

\caption{CMD (\textcolor{black}{F606W vs $\rm F450W - F606W$}, left) and CCD (\textcolor{black}{$\rm F450W - F606W$ vs $\rm F606W - F814W$}, right) of the star-forming knots inside (circles) and outside \textcolor{black}{(squares)} the ring of Arp 147. \textcolor{black}{Solid triangles} represent the knots in the dusty S-SE region. Density contour plots are overlaid to show the distribution of the clusters. \textcolor{black}{The greyscale bars represent F606W magnitude ($\rm F450W - F606W$ colour index) where darker shades correspond to brighter (bluer) knots and lighter shades indicating fainter (redder) ones}. Grey lines show {\tt Yggdrasil} SSP models of synthetic clusters with $Z = 0.008$, $f_{\rm cov}$ = 0.5, and masses of $ 10^5$ (dash-dotted), $ 10^6$ (dashed), and $ 10^7 \ {\rm M}_{\odot}$ (solid) overplotted on the CMD for comparison. The solid and dashed lines on top of the CCD respectively represent the models with $A_V = 0$ and $A_V = 1$ assuming a starburst attenuation law. Few key ages between 5 Myr and 1 Gyr of the predicted evolutionary tracks are also indicated as multicolour squares. 
The arrows indicate the directions in which the data points will move with a reddening $A_V = 1$. The line plots on the top right of each panel denote the median values of the associated error bars to the data points.}
\label{fig:cd}
\end{figure*}

Figure \ref{fig:cd}, left panel, shows the \textcolor{black}{F606W vs $\rm F450W - F606W$} colour magnitude diagram (CMD)  for the selected knots. Sources flagged as 0 and 1 in Table \ref{tab:knots-cat} are represented by the solid circles and the \textcolor{black}{solid squares}, respectively.
Knots embedded in the reddish area of the ring are shown as \textcolor{black}{solid triangles}. Dash-dotted, dashed and solid lines overplotted on the CMD denote {\tt Yggdrasil} models assuming synthetic clusters with masses of $ 10^5$, $ 10^6$ and $10^7 \ \text{M}_{\odot}$, respectively. They were retrieved using $Z$ = 0.008 and $f_{\rm cov}$ = 0.5. Different ages of the synthetic cluster are marked with multicolour squares 
between 5 Myr to 1 Gyr. The data points are mostly enclosed in between the evolutionary tracks
associated with masses M$_{\rm cl} = 10^5 - 10^7~\text{M}_{\odot}$. From this first order approximation, Arp 147 seems to host the most massive form of SF (M$_{\rm cl} > 10^4~\text{M}_{\odot}$),  even if we were to consider SSP models with $A_V = 1$. 

The right panel of Fig. \ref{fig:cd} displays  \textcolor{black}{$\rm F450W - F606W$ vs $\rm F606W - F814W$} colour-colour diagram (CCD). {\tt Yggdrasil} models with the same input parameters as in the CMD are overplotted on the CCD for an extinction $A_V = 0$ (solid line) and $A_V = 1$ (dashed line). \textcolor{black}{$\rm F450W - F606W$} colours of the sources lie between 0 and 1.7 with a mean value of $\sim 0.5$ as illustrated by the solid contour plots. The CCD plot suggests that blue knots with ages 
as young as $\approx 5 - 10$ Myr mostly dominate the cluster population of the CRG. In addition, the \textcolor{black}{solid triangles} are concentrated in the \textcolor{black}{upper} right quadrant of the CCD ($\rm 0.86 < \textcolor{black}{F450W - F606W} < 1.71$ and $\rm 0.57 <\textcolor{black}{F606W - F814W} < 1.05$), which suggest that the reddish area hosts highly extinguished knots with $A_V > 1$. 
The consistency with which the data agree with SSP models is generally the same regardless of the set of {\tt Yggdrasil} models used. 
The main difference occurs at $\tau \lesssim  5 - 10$ Myr. This is not surprising given the sensitivity of {\tt Yggdrasil} to the relative contribution of nebular line + continuum emission 
from bright massive stars of the youngest clusters (see Section \ref{sec:which-Ygg}). 

\textcolor{black}{We note that using optical F450W, F606W, and F814W filters alone cannot break age-extinction degeneracy.} Estimated ages of the knots based on CCD and/or SED fitting tend to be younger or older than their real ages. One would require UV observations to derive more accurate results, especially for cases where $\tau <$ 10 Myr \citep{2015AJ....149...51C}.
\textcolor{black}{They are sensitive to young light primarily emitted by newly formed massive OB stars still embedded in their dust cocoons. UV data can thus help distinguish between highly extinguished young clusters and any older population with low reddening. There are publicly available UV data e.g. from {\it Galex} and SWIFT/UVOT in MAST. However, with an angular resolution of $\sim 5$ arcsec, which translates to a physical scale of {\bf $\sim 3$} kpc at the distance of Arp 147, they cannot resolve our individual knots as in the high-resolution HST observations. Although we could not extract the UV magnitude of each detected knot, we still use these archival data to cross-validate our derived ages with the spatial distribution of the UV emission imaged across the ring (see Section \ref{sec:sed-res}). For instance, clusters with derived ages younger than 10 Myr are expected to inhabit regions with strong UV emission \citep{2015AJ....149...51C}.} 

\textcolor{black}{In addition to broadband UV observations, including a narrowband filter that traces H$\alpha$ emission line is also highly recommended to get robust results during SED fitting. H$\alpha$ emission originates from massive and short-lived O-stars by ionising the surrounding hydrogen gas clouds.  \citet{2020ApJ...889..154W} have shown that H$\alpha$ photometry is more important than broadband UV measurements to resolve age-extinction degeneracy when age dating old red clusters ($>$ 100 Myr) in actively star-forming galaxies. The inclusion of H$\alpha$ data prevents these sources from being misclassified as dusty knots younger than 10 Myr, especially if the extinction is a free parameter that cannot be constrained in the least-square fitting. Red clusters with concentrated or partially exposed H$\alpha$ are expected to have young ages (5 - 10 Myr) given that their redness arises from extinction. Because of the low resolution of available H$\alpha$ observations to date (with a physical scale as large as $\sim$ 400 pc), which were used by \citet{2008AJ....136.1259R} and 
\citetalias{2011MNRAS.417..835F}, it is not possible to conduct a one-to-one identification of knots with or without H$\alpha$ emission. Nevertheless, the published spatial distributions of H$\alpha$ on a global scale can also serve as a benchmark 
for validating the fitted ages reported in Section \ref{sec:sed-res}. For instance, we expect a substantial number of knots younger than 10 Myr in the quadrant of the ring with the largest EW(H$\alpha$) (see Table 3 in \citetalias{2011MNRAS.417..835F})}. 

The \textcolor{black}{$\rm F450W - F814W$} colour range varies from $-$0.4 to 2.6. Fig. \ref{fig:Av-map} also displays the spatial distribution of the detected knots as a function of  their \textcolor{black}{$\rm F450W - F814W$} colour. Blue knots with \textcolor{black}{$\rm F450W - F814W$} $\leq 0$ dominate the northern part of the ring while the red ones (\textcolor{black}{$\rm F450W - F814W$} $> 1$) are mostly found in the eastern region.  The trend is still prominent even if we adopt slightly different cutoffs while defining the colour-coded labels.
If not caused by the reddening effect, the distribution in the northern region is a reflection of ongoing SF activity, \textcolor{black}{and/or a combination of both}. The prominence of dust in the eastern region, however, explains the concentration of red sources in that area, especially in the S-SE. We explore further the reasons behind the colour gradient in Section \ref{sec:sub-gal}.

\begin{figure}
\begin{center}
\begin{tabular}{c}
 \resizebox{.8\hsize}{!}{\includegraphics[trim= 4cm 0cm 4cm 0cm]{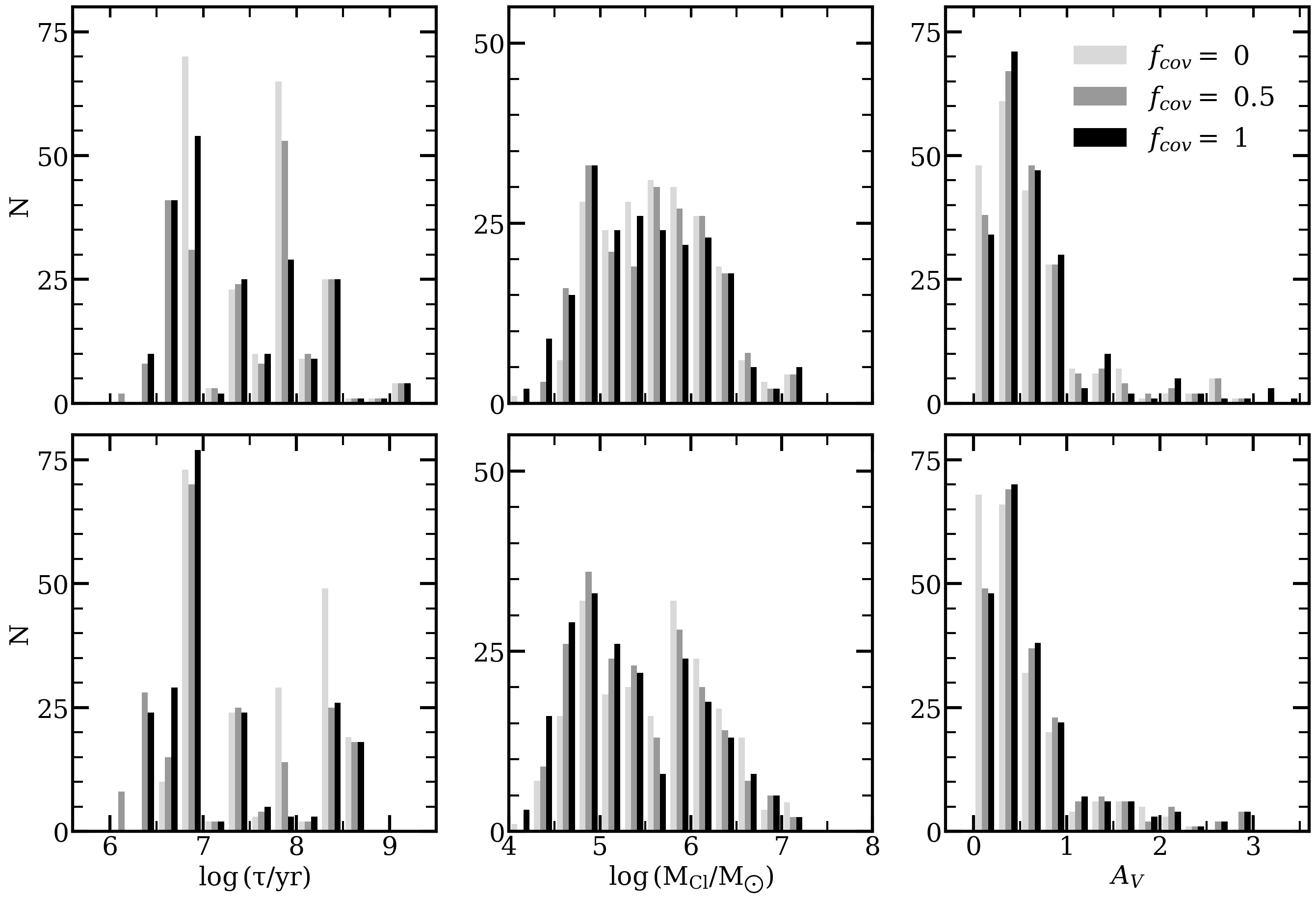}} 
\end{tabular}
\end{center}
\caption{Distributions of the derived cluster ages (left), masses (middle), and extinctions (right) depending on the assumed metallicity $Z$ and the gas covering factor $f_{\rm cov}$. Top and bottom panels correspond to metallicities of $Z = 0.004$ and  $Z = 0.008$, respectively.} 
\label{fig:compare-out}
\end{figure}

\subsection{Ages, masses and extinctions of the blue knots}\label{sec:res-age}

\subsubsection{Which {\tt Yggdrasil} models to consider?}
\label{sec:which-Ygg}
Figure \ref{fig:compare-out} compares the derived ages (left panel), masses (middle), and extinctions (right) assuming {\tt Yggdrasil} SSP models with a metallicity of $Z = 0.004$ (top panel) and $Z = 0.008$ (bottom). Outputs using different values of $f_{\rm cov}$ are also overplotted for each distribution. We find that, irrespective of the input parameters $Z$ and $f_{\rm cov}$, more than one third of the cluster population are younger than 10 Myr. 
The SSP model with $Z = 0.008$ and $f_{\rm cov} = 1$ returns the highest number (62 per cent) of clusters with $\tau <$ 10 Myr.  
We make note, however, of a narrow  "chimney" near log\,$(\tau/{\rm yr}) \sim 6.8$, instead of the cluster ages being evenly spread out over a wider time interval.
This high concentration can be explained by the appearance of RSGs which dominate the stellar population ages around this time \citep[e.g.][]{2013MNRAS.431.2917D,2019MNRAS.482.2530R}. In terms of mass, the mass-metallicity degeneracy does not seem to have a strong influence on the derived values for all $f_{\rm cov}$ model parameters. There is no prominent trend of increasing cluster mass for higher metallicities. In fact, the median values of the derived cluster masses for $Z = 0.008$ tend to be smaller than the ones for $Z = 0.004$, regardless of the values of $f_{\rm cov}$ and whether or not we apply an upper mass cutoff. 
Finally, distributions of the derived extinctions are very similar for all models. At least half of the population have $A_V < 0.5$.

It is not surprising to observe different age distributions for clusters younger than 10 Myr. 
Cluster SEDs are sensitive to the relative contribution from nebular continuum and emission at the youngest ages 
 \citep{2001A&A...375..814Z}. 
SSP models that take into account this factor ($f_{\rm cov} = 1$) thus return more objects younger than $\approx$ 3 Myr than the ones without nebular treatment ($f_{\rm cov} = 0$).  For this reason, our main analysis and discussion are based on results considering models with a gas covering factor of $f_{\rm cov}$ = 0.5 which is a good compromise between the two cases. 
\textcolor{black}{As for metallicity, we consider, hereafter, the best fit results (that correspond to the minimum $\chi^2$ value) from the two SSP models of $Z = 0.004$ and $Z$ = 0.008. Given that these values were chosen to be consistent with \citetalias{2011MNRAS.417..835F}'s stellar population analysis of the CRG, we can then minimise bias while comparing our findings with their work in Sections \ref{sec:sed-res} and \ref{sec:az-age}.  
We note that adopting an SSP model with $Z =Z_{\odot}$ returns a slightly older population with a median age of $\sim$ 12 Myr as opposed to $\sim$ 8 Myr assuming the chosen metallicities. However, we find that the discrepancy in the results 
does not significantly affect our conclusions in Section \ref{sec:conclusion}.} 

\subsubsection{Output results from the SED fitting}\label{sec:sed-res}
\begin{figure}
\begin{center}
\begin{tabular}{c}
 \resizebox{1.\hsize}{!}{\includegraphics{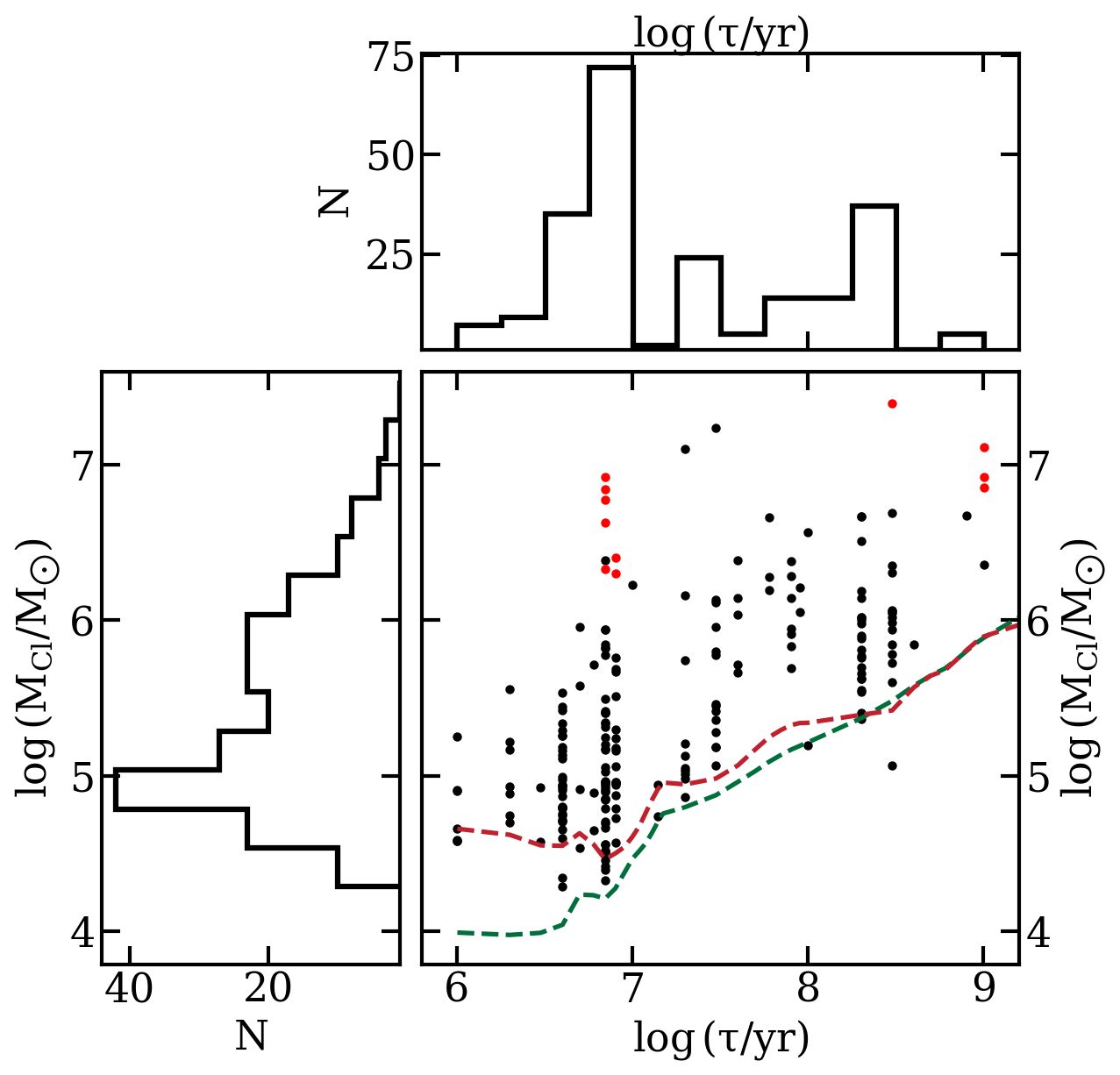}} 
\end{tabular}
\end{center}
\caption{Mass vs age distribution of the cluster population. Red circles represent the knots embedded in the reddish area of the ring. The dashed lines mark the mass limits assuming 80 per cent completeness limits in the \textcolor{black}{F606W} (green) and \textcolor{black}{F814W} (red) filters.}
\label{fig:Agevsmass}
\end{figure}

\begin{figure*}
\begin{center}
\begin{tabular}{ccc}

 \resizebox{0.33\hsize}{!}{\includegraphics{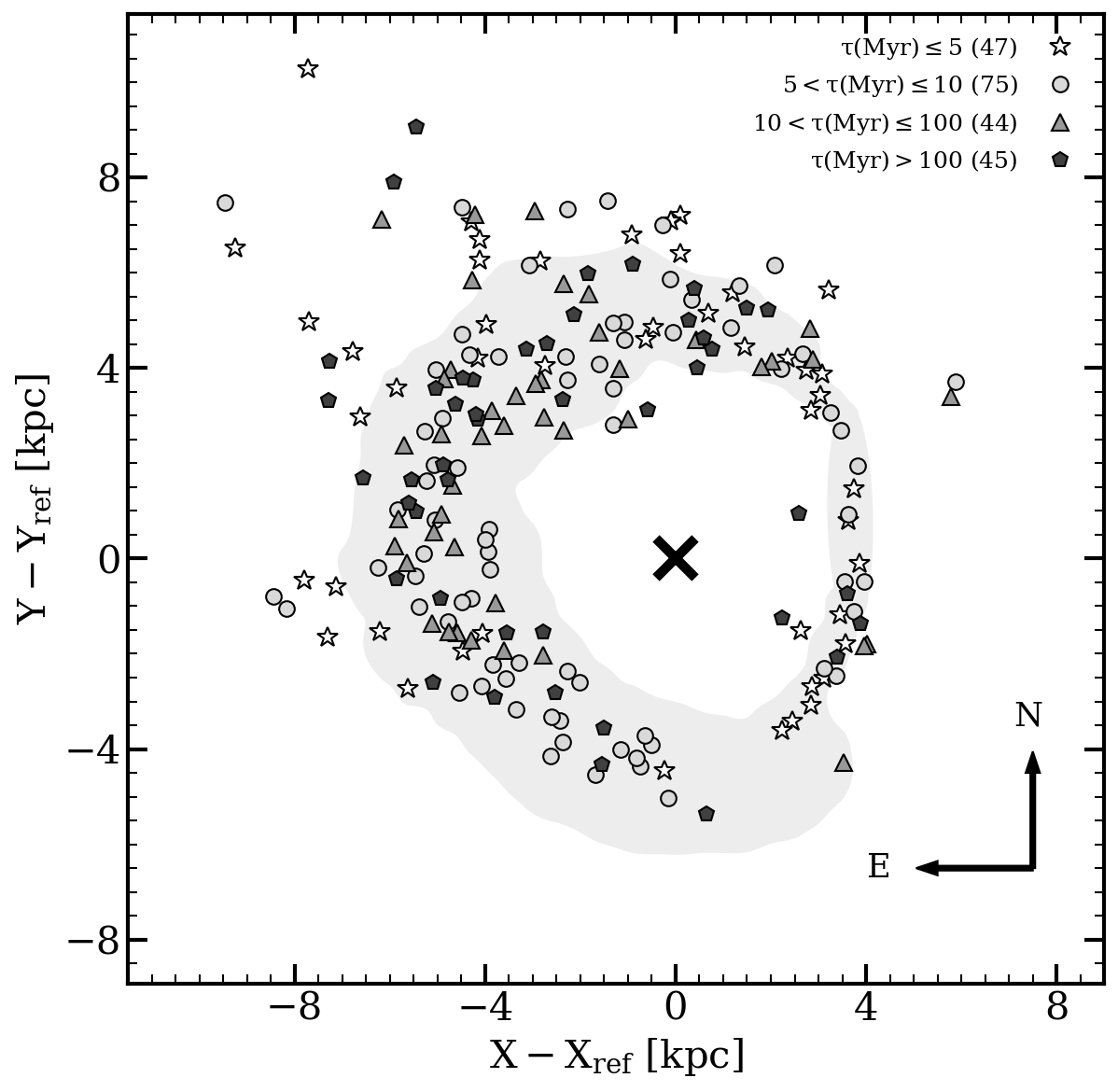}}   
  \resizebox{0.33\hsize}{!}{\includegraphics{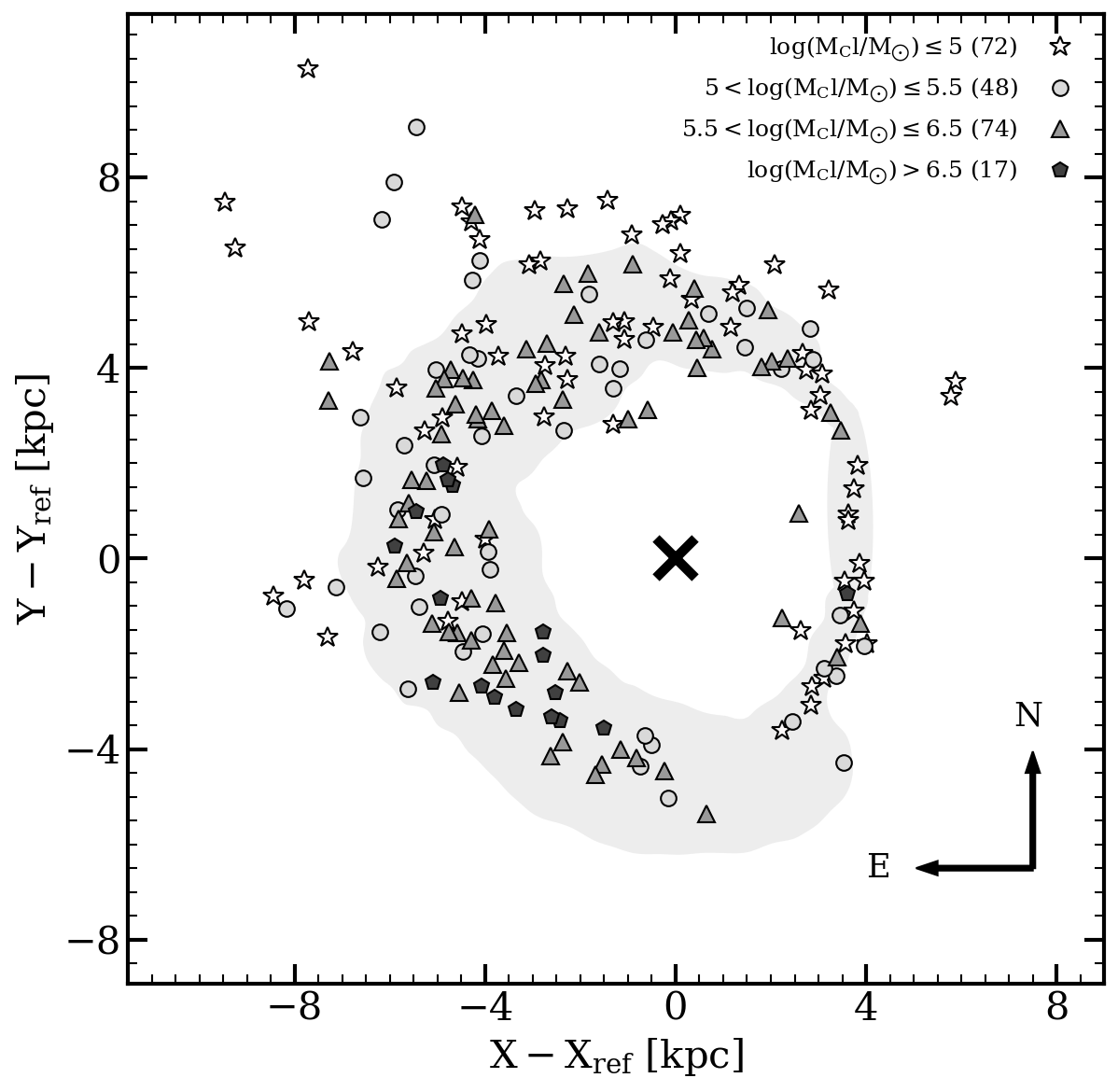}}   
  \resizebox{0.33\hsize}{!}{\includegraphics{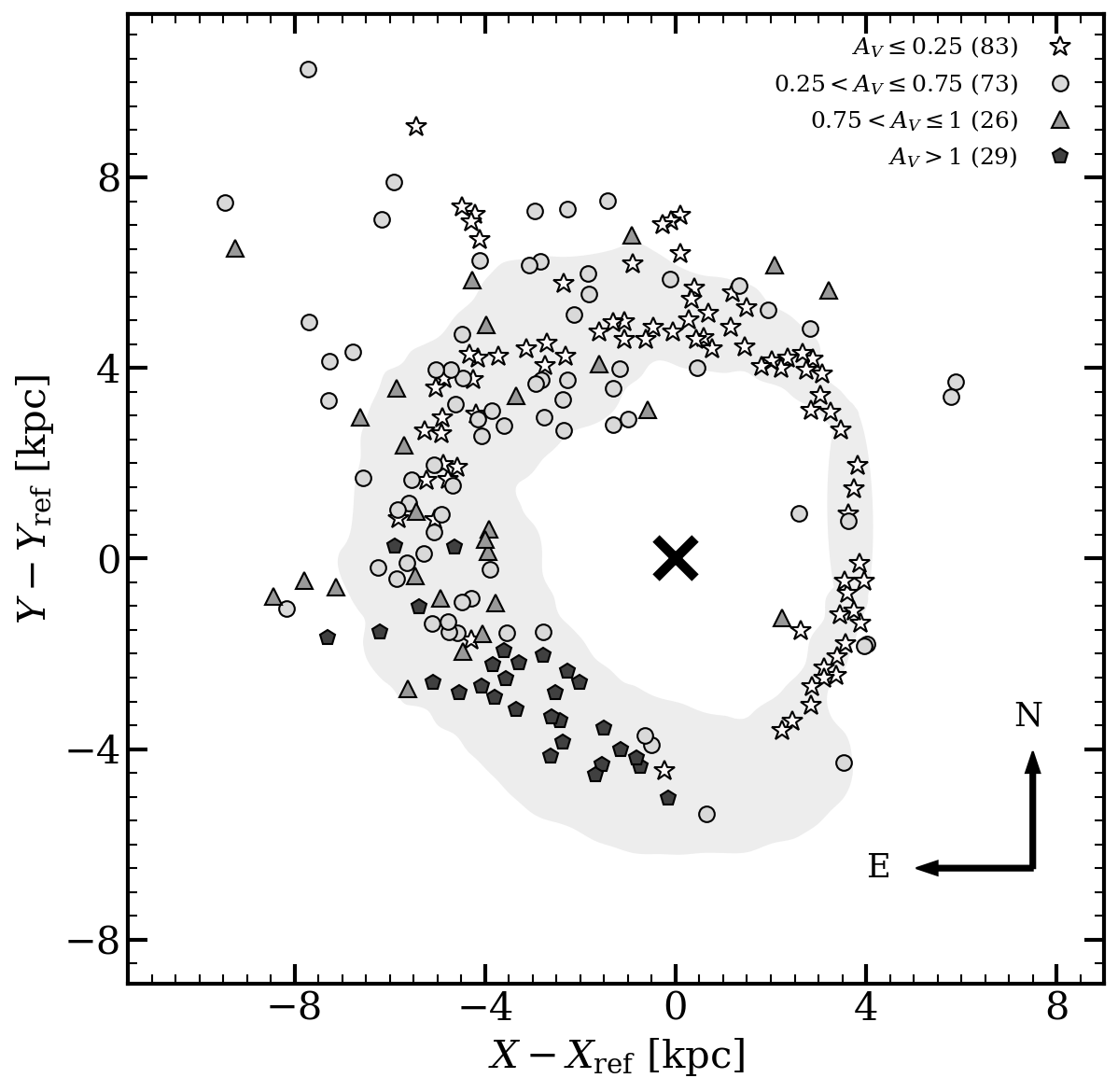}}  
\end{tabular}
\end{center}
\caption{The spatial distributions of the star-forming knots as a function of their age (left), mass (middle), and extinction (right). The different labels correspond to four different bin ranges as shown in the legend of each panel. The cross marks the centre of the ring \textcolor{black}{which is defined by the grey area.}}
\label{fig:spatial}
\end{figure*}

Figure \ref{fig:Agevsmass} shows the resulting mass versus age distribution of the blue knots considering SSP models of both $Z  =  0.004$ and $Z  = 0.008$, and $f_{\rm cov}$ = 0.5.  To help investigate where the knots form and how they move along the expanding ring, we also plot in Fig.\,\ref{fig:spatial} the spatial distributions of the blue knots split into different age (left), mass (middle) and extinction (right) bins with the derived values of the parameters listed in Table \ref{tab:knots-cat}. 

Overall, Arp 147 has a young cluster population with an age range spread between $\approx 1 - 1000$ Myr and a median age of $\sim$ 8 Myr. More than half (58 per cent) of the sources have ages below 10 Myr. 
\textcolor{black}{Before interpreting further the values of the fitted parameters, we should keep in mind that our photometric analyses lack UV and H$\alpha$ observations.  It implies that age-extinction degeneracy is only minimised but not entirely broken, though effort has been made to constrain the extinction range during the SED fitting procedure (see Section \ref{sec:fit}). In that case, it is possible that the fraction of clusters younger than 10 Myr is overestimated and the median value of $\sim$ 8 Myr is biased toward a younger age, i.e. the fitting algorithm wrongly assigns ages younger than 10 Myr to old knots with low extinction. The degeneracy will overestimate as well critical parameters such as the cluster formation efficiency $\Gamma$ (Section \ref{sec:CFE}). Nevertheless,} because of the overabundance of blue colours in Fig. \ref{fig:Av-map} \textcolor{black}{and especially the bright UV (see Fig.1 bottom left panel in \citetalias{2010ApJ...721.1348R}) and H$\alpha$ (see Fig.7 top panel in \citetalias{2011MNRAS.417..835F}) emissions mostly spread across the ring, we expect the majority of the fitted ages to be younger than 10 Myr.}  
This is irrespective of the strong peak 
near log\,$(\tau/{\rm yr}) \sim$ 6.8 in the age-mass plane in Fig. \ref{fig:Agevsmass}. The youngest candidates ($\tau <$ 5 Myr, open stars) are mainly located in the NE quadrant and the western region of the CRG, \textcolor{black}{which are in agreement with the detection of bright hotspots of UV emission in these regions \citepalias{2010ApJ...721.1348R}}. Knots in the age range $5 - 10$ Myr (grey circles) are found everywhere, especially the SE quadrant of the ring (see Fig. \ref{fig:spatial} left panel). \textcolor{black}{Such distribution is consistent with that of H$\alpha$ emission (which is more or less uniformly distributed across the ring; see \citealt{2008AJ....136.1259R} and \citetalias{2011MNRAS.417..835F}) and the EW(H$\alpha$)-based starburst age range of each quadrant by \citetalias{2011MNRAS.417..835F}. The authors record as well the youngest age range of the recent SF of $\approx 1 - 2$ Myr in the SE quadrant of the ring.} 
Clusters in the age range 10 $-$ 100 Myr (solid triangles) that constitute \textcolor{black}{21} per cent of the whole sample are also concentrated in the NE quadrant. The rest of the population, i.e. $\tau > 100$ Myr (solid diamonds) accounting for 21 per cent of the sample, mostly reside along the northern part of the ring.  
The spatial concentration of knots with similar ages suggests that they must have been born in the same SF episode. 
Even if the low number of clusters with ages $\tau >$ 10 Myr in the S - SE dusty region can be explained by the challenges of detecting these fading clusters in such highly extinguished regions, the use of only three filters 
and the lack of UV and H$\alpha$ data to break age-extinction degeneracy should also be kept in mind  
\citep[see e.g.][]{2015AJ....149...51C,2023ApJ...949..116C}.

The masses of the blue knots span a range between \textcolor{black}{${\rm 4.29 < log\,(M_{cl}/M_{\odot})< 7.40}$}  with a median value of $ \sim 2 \times 10^5 \ {\rm M}_{\odot}$. 
Arp 147 hence appears to have a slightly more massive cluster population compared to the  usual mass range of $\approx 10^3 - 10^7\,{\rm M}_{\odot}$ for clusters in nearby starburst environments  
\citep[e.g.][]{2003dhst.symp..153W}. 
In spite of a rigorous selection of {\tt SExtractor} detection parameters and a small aperture radius to perform photometry, resolution bias cannot be avoided at the distance $D_L > 100$ Mpc of Arp 147 \citep{2013MNRAS.431..554R}.
In particular, sources with masses ${\rm M_{cl} > 10^{6.5}\,M_{\odot}}$, labelled as solid diamonds in Fig. \ref{fig:spatial} middle panel and which represent 8 per cent of the sample, should be interpreted with caution. 
Both blending effects and age-extinction degeneracy tend to overestimate the derived mass. The latter bias 
applies especially to those hosted by the highly extinguished reddish area of the ring. 
However, the cluster sample should be 
less susceptible to stochastic IMF sampling which primarily affects derived ages of sources with masses ${\rm M_{cl} \lesssim a ~few~\times 10^{4} \ M_{\odot}}$  \citep[e.g.][]{2010A&A...521A..22F}. This is because our least massive source has an estimated value of ${\rm M_{cl} \sim 2 \times 10^{4} \ M_{\odot}}$. 
Finally, we notice a clear trend in the cluster position as a function of mass. While 69 per cent of sources with  ${\rm M_{cl} \leq 10^{5} \ M_{\odot}}$ (open stars) reside in the northern region, the SE quadrant alone hosts 36 per cent of clusters with masses above this value.  
There is also a distinct trail of sources with ${\rm M_{cl} \leq 10^{5} \ M_{\odot}}$ across the thin component in the western part of the ring. The clustering could indicate that there may be a physical connection between the knot properties and its host environment, assuming statistical bias and uncertainties do not dominate over plausible external factors. 

The fitted extinctions of the knots have a median value \textcolor{black}{$A_V = 0.46$ mag or $E(B-V) = 0.11$} mag where the majority (\textcolor{black}{86} per cent) of the population 
has extinction values $A_V < 1$ mag. While 73 per cent of the least extinguished sources ($A_V < 0.25$ mag, open stars) reside in the northern side of the CRG, 29 sources with $A_V > 1$ mag (solid diamonds) are mainly found in its eastern part. In particular, the most obscured ones ($A_V > 1.5$ mag) are highly concentrated in the S-SE reddish area, as expected. Besides age-extinction degeneracy, the observed clustering could be evidence linking the extinction and the cluster host environment.

Why do we find sources older than the $\sim 50$ Myr dynamical age of the ring? Are the higher mass values of the detected sources solely caused by systematic bias? What is (are) the main reason(s) behind the distinct clustering in the age and mass spatial distributions? 
To address these questions and get more insight into the CRG SFH, we conduct more thorough investigations in Section \ref{sec:more-res}.

\subsubsection{{\tt Starburst99} vs {\tt Yggdrasil} fitted parameters}\label{sec:sb99-comp}
\textcolor{black}{We also assess any systematic bias introduced by our choice of the evolutionary track models when performing SED fitting to the source photometric catalogues. 
This is done by comparing the least-square fitted parameters based on {\tt Yggdrasil}  and {\tt Starburst99} SSP models \citep{1999ApJS..123....3L,2014ApJS..212...14L} of a subset of 70 star-forming knots.
With an initial extinction estimate of $A_V^0 \leq$ 0.25 mag, these 70 sources lie in regions less affected by dust, specifically the western and northern parts of the ring. They serve as our testbed for assessing the dependence of the output results on the chosen SSP models.}

\textcolor{black}{{\tt Starburst99} models with ages between 1 Myr and 3 Gyr were retrieved assuming similar input parameters as for {\tt Yggdrasil}, i.e. a single instantaneous burst of SF with a Kroupa IMF, PADOVA AGB SSPs, and the same metallicities of $Z = 0.004$ and then $Z = 0.008$.  We select {\tt Starburst99} among many others for comparison, since the SSP model is also considered appropriate in the context of star cluster analysis \citep[see e.g.][]{2005A&A...431..905B,2013MNRAS.431.2917D,2019MNRAS.482.2530R,2023MNRAS.522..173W}.  We note, however, that {\tt Starburst99} does not take into account nebular emission from the warm and ionised gas that is mostly associated to star clusters younger than 10 Myr.}

\begin{figure}
\begin{center}
 \resizebox{1.\hsize}{!}{\includegraphics[trim= 0cm 0cm 0cm 0cm]{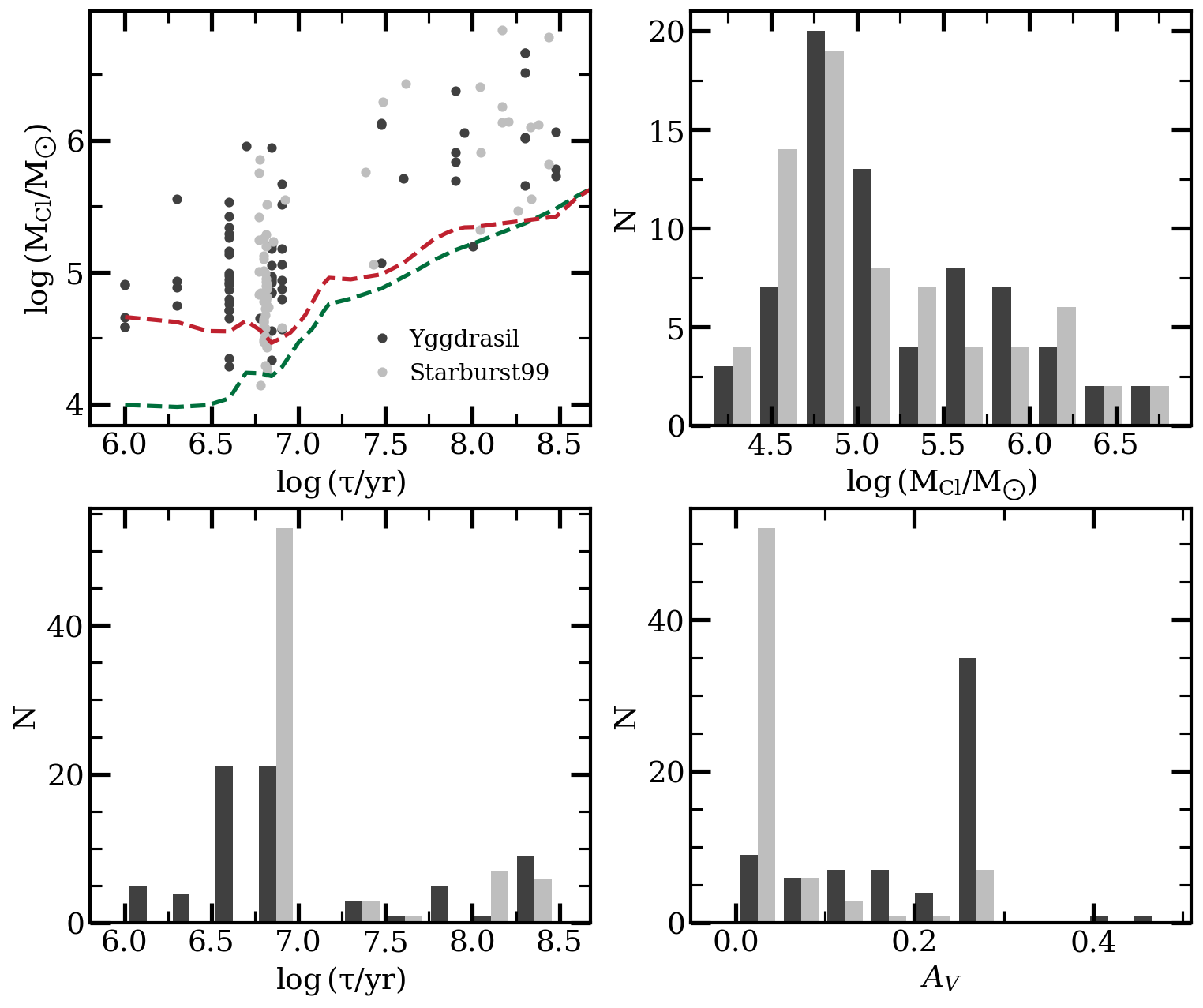}}   
\end{center}
\caption{Comparisons of the knot parameters obtained from fitting {\tt Yggdrasil} (black markers and histograms) and {\tt Starburst99} (grey) SSP models. Displayed are the age-mass plane (top left panel) and the distributions of the cluster mass (top right), age (bottom left), and extinction (bottom right) parameters. The dashed lines in the age-mass plane represent the mass limits assuming 80 per cent completeness limits in the \textcolor{black}{F606W} (green) and \textcolor{black}{F814W} (red) filters.}
\label{fig:sb99-ygg}
\end{figure}

Black and grey markers (or histograms) in Fig. \ref{fig:sb99-ygg} represent results assuming {\tt Yggdrasil} and {\tt Starburst99} models, respectively. The distribution of the data points in the age-mass plane (top left panel) reveals that {\tt Starburst99} ages mainly converge into a narrow range creating a prominent chimney at log$(\tau/{\rm yr}) \sim$ 6.8. Although such a distribution (due to the appearance of RSGs at this age) also exists for {\tt Yggdrasil} ages, it is, however, less pronounced as the data points extend to ages as young as 1 Myr. In fact, \textcolor{black}{29 out of the 70 knots (41 per cent)} have {\tt Yggdrasil} ages below 5 Myr versus none for {\tt Starburst99} ages. 
\textcolor{black}{With median values of log\,$\rm (M_{cl}/M_{\odot}) \sim $ 4.90 ({\tt Starburst99}) and  $\sim$ 5.06 ({\tt Yggdrasil})}, the ranges of the resulting masses (top right panel) are nevertheless generally consistent with each other.
As for the derived $A_V$ shown on the bottom right panel, the values peak at \textcolor{black}{$A_V \sim 0.01$}  and $\sim$ 0.27 mag considering {\tt Starburst99} and {\tt Yggdrasil} models, respectively. The discrepancy, which correlates with the age distributions, is a signature of the effects of the age-extinction degeneracy, i.e. reddening is underestimated while age is overestimated, and vice versa. The bias seems to be more prominent for {\tt Starburst99} results where more than \textcolor{black}{65 per cent of the fitted data points have $A_V \sim$ 0} leading to the age overestimate of the youngest clusters. Finally, the data points are better reproduced by the latter model based on the $\chi^2$ statistic of the least-square fitting.

Models with ({\tt Yggdrasil}) and without ({\tt Starburst99}) emission line treatments in the early stages of cluster evolution indeed return different values of the fitted parameters. The former is expected to better reflect the evolutionary track of clusters younger than 10 Myr, not yet free from extinction, since they are still associated with the surrounding nebular emission from their massive OB-type stars.
Based on these comparisons and given that Arp 147 likely consists of a young cluster population, the fitted results from {\tt Yggdrasil} models are more suitable as reference values in this work. 

\subsection{Properties of the clumps}
\label{sec:clumps-res}
The estimated values of $A_V \leq 0.25$ indicate that the sources are relatively free from extinction, except C4 where $A_V = 0.83$. This is expected given its location, and hence the inferred $A_V^0$ from the extinction map.
The bright kpc-sized clumps have areas between \textcolor{black}{0.20 and 0.88 kpc$^2$ (with a diameter range from $\approx$ 510 to 1060 pc)} and masses larger than $\rm 10^6 \ M_{\odot}$, except C5 having a slightly smaller mass of ${\rm \sim 8.9 \times 10^{5} \ M_{\odot}}$. 
The size of the largest \textcolor{black}{($\approx$ 1100 by 780 pc)} and most massive (${\rm \sim 6.8 \times 10^{6} \ M_{\odot}}$) clump, C4,  follows the relation established by \citet{1994ApJ...425...57E}, i.e. the size of the largest clump 
is expected to scale with the $B-$band absolute magnitude of its host galaxy. This implies that clumps form from giant clouds, which are the product of gravitational instabilities in the ambient interstellar medium. 
The masses of clumps observed in Arp 147 are comparable with those (a dozen) of the CRG NGC 922 with masses between ${\rm \sim 9 \times 10^{5} - 3 \times 10^{7} \ M_{\odot}}$ \citep{2010AJ....139.1369P} taking into account the $\sim$1 mag difference in their absolute magnitudes. Both CRGs have larger and more massive clumps than the ones of local spirals with similar absolute magnitudes because of their original galaxy's disk stirred up by the collision, which in turn triggers a much larger scale for gravitational collapse \citep{2010AJ....139..545G}.
In other words, clump sizes and masses are tightly related to the local velocity dispersion in the gas. Finally, our clump ages are all younger than 10 Myr. Hence, these star-forming regions must have formed within the ring, not being ejected into it from the field of the distorted spiral galaxy. Unlike Arp 147, both young ($< 10$ Myr) and relatively older ($45 - 90$ Myr) clumps are found in the field of NGC 922. The {\it N}-body simulations by \citet{2006MNRAS.370.1607W} have suggested that the drop-through collision of an intruder through the disk of the larger spiral occurred $\sim$ 330 Myr ago to form NGC 922. It is thus normal to detect intermediate age clumps given its dynamical age.

The clumpy structure of local CRGs 
is not only ideal to help establish the differences and similarities of SF of galaxies as a function of redshift. These bright sources, which are large scale SF sites of YMCs, are also useful for investigating the SF efficiency (SFE) on non-nuclear sub-galactic scales. Unfortunately, we could not derive accurate SFRs at the spatial scale of our clumps. 
At the distance of Arp 147, archival {\it Spitzer}/IRAC data can only give rough estimates of the parameter associated to areas larger than $\sim$ 15 kpc$^2$ (see Section \ref{sec:sub-gal}). 

\subsection{Any association with ULXs?}\label{sec:ulx}
CRGs are recognised as ideal ULX nurseries because of the enhanced SF triggered by the drop-through collision \citep[see e.g.][]{2009ApJ...703..159S,2017ARA&A..55..303K,2018ApJ...863...43W}. In the case of the Cartwheel, \citet{2023MNRAS.522.1377S} have recently detected at least 29 ULXs across its rings. This is the largest number ever discovered in a single galaxy. 
For Arp 147, the excellent angular resolution of Chandra X-ray observations ($\sim$ 0.5 arcsec) has led \citetalias{2010ApJ...721.1348R} to identify 8 ULXs around its ring and another belonging to the plume.
The powerful sources have unabsorbed X-ray luminosities in the range of $L_X \sim 1.5 - 7.5 \times 10^{39}\,{\rm erg\,s^{-1}}$ \citep{2018ApJ...863...43W} and they reside in the eastern part of the ring, except ULX-1 which is located in the NW region. 
This spatial distribution could either merely be random, or a reflection of the post-collision SFH of the galaxy.   
According to \citetalias{2010ApJ...721.1348R}, the high X-ray activity of Arp 147 in the eastern region suggests a most recent (and possibly current) SF activity occurring in that area as a result of the collision. In fact, with typical ages $\lesssim$ 10 Myr \citep{2003ApJ...596L.171G,2004MNRAS.348L..28K},  ULXs are good tracers of the most recent SF activity of their host galaxies. The spatial distribution is consistent with the findings from our work, where more than half of the sources are younger than $\tau < 10$ Myr in the same region. 

We also searched for possible ULX counterparts of blue stellar clusters considering a search radius of $r$ = 0.5 arcsec ($\sim$ 300 pc) after performing {\tt IRAF/CCMAP} astrometry calibration to the HST images using Guide Star Catalogue II.
The value of $r$ is chosen to take into account the relative uncertainty in the ULX positions derived from
Chandra coordinate systems ($\sim 0.3 - 0.5$ arcsec as per \citetalias{2010ApJ...721.1348R}) while minimising the inclusion of spurious counterparts. It is also possible that an ULX has been ejected up to $\sim$ 300 pc from a nearby cluster that could have been its birthplace  \citep{2013MNRAS.432..506P}. In that case, any separation below our chosen value of $r$ can still be treated as a true ULX - cluster association.
Each X-ray source has at least one nearby cluster younger than 10 Myr within the search radius, except ULX-3 and ULX-4 with much older neighbouring clusters of $\approx$ 200 Myr old. In all cases, the immediate 
neighbouring clusters are located at least $\sim$ 0.15 arcsec away, i.e. not coincident to the ULX positions. 
ULX-1 and ULX-5 are also associated with clumps C3 ($\tau \sim 4$ Myr) and C6 ($\tau \sim 7$ Myr), respectively.
Table \ref{tab:ULXs} summarises the properties of the clusters near the positions of the ULXs. 
\textcolor{black}{We note however that at the adopted 0.5 arcsec matching radius, a simple Poisson estimate following the standard approach used in counterpart-identification studies \citep[e.g.][]{2004MNRAS.349..135G} and based on the surface density of catalogued knots yields a per-source chance alignment probability of $p_{\rm chance}\sim 0.25 - 0.35$, implying that $\sim 2 - 3$ of the $9$ ULX--knot associations may be spurious.}
 The knot and clump ages below 10 Myr are comparable with the young ages of clusters ($\lesssim 6$ Myr) associated with X-ray sources in the Antennae galaxies \citep{2013MNRAS.432..506P}. However, source ages around 200 Myr imply that the relatively old knots are less likely to be related with the X-ray sources unless there is an overestimate in the cluster age from age-extinction degeneracy. \textcolor{black}{In fact, the colour of a dusty young cluster appears red and it can be misidentified as an old one without extinction because of the degeneracy.}

We did not find any prominent correlation between the intrinsic luminosity of the ULX sample and the properties of their potential cluster counterparts. Nevertheless, the detection 
of young blue knots and clumps within the search radius from the X-ray source positions could suggest that these objects are none other than the ULX birth sites  \citep{2023MNRAS.519.5492M,2024MNRAS.529.4369G}. 
The nature of any ULX - knot association is still a matter of debate and thus requires sufficient evidence involving spectroscopic observations, especially at the distance of Arp 147 where the individual knots have finite physical sizes of \textcolor{black}{$\sim 350$ pc and given the relatively low value of $p_{\rm chance}\sim 0.25 - 0.35$}.

\begin{figure*}
\begin{center}  
  \resizebox{1.\hsize}{!}{\includegraphics[trim= 0cm 0cm 0cm 0cm]{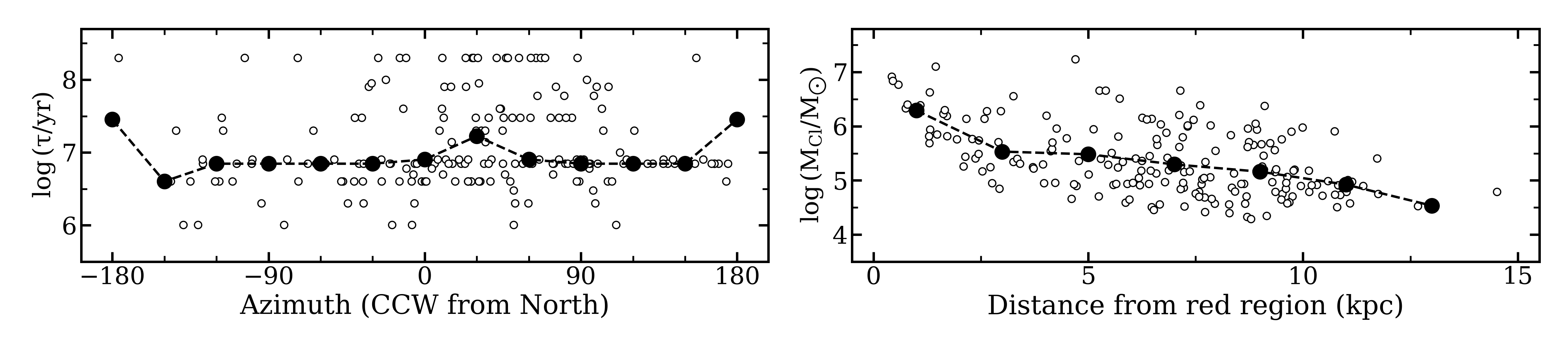}}  
\end{center}
\caption{The azimuthal age (left) and radial mass (right) distributions of the blue knots. The solid points represent the mean ages and mean masses per 30$^{\circ}$ AZ bin size and per 2.5 kpc bin size, respectively. The major axis of the ring refers to the $0^{\circ}$ AZ.  The reddish S-SE area is the reference point used to estimate the radial distance.}
\label{fig:azrad}
\end{figure*}

\section{Various diagnostic tools and their implications}\label{sec:more-res}

\subsection{Azimuthal age distribution}
\label{sec:az-age}
In Figure \ref{fig:azrad} left panel, we plot the azimuthal (AZ) distribution of the blue knots against their computed ages to assess the dominant SF mechanism across the CRG and to check whether its ring-like structure is long-lived or a transient feature.
We exclude sources older than 200 Myr to avoid statistical bias, since the cluster population is significantly incomplete beyond that age. 
The azimuthal distance is measured counterclockwise from the 
ring's fitted major axis which is the 0$^\circ$ AZ in this work. The black solid points represent the mean age per 30$^\circ$ AZ bin size. 
Unlike clusters in the circumnuclear ring of NGC 2328  \citep{2014ApJ...797L..16V} or the ones along the spiral arms of NGC 1566 \citep{2018MNRAS.478.3590S}, there is no evidence for any prominent age sequence across the ring of Arp 147 apart from a bump at AZ distances of $\sim 0 - 60^\circ$. The absence of data points at AZ distances between $-150^\circ$ and $-180^\circ$ is explained by the \textcolor{black}{poor} knot detection in the most diffuse region near the reddish area of the ring.

The most common explanation for the lack of an AZ age (colour) gradient, \textcolor{black}{assuming the observed pattern is real,} arises from the occurrence of an outwardly self-propagating SF mechanism across the ring as a result of the drop-through collision. A gradual increase of the gas density in the ring would then trigger a gravitational collapse and massive SF once it reaches  critical density at some stage. In this scenario, the knots can form anywhere throughout the ring due to local gravitational instabilities 
\citep[e.g.][]{1976ApJ...209..382L,1977ApJ...212..616T,1992ApJ...399L..51G}. 
The bump within $\sim 0 - 60^\circ$ AZ 
is \textcolor{black}{most likely} real given that this AZ bin coincides with the locations of the point of impact \citepalias{2011MNRAS.417..835F} and an extending plume N-NE side of the ring. This
implies that although a self-propagating SF might trigger the formation of knots across the ring, there is a possibility that the process is coupled with some external factors in this particular region (e.g., any residual tidal interaction with the companion) \textcolor{black}{or even further beyond.}

\textcolor{black}{In fact, ULX properties of Arp 147 do not support a radially self-propagating SF. The relatively high number of ULXs with a population lacking extremely bright ones, i.e. $L_X \gtrsim 10^{40}\,{\rm erg\,s^{-1}}$, rather defines the large rate of SF as a collision-driven intense starburst that likely reached its peak $\approx 10 -40$ Myr ago \citepalias{2010ApJ...721.1348R}. These bright sources are believed to emerge from the same burst that produced the young blue knots across the ring, and hence the detection of the latter in their vicinity (see Sec. \ref{sec:ulx}). Furthermore, previous works focusing on the SF activity of Arp 147 by \citetalias{2011MNRAS.417..835F} have reported the presence of a colour (age) gradient in the recent SF across the CRG's ring. They found that the W-SW quadrant of the ring is bluer compared to the others with an increasingly redder colour counterclockwise. The existence of such pattern can be interpreted as a direct visual record of a coherent and large-scale collision-triggered starburst event happening all around the ring. The triggering of a synchronised recent SF is known to be a strong evidence of a radially propagating star-forming density wave 
as opposed to a self-propagating SF mechanism \citep{1997AJ....113..201A,1998AJ....116.2757B}. 
This is consistent with the combined stellar and gas-dynamical model by \citet{1992ApJ...399L..51G} to predict the formation of a system resembling Arp 147. Their numerical works suggest that an outward-moving density wave, which is induced by an off-center collision, sweeps up the surrounding gas along way. The process enhances SF by triggering a strong starburst activity that creates the expanding ring. Thereafter, bursts of SF are found in high gas density regions furthest away from the nucleus. 
Recent hydrodynamical simulations of a Cartwheel-like CRG by \citet{2018MNRAS.473..585R} support as well an enhanced SF being triggered by compression from high-pressure shock waves, expanding outward at a constant speed. Their model predicts a reversed radial age gradient which indicates a distinct SF sequence across the post-collision features of the simulated CRG. Note that their proposed material wave model differs from the classical density wave scenario whereby the expanding ring drags gas and the stars it forms to its current location instead of simply passing through and not carrying them along. }

\textcolor{black}{Based on these observational and theoretical works, the lack of AZ age gradient likely stems from possible bias. The internal effects of light absorption may be responsible for the absence or the weakness of an AZ age (colour) gradient \citep{1976ApJS...31..313S}.} 
Such a process can smear out the colour gradient as the bluer colour indices towards the inner edge of the ring are compensated by increasing light absorption. \textcolor{black}{Alternatively, the age sequence can also be smeared out if the collision-triggered starburst event gets superimposed with an apparent stochastic SF process. Driven by stellar feedback (e.g. SN explosions, stellar winds), stochastic SF is a self-propagating random process that compresses surrounding molecular clouds and thereby igniting a chain reaction mechanism of more massive star birth
\citep{1978ApJ...223..129G}. 
Robust age dating including UV and H$\alpha$ observations will hopefully help to resolve these conflicting scenarios behind the azimuthal age distribution across the ring. In any case, both radially expanding star-forming wave and self-propagating SF are} associated with a ring that is short-lived, making it a transient feature that will fade away in one galactic year. It is predicted that the expanding ring will fragment into large clumps by strong perturbations to finally disappear almost completely in about $\sim 0.2 - 1$  Gyr \citep{1977ApJ...212..616T,2018MNRAS.473..585R,2021MNRAS.507.6140I}. \textcolor{black}{Otherwise, differential rotation (which causes the gas to move at different speeds) could also be the underlying factor for stretching and breaking apart a ring that formed by self-propagation within $\sim 100 - 200$ Myr \citep{1978ApJ...223..129G}.} 

\subsection{Radial mass distribution}
\label{sec:rad-mass} 
The right panel of Fig. \ref{fig:azrad} shows the cluster masses plotted against their distances from the  highly extinguished S-SE side of the ring. Black solid points denote the mean mass per 2.5 kpc bin size. We find a decreasing mass with an increasing distance from the reddish area. Such a trend is similar to the cluster mass-galactocentric radius relation based on observational works by e.g. \citet{2015MNRAS.452..246A}, \citet{2018MNRAS.477.1683M}, and \citet{2019MNRAS.482.2530R}, 
except that the reference point to estimate the radial distance is not at centre of the galaxy but rather within the most extinguished region of the empty ring. The location is chosen under the assumption that it hosts the displaced remnant nucleus of the distorted spiral galaxy \citep[e.g.][]{1992ApJ...399L..51G,2012MNRAS.420.1158M}. We still observe a radially varying cluster mass even if we were to exclude objects more massive than \textcolor{black}{$\rm 10^{6} \ M_{\odot}$}. 

\citet{2016ApJ...816....9S} suggested a stochastic cluster formation process to explain the observed relation as a result of size-of-sample effect. Galaxies with higher SFR 
stand a higher chance to form a larger cluster population and, thus, they are more likely to sample the high-mass bins of the cluster IMF. 
On the other hand, given that the collision occurred $\sim$ 50 Myr ago, and that two galaxies are still relatively close with a projected separation of $\sim 13$ kpc,  
the elliptical intruder is likely to be still deep within the potential well of the distorted spiral, and hence can bear some influence on the gas dynamics in the empty ring. The ongoing interaction would trigger the concentration of high-gas density in the nearby S-SE side of the ring, 
which is a necessary key ingredient for the birth of the most massive knots in an environmentally-dependent cluster formation scenario \citep[e.g.][]{2015MNRAS.452..246A,2019MNRAS.482.2530R}.

\textcolor{black}{This second scenario is in agreement with numerical predictions by \citet{2018MNRAS.473..585R}. The authors have reported that clusters that formed after the collision are predominantly more massive than the ones before the collision. In particular, the most massive clusters inhabit regions with high gas densities and low shear,
and thus favouring the concentration of intense SF.
For a Cartwheel-like system, these regions are found in the opposite side of the nucleus, across the outer ring, where SF is less efficient because of high shear-driven turbulence that stretches and heats gas clouds. This distinct distribution of the SF activity generates a strong asymmetry in the azimuthal distribution of the stellar mass formed in the ring. It is then reasonable to use the azimuthal and/or radial mass distribution to trace the SFH of the CRG as well as the epoch when the SF peaked.} 

\subsection{Luminosity and mass functions}
\label{sec:LMFs}
Cluster luminosity function (CLF) and its corresponding mass function (CMF) are measurable traits used to better understand the physical properties of a cluster population. Their shape and any distinct features encode past and ongoing processes triggering the YMC birth and subsequent evolution. Both functions are generally consistent with power-laws (PL) of the form $dN/dL \propto L^{-\alpha}$ with $\alpha \sim 2$ and $dN/d{\rm M} \propto {\rm M}^{-\beta}$ with $\beta \sim 2$, respectively (see e.g. \citealt{1997ApJ...480..235E,1999AJ....118.1551W} for CLFs and \citealt{1999ApJ...527L..81Z,2003A&A...397..473B} for CMFs). 

 \begin{table}
\caption{Fitted slopes of the \textcolor{black}{completeness-corrected} cumulative CLFs.}
\begin{center}
\begin{tabular}{l@{\hskip 0.3in} c@{\hskip 0.3in} c@{\hskip 0.3in} c}
 \hline \hline
  Filter & Cutoff (mag) &  $\rm \alpha_{LS}$ & $\rm \alpha_{Bayes}$  \\ 
  \hline 

    \textcolor{black}{F450W} & \textcolor{black}{\m10.43} & 2.19 & 2.01 \\   
    \textcolor{black}{F606W} & \textcolor{black}{\m10.52}  & 2.15 & 1.94 \\
    \textcolor{black}{F814W} & \textcolor{black}{\m10.83} & 2.21 & \textcolor{black}{2.01} \\
  \hline
  \multicolumn{4}{@{} p{6.5cm} @{}}{\footnotesize{{\it Notes. } For each filter, we derive the slope of a pure PL fit to the completeness-corrected CLF up to the 80 per cent completeteness level listed in the second column. Least-square (LS) and Bayesian (Bayes) fitting techniques are used to estimate the slopes  $\rm \alpha_{LS}$ and $\rm \alpha_{Bayes}$, respectively.}}

\end{tabular}
\label{tab:alpha-lf}
\end{center} 
\end{table}

\begin{figure}
\centering
\includegraphics[width=1.\linewidth]{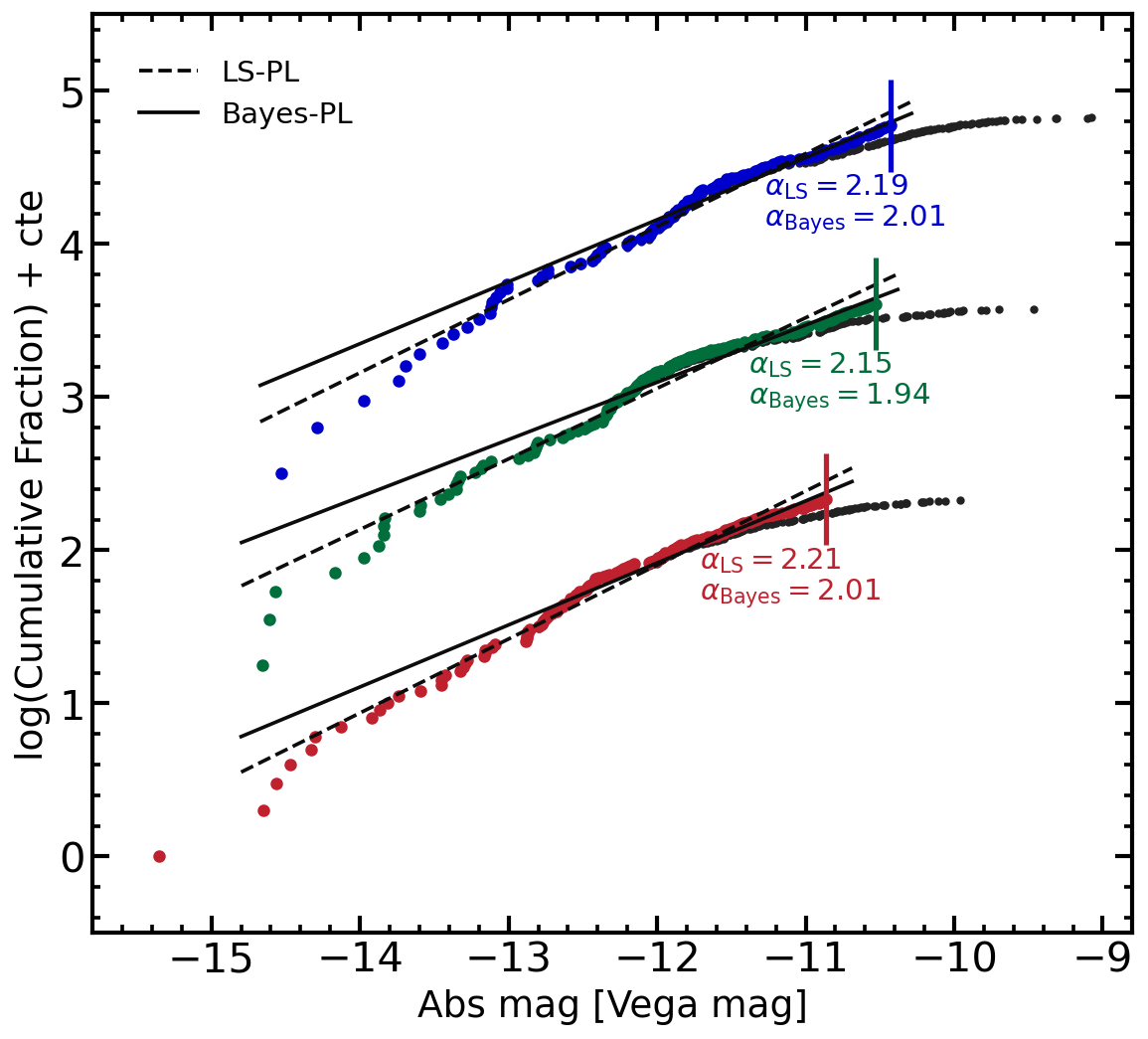}
\caption{Cumulative CLFs corrected from observational incompleteness using the \textcolor{black}{F450W} (blue curve), \textcolor{black}{F606W} (green), and \textcolor{black}{F814W} (red) photometric data points. The black curves represent the original CLFs before applying completeness correction. The black dashed and solid lines denote a single PL fit to the sources brighter than the 80 per cent completeness limits (vertical lines) using least-square and Bayesian fitting techniques, respectively.} 
\label{fig:cumlf}
\end{figure}

\subsubsection{Multiband CLFs}
 In this work, we use cumulative distribution to construct unbinned CLFs of Arp 147 \textcolor{black}{using the F450W, F606W, and F814W observations}. Cumulative methods are deemed more appropriate to test the hypothesis of any truncated distribution (especially for galaxies with a small number of clusters; \citealp[see e.g.][]{2012MNRAS.419.2606B,2018MNRAS.473..996M,2021ApJ...923..278L}). We fit a single PL function to the data using least-square method (LS) and Bayesian (Bayes) fitting technique. Refer to e.g. \citet{2018MNRAS.477.1683M,2018MNRAS.473..996M} for a comprehensive description of the Bayes fitting method. Before fitting a single PL to the CLFs, we first apply a completeness correction to the data using the predicted completeness fractions in Section \ref{sec:comp}. The process is necessary to avoid any spurious turnover at the faint end of the distribution due to the instrumental detection limit from the observations. Table \ref{tab:alpha-lf} lists the 80 per cent completeness limits, until which we perform the fit, and the resulting PL slopes of $\rm \alpha_{LS}$ and $\rm \alpha_{Bayes}$ using least-square and Bayesian methods, respectively. 

 For both fitting techniques and for all three filters, the values of the PL slopes are $\alpha \gtrsim 2$ (with their uncertainties ranging between $0.03 - 0.10$), except for the Bayesian-fitted CLF \textcolor{black}{in F606W}  with $\rm \alpha_{Bayes} = 1.94$. Nevertheless, the values of \textcolor{black}{$\rm \alpha_{LS} = 2.15 - 2.21$} are steeper by $\sim 0.2$ than those of  \textcolor{black}{$\rm \alpha_{Bayes} = 1.94 - 2.01$}. This discrepancy is due to LS giving more weight to data points at the upper end of the CLFs, and hence a steeper slope, as opposed to the Bayesian fitting process. Since the latter is deemed more robust when fitting a PL (or other functions) to the cluster distribution, our discussions are mainly based on Bayesian-fitted parameters of the CLFs (and CMFs) hereafter. 
 
A deviation from a PL can be observed in the cumulative competeness-corrected CLFs shown in Fig. \ref{fig:cumlf}, especially for \textcolor{black}{the F606W data points}. Given that unbinned CLFs are less sensitive to statistical bias compared to binned ones, other factors could also play a role in shaping this truncated feature,  especially if we were to detect the same trend in the CMFs (see Section \ref{sec:sub-CMF}).

\subsubsection{CMFs of the knots in different age intervals} \label{sec:sub-CMF}
\begin{figure}
\begin{tabular}{c}
\includegraphics[width=1.\linewidth]{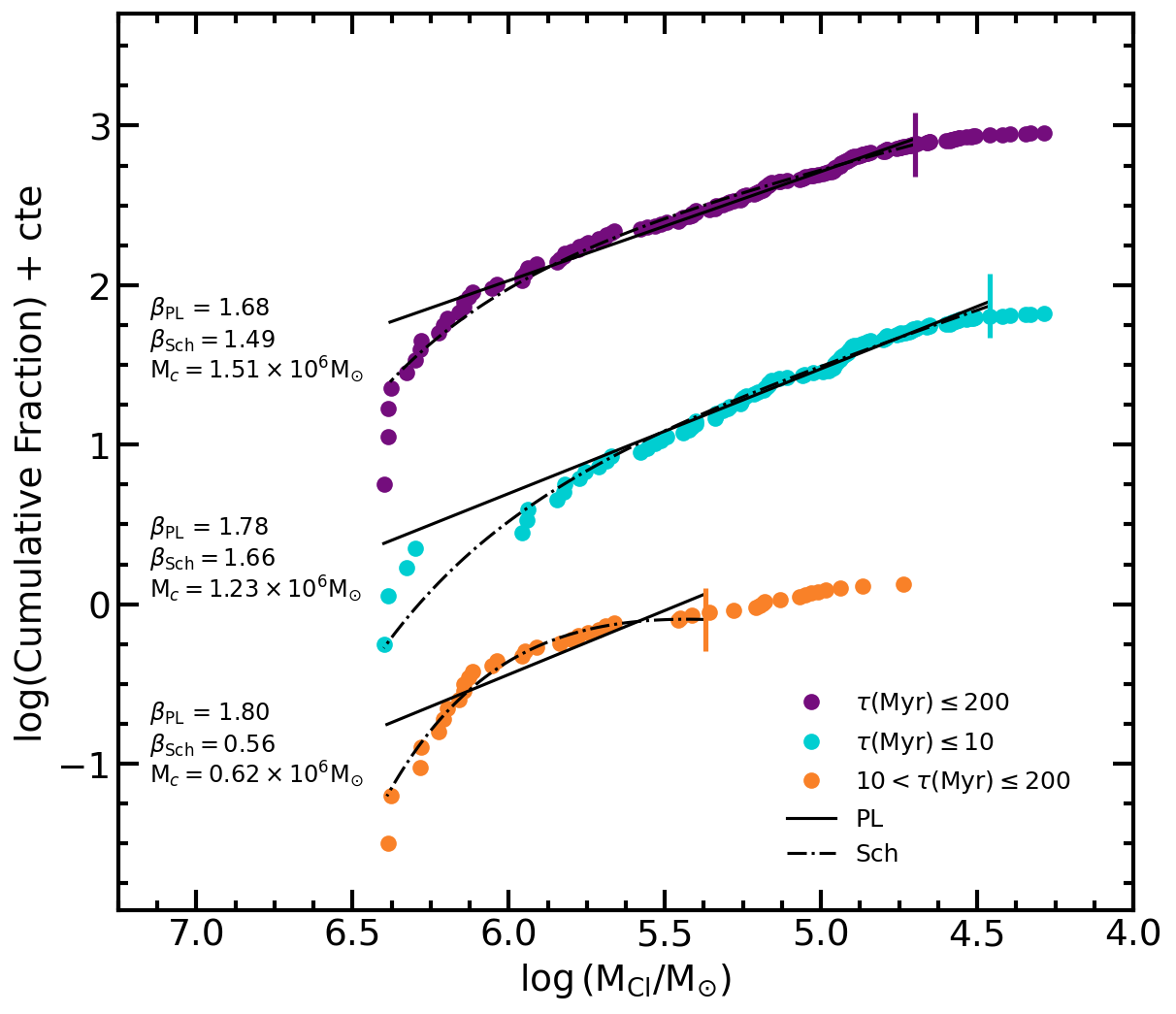}
\end{tabular}
\caption{Cumulative CMFs of the blue knots as a function of the age interval (see the insets). Solid and dash-dotted lines respectively represent Bayesian PL and Sch fits to the CMFs for knots with $\rm M_{cl} \geq M_{cut}$. The vertical lines denote log$\rm ( M_{cut}/M_{\odot})= 4.46$ (dotted line) and 5.37 (dashed) to get mass-limited samples for $\tau <$ 10 Myr and $\tau <$ 200 Myr, respectively.}
\label{fig:CMF}
\end{figure}

 The purple points in Fig. \ref{fig:CMF} represent a cumulative CMF for all clusters younger than 200 Myr. Besides excluding sources with $\rm M_{cl} > 10^{6.5}~\text{M}_{\odot}$ (to minimise resolution bias), we also exclude the old ones ($\tau > 200$ Myr in this work) to minimise statistical bias from incompleteness. We did not apply a completeness correction to the CMF since deriving cluster mass completeness limits is \textcolor{black}{highly uncertain because of the limited availability of filters in this work.} The solid line represents a PL fit to data points more massive than $\rm M_{cut} =10^{4.7}\,M_{\odot}$, which is the highest peak of the CMF if we were to use a binned distribution. We find a slope \textcolor{black}{$\beta_{\rm PL} = 1.68$}, a smaller value than the canonical slope of $\beta \sim 2$. We also perform a Schechter (Sch) 
 fit to the cumulative CMF. 
 The two component function (dash-dotted line) is of the form $dN/d{\rm M} \propto ({\rm M/M_{\it c}})^{-\beta_{\rm Sch}}\,e^{(\rm M/M_{\it c})}$ with \textcolor{black}{$\beta_{\rm Sch}$ = 1.49} the PL slope at the low-mass end, and $\rm M_{\it c} = 1.51\,\times 10^{6} \, M_{\odot}$ the characteristic truncation mass at high-masses. Based on the posterior probability distributions (PDFs) in Appendix \ref{fig:PDF-lt200}, the CMF of Arp 147 is better represented by a pure PL than a Schechter function in spite of a turnover observed at the upper end of the CMF, i.e. around $\approx 10^{6}~\text{M}_{\odot}$.

To check whether the turnover is real,
 we split the cluster population in two subsamples based on the source ages: $\Delta t_1 = 1 - 10$ Myr (cyan) and $\Delta t_2 = 10 - 200$ Myr (orange). 
  We consider \textcolor{black}{conservative} mass-limited samples to mitigate observational incompleteness when fitting the CMFs. Based on the age-mass plane in Fig. \ref{fig:Agevsmass}, we thus perform the fit up to $\rm log\,(M_{cut}/M_{\odot}) = $ 4.46 and 5.37 for the corresponding MFs of sources of ages within $\Delta t_1$ and $\Delta t_2$, respectively. 
 Again, the solid and dashed lines respectively represent PL and Sch fits to the data. 
While the CMF of sources younger than 10 Myr can still be represented by a PL with a slope $\beta_{\rm PL} = 1.78$, especially if we were to omit the data points above $\rm log\,(M_{cl}/M_{\odot}) \sim $ 6.2,   
 the CMF of the older age interval $\Delta t_2$ clearly deviates from the PL function well below that value. 
In fact, it is best described by a Schechter fit where \textcolor{black}{$\beta_{\rm Sch}$ = 0.56 and M$_{\it c}$ $\rm \sim 6.17 \times 10^{5} \, M_{\odot}$} (refer to the PDF distributions in Appendix \ref{fig:res-Bayes-CMFs}). Table \ref{tab:beta-mf} summarises the fitted parameters of the CMFs. 

 Resolution bias is a known factor behind an artificial upper cutoff in the CMF. If the observed turnover at the high-mass end of our CMFs is due to blending effects, then the chosen upper cutoff of $\rm 10^{6.5} \, M_{\odot}$ to minimise such bias is not enough to exclude most of the blended sources. \textcolor{black}{To explore further whether the apparent turnover is real, we then applied an upper cutoff value of $\rm 10^{6} \, M_{\odot}$. Again, a Schechter function with $\rm M_c > (5 - 6) \times 10^5\,M_{\odot}$ fits better the CMF than a PL distribution. 
Furthermore, numerical simulations by \citet{1992ApJ...399L..51G} have shown that an off-center collision between a gas-rich disk galaxy and its equal-mass companion produces an Arp 147-like CRG with an empty ring, of which presents some generic features. In particular, they have predicted the appearance of hot spots with elevated gas densities and shocks across the ring, especially on the opposite sides away from the displaced nucleus. These regions are primed of strong starburst events, which explain enhanced SF and thereby the extensive birth of young massive blue knots in the empty ring. 
In that case,} it should not be a surprise to detect knots more massive than $\rm 10^{6} \, M_{\odot}$ in Arp 147 regardless of its distance. In fact, nearby LIRGs such as NGC 4149 (39 Mpc, \citealt{2020MNRAS.499.3267A}), Arp 299 (45 Mpc, \citealt{2019MNRAS.482.2530R}), and NGC 1614 (70 Mpc, \citealt{2024AJ....168..259C}) have been reported to host star clusters as massive as $\rm 10^{7} \, M_{\odot}$.
In that case, it is plausible that the high-mass truncation of the CMF that corresponds to $\Delta t_2 = 10 - 200$ Myr is real and thus worth investigating, especially that the mass-limited sample should be free from rapidly-expanding young associations and less prone to incompleteness. 
The measured truncation of the CMF is believed to be environment-dependent \citep[see e.g.][]{2009A&A...494..539L,2015MNRAS.452..246A}. High-SFR host galaxies such as interacting systems have their CMF truncation at the high-mass end (as massive as  $\rm 10^{6} \, M_{\odot}$) compared to their normal star-forming galaxy counterparts (with a mass turnover at  $\rm 10^{5} \, M_{\odot}$). Given the nature of Arp 147 (a CRG with enhanced SF, \citealt{1992ApJ...399L..51G}) and the measured value of its characteristic truncation mass ($\rm M_{\it c} = 1.51 \times 10^{6} \, M_{\odot}$), our CMFs tend to follow this observed trend in the literature, i.e. an increasing $\rm M_{\it c}$ for higher SFRs. In other words, SF intensity in Arp 147 is high enough to allow very massive knots to form.

The CMFs of young (cyan) and intermediate-age (orange) knots have distinct trends and values of \textcolor{black}{$\rm M_{\it c}$ ($\rm 1.23 \times 10^{6} \, M_{\odot}$ vs. $\rm 0.62 \times 10^{6} \, M_{\odot}$)}. This could be evidence for an evolving CMF where the two populations undergo different dissolution rates over time \citep{2015ApJS..216....6L}.
Nevertheless, we should note the small number statistics of \textcolor{black}{27} used to fit PL and Schechter functions to the CMF for the age interval of $10 - 200$ Myr (see Table \ref{tab:beta-mf}). \textcolor{black}{We also emphasize that age-extinction degeneracy could also lead to cluster mass overestimates besides blending, and hence should be taken into account when interpreting the CMF shape and especially the value of $\rm M_c$. Furthermore, Monte Carlo mass completeness simulations should be conducted once more multiband data become available to accurately constrain the lower end of the CMF.}

\begin{table}
\caption{Fitted parameters of the \textcolor{black}{completeness-uncorrected} CMFs assuming pure PL and Schechter functions.}
\begin{center}
\begin{tabular}{l@{\hskip 0.1in} c@{\hskip 0.1in} c@{\hskip 0.1in} c@{\hskip 0.15in} c@{\hskip 0.15in} c@{\hskip 0.15in} c}
 \hline \hline
     \noalign{\smallskip}  
  Region & $\Delta t$ & $\rm log(M_{cut})$ &$\rm N_{cl}$ & $\beta_{\rm PL}$ & $\beta_{\rm Sch}$ & ${\rm M}_{c}$  \\
    \noalign{\smallskip}  
   & (Myr) & $\rm (M_{\odot})$ & & & & $(\rm \times \ 10^6\,M_{\odot})$ \\ 
       \noalign{\smallskip} 
  \hline 
    \noalign{\smallskip} 
      & 1 \m~200 & 4.70	& \textcolor{black}{137} & \textcolor{black}{1.68} & \textcolor{black}{1.49} & 1.51 \\   
    Arp 147 & 1 \m~10  & 4.46	& \textcolor{black}{113} & 1.78 & \textcolor{black}{1.66} & \textcolor{black}{1.23} \\
     & 10 \m~200 & 5.37	& \textcolor{black}{27} & \textcolor{black}{1.80} & \textcolor{black}{0.56} &  \textcolor{black}{0.62}  \\
    \hline 
    
     & 1 \m~200 & 4.50     & \textcolor{black}{77} & \textcolor{black}{1.73} & \textcolor{black}{1.60} & 1.48   \\    
    North & 1 \m~10  & 4.46	    & \textcolor{black}{53} & \textcolor{black}{2.15} & 1.93 & \textcolor{black}{0.36}  \\
    & 10 \m~200 & 5.37    & 14 & \textcolor{black}{1.89} & \textcolor{black}{1.05} & 1.00 \\
    \hline 
    
    & 1 \m~200 & 4.54 & 77 & \textcolor{black}{1.59} & \textcolor{black}{1.36} & \textcolor{black}{1.29} \\    
    South & 1 \m~10 & 4.46 & \textcolor{black}{60} & \textcolor{black}{1.64} & \textcolor{black}{1.41} & \textcolor{black}{0.89} \\
     &10 \m~200 & 5.37 & 12 & 1.68 & \textcolor{black}{0.60} & \textcolor{black}{0.66} \\
    \hline 
    
     & 1 \m~200 & 4.53 & \textcolor{black}{75} & \textcolor{black}{1.56} & \textcolor{black}{1.35} & \textcolor{black}{1.44}  \\    
    East & 1 \m~10 & 4.46 & \textcolor{black}{50} & \textcolor{black}{1.77} & \textcolor{black}{1.66} & \textcolor{black}{1.32}  \\
     &10 \m~200 & 5.37 & \textcolor{black}{19} & \textcolor{black}{1.75} & \textcolor{black}{0.51} & \textcolor{black}{0.62}  \\
    \hline 
    
    & 1 \m~200 & 4.50 & 79 & \textcolor{black}{1.64} & \textcolor{black}{1.48} & \textcolor{black}{1.05}  \\    
    West & 1 \m~10 & 4.46 & 63 & 1.77 & \textcolor{black}{1.59} & \textcolor{black}{0.91} \\
    & 10 \m~200 & 5.37 & 7 & \textcolor{black}{1.87} & 0.44 & 0.54\\
       \noalign{\smallskip} 
  \hline
         \noalign{\smallskip} 
  \multicolumn{7}{@{} p{8.5cm} @{}}{\footnotesize{{\it Notes.} {For each region (whole field and per sub-galactic area) and age interval $\Delta t$, we derive the pure PL slope $\beta _{\rm PL}$ of the cumulative CMF by considering $\rm N_{cl}$ data points more massive than the cutoff mass $\rm M_{cut}$. \textcolor{black}{The cutoff value is necessary given that the function is not completeness-corrected}. We also fit a Schechter function to the CMF to derive the PL slope $\beta _{\rm Sch}$  and the truncated mass $\rm M_{\it c}$.}}}  

\end{tabular}
\label{tab:beta-mf}
\end{center} 
\end{table}

\subsection{Cluster age distribution function}\label{sec:caf}
The cluster age distribution function (CAF) is defined as the differential number of clusters per time interval of a mass (or luminosity)-limited sample. CAFs can be approximated by a PL of the form $dN/d\tau \propto \tau^{-\delta}$ with $\delta$ varying between 0 and 1 and where a steeper PL index indicates a more rapid cluster dissolution \citep[see e.g. ][]{2005ApJ...631L.133F,2007AJ....133.1067W,2014ApJ...786..117F,2019MNRAS.482.2530R,2021ApJ...923..278L}.

\begin{figure}
\centering
\includegraphics[width=1.\linewidth]{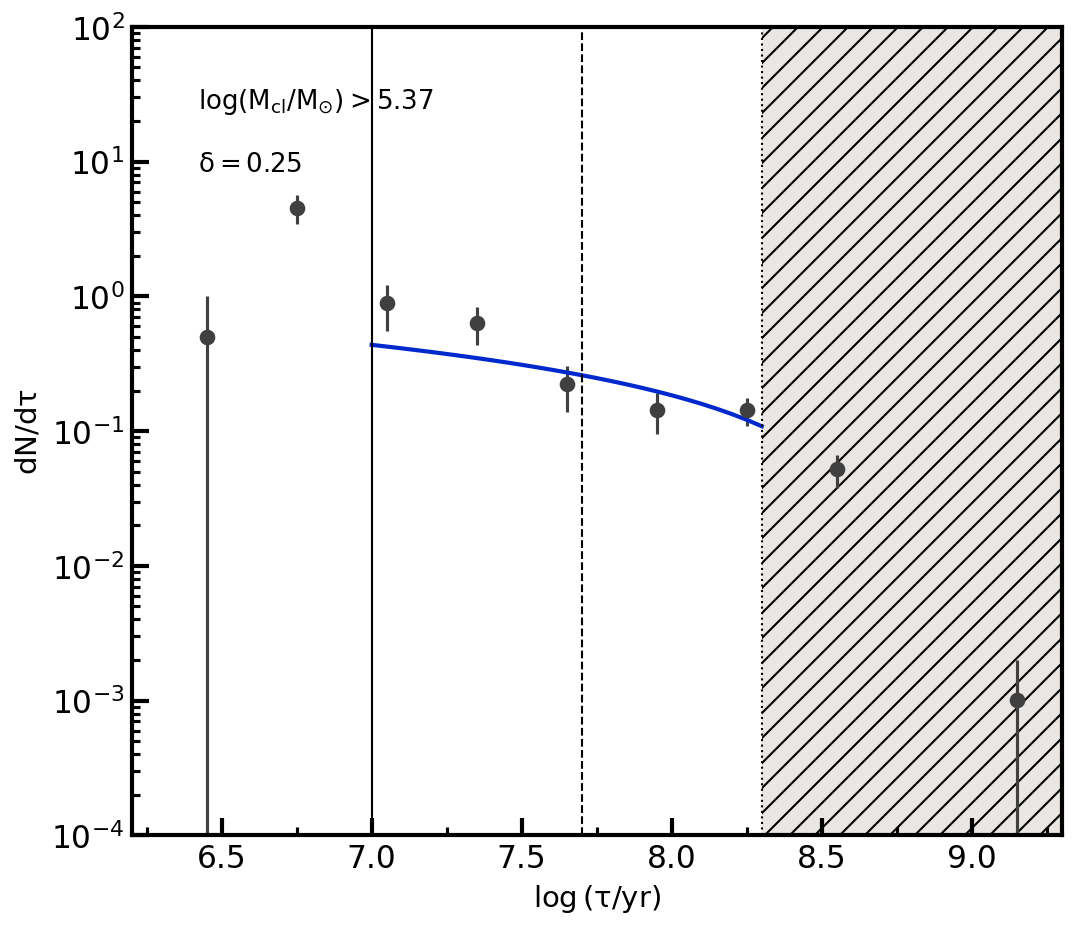}
\caption{Cluster age function of a mass-limited sample more massive than $\rm log(M_{cl}/M_{\odot}) = 5.37$. 
Solid, dashed and dotted lines mark the ages at 10, 50 and 200 Myr. The blue line represents a PL fit to the distribution within the age range $10 - 200$ Myr with the recovered slope $\delta$ included in the inset. The dashed area marks the ages above 200 Myr at which incompleteness artificially steepens the distribution.}
\label{fig:caf}
\end{figure}

Figure \ref{fig:caf} shows the CAF of knots with masses above log$\rm (M_{cl}/M_{\odot}) = 5.37$ assuming equally space temporal bins of 0.3 dex in widths. The mass cut leads to a complete mass-limited sample of sources with log$\rm (M_{cl}/M_{\odot}) > 5.37$ and  $\tau <$ 200 Myr.  
We exclude knot candidates with $\rm M_{cl} > 10^{6.5} \ M_{\odot}$ which are likely affected by imperfect extinction correction and/or the result of a significant resolution bias. The cluster age distribution has a PL index of \textcolor{black}{$\delta = 0.25$} between the age interval of $10 - 200$ Myr. The value of $\delta$ becomes \textcolor{black}{0.42 and 0.37} if we bin the data by 0.35 and 0.4 dex in widths, respectively. 
The age interval of $1 - 10$ Myr is most likely contaminated by gravitationally unbound clusters like OB associations \citep{2012MNRAS.419.2606B,2020MNRAS.499.3267A} and a mass-independent loss of very young objects known as infant mortality \citep{2005ApJ...631L.133F}, and thus is not taken into account in order to derive a more accurate value for the CAF slope. 
  
The CAF shape between $10 - 50$ Myr appears to be steep and can be interpreted as an increase in the formation rate of clusters (CFR, the total mass formed
in clusters per unit time over a given age interval) since the collision (i.e. $\approx$ 50 Myr ago) to the present day. In that case, it is a piece of evidence for a strong starburst event that likely happened in the ring  within that timeframe as predicted by the simulated collision of the CRG \citep[e.g.][]{1992ApJ...399L..51G,2012MNRAS.420.1158M}. 
However, the age distribution is rather flat between $50 - 200$ Myr implying a roughly constant CFR within this age interval, prior to the collision. 
The observed trends (steep and then flat distributions) persist even if we use other bin widths. 
We prefer not to attempt interpretations of the steep decline of the distribution beyond 200 Myr (log $\tau < 8.3$) since it is strongly affected by incompleteness. \textcolor{black}{We note that the detection of knots older than the dynamical age of the ring ($>$ 50 Myr) is rather unusual. Recent works on star clusters and stellar populations in merging systems \citep[e.g.][]{2023MNRAS.526.2341R,2025ApJ...994...90A} and CRGs \citep[e.g.][]{2011MNRAS.417..835F,2024A&A...688A..89D} have suggested that massive objects older than the epoch of interaction or collision, if present, could have survived the cluster dissolution mechanisms triggered by the violent interaction between the galaxies. They would then migrate away from their birth sites rather than having formed in-situ in their current location. However, the authors have highlighted as well that the detection of such an older population is by no means certain and hence, some caution must be taken in their interpretations. In fact, the presence of older knots in our analysis can be mainly attributed to age-extinction degeneracy (the derived ages of reddened knots are biased toward larger values), or other artefacts such as Milky Way foreground star contamination or misidentification of remnants of the disturbed spiral galaxy's nucleus that resemble knotty-like objects.} 

The decreasing rate of sources as a function of time in Fig. \ref{fig:caf} is expected due to the gradual mass-loss through a variety of internal processes (e.g., evolutionary fading, two-body relaxation). 
However, the PL slope $\delta > 0$ does not support the hypothesis of a constant disruption rate in the age range $10 - 200$ Myr. The value of \textcolor{black}{$\delta = 0.25 - 0.42$} is comparable to the CAF slopes of galaxies undergoing mild disruption from internal effects and/or outside influence \citep[e.g.][]{2011A&A...529A..25S,2018ASSL..424...91A}. In fact, any residual tidal forces between the distorted spiral and the hit-and-run companion galaxy likely contribute to the cluster dissolution across the ring over the last 50 Myr. Finally, if the star-forming ring is indeed a transient feature, which only lasts for $\sim$ 500 Myr, then its fragmentation into large clumps would also be an external factor affecting the long-term evolution of its cluster population, especially the low-mass ones.

Due to the limitations of this work and because we did not disentangle the contributions of cluster formation and disruption rates to the overall cluster age distribution, we prefer not to explore further the ongoing disruption model at play. We need more filters for a more robust mass estimation from our cluster age-dating analysis. 

\subsection{Cluster formation efficiency}\label{sec:CFE} 
CFE can be expressed as the ratio between CFR and the host SFR \citep{2008MNRAS.390..759B}. 
In this work, we constrain the value of the CFE (or $\Gamma$) by comparing the mass of clusters above the observational mass limit and within a certain age range to the galaxy SFR.
We remind the reader that blue knots with masses  M$_{\rm cl} >  10^{6.5}~{\rm M}_{\odot}$ are excluded to minimise overestimation of the CFR (and hence the CFE) from resolution bias. 
In this work, we derive $\Gamma$ assuming SFR\,(FIR, \citealp{2008AJ....136.1259R}) = $8.6\,{\rm M_{\odot} \,yr^{-1}}$, a value comparable to $\overline{\rm SFR} \sim 7\,{\rm M_{\odot} \,yr^{-1}}$ which is the mean of the parameter taken from the literature.
We do not use H$\alpha$-based SFR to avoid overestimating the CFE because of nearby and/or unbound clusters usually younger than 10 Myr easily contaminating the ${\rm H\alpha}$ emission \citep[e.g.][]{2016MNRAS.460.2087H, 2020MNRAS.499.3267A}. In addition, FIR luminosity provides a more reliable measure of the SFR since it is less affected by the dust extinction than H$\alpha$ emission is. 

  To match our age intervals to previous observations and to assess the evolution of $\Gamma$ over time, we consider two different intervals of $1 - 10$ Myr ($\Delta t_1$) and $10 - 100$ Myr ($\Delta t_3$), implying mass limits down to log$\rm (M_{cl}/M_{\odot}) = 4.46$ and 5.21, respectively.  We assume the cluster catalogues to be complete above these mass limits. We also include $\Delta t_2 = 10 - 200$ Myr to be consistent with our CAF and CMF analyses. 
  We find that \textcolor{black}{$\Gamma_{\Delta t_1} \sim 35.6$} per cent, \textcolor{black}{$\Gamma_{\Delta t_2} \sim 3.4$} per cent and \textcolor{black}{$\Gamma_{\Delta t_3} \sim 3.9$} per cent. 
   Had we considered an H$\alpha$-based SFR instead,  the value of \textcolor{black}{$\Gamma_{\Delta t_1}$ is equal to $\sim 65$} per cent which is almost a factor of 2 higher than that of a FIR-based SFR. 
   Finally, to quantify the impact of the head-on collision 
   on the SFH, the values of $\Gamma$ that correspond to the time intervals $\Delta t_4 = 50 - 100$ Myr and $\Delta t_5 = 1 - 50$ Myr are also estimated, i.e. around 50 Myr before and then after the collision. We find  \textcolor{black}{$\Gamma_{\Delta t_4} \sim 6.7$} per cent and \textcolor{black}{$\Gamma_{\Delta t_5} \sim 9.5$} per cent. 
    The decreasing fraction of stars forming in bound clusters is consistent with the declining shape of the CAF in Section \ref{sec:caf}. 
    
\begin{figure}
\centering
\resizebox{1.\hsize}{!}{\includegraphics[trim= 0.7cm 0.cm 0.2cm 2.5cm]{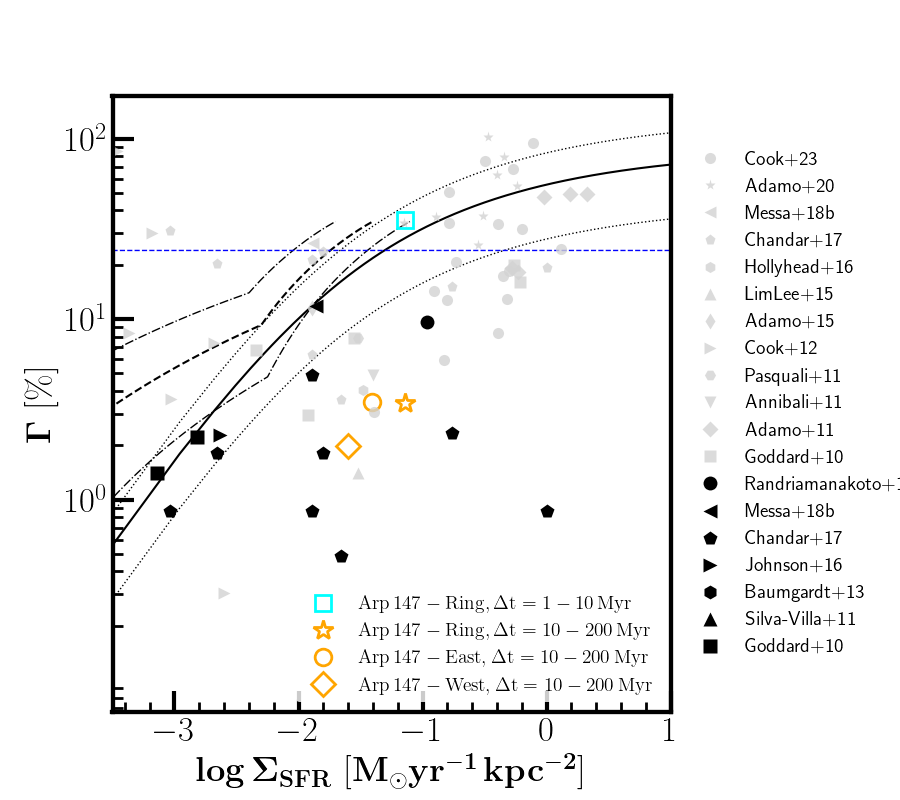}}\\ 
  
\caption{The cluster formation efficiency ($\Gamma$) plotted against the SFR density ($\Sigma_{\rm SFR}$). The open symbols denote the data points from this work. The grey and black solid markers respectively represent data points taken from the literature assuming age intervals of $1 -10$ Myr and $10 - 100$ Myr when deriving the value of $\Gamma$. Solid and dashed black lines represent the predicted fiducial model from \citet{2012MNRAS.426.3008K} assuming the global Kennicutt-Schmidt law and then sub-kpc \citet{2008AJ....136.2846B} conversion between $\Sigma_{\rm gas}$ and $\Sigma_{\rm SFR}$, respectively. The $2\sigma$ uncertainty of the theoretical models are enclosed by the dotted and dashed-dotted lines. The horizontal line marks the constant value of $\Gamma = 24$ per cent from \citet{2017ApJ...849..128C}.}
\label{fig:cfe-sfr}
\end{figure}

Figure \ref{fig:cfe-sfr} shows the positions of some of our derived CFEs (open symbols) in one of the most up-to-date versions of the $\rm CFE - \Sigma_{SFR}$ relation established for the first time by \citet{2000A&A...354..836L} and \citet{2010MNRAS.405..857G}.
The solid grey ($\Delta t_1 = 1 - 10$ Myr) and black ($\Delta t_3 = 10 - 100$ Myr) symbols denote previous literature data compiled by \citet{2023MNRAS.519.3749C} and references therein. In our case, $\Sigma_{\rm SFR} = 0.07$ ${\rm M_{\odot} \,yr^{-1}\,kpc^{-2}}$ 
using the same method as \citet{2012ApJ...751..100C}, i.e., by normalising SFR  over a projected area of \textcolor{black}{$A \sim 118~{\rm kpc^2}$} that coincides with the spatial distribution of the blue knots in the CRG. 
Theoretical predictions of $\Gamma$ with respect to the SFR density are also shown as black lines with the 2$\sigma$ lower and upper limits indicated by the dotted and dashed-dotted lines. The fiducial models are built based on a formulation from \citet{2012MNRAS.426.3008K} with the solid line adopting the global Kennicutt-Schmidt law \citep{1998ARA&A..36..189K}, i.e. a universal relation between $\Sigma_{\rm SFR}$ and $\Sigma_{\rm gas}$ at all scales. The dashed line, however, represents a modified version of the original \citet{2012MNRAS.426.3008K} model as suggested by \citet{2016ApJ...827...33J}. They have considered the assumptions from  \citet{2008AJ....136.2846B} of a varying relation at sub-kpc scales 
between the two parameters, i.e. a decreasing $\Sigma_{\rm SFR}$ for lower $\Sigma_{\rm gas}$, where HI dominates the gas environment over H2. 

Finally, work by \citet{2017ApJ...849..128C} proposed a constant value of $\Gamma \sim$ 24 per cent in all host galaxies when using $\Delta t_1 = 1 - 10$ Myr. This would indicate that the efficiency to form stars in clusters is independent of the galactic environment and that any observed $\Gamma - \Sigma_{\rm SFR}$ relation is merely a result of mixing data sets derived from different age intervals.  
The constant value is represented by the horizontal dashed line in the CFE - SFR density plot.

Compared to various host environments (e.g. LMC and SMC, normal spirals, dwarf irregular galaxies, starburst and interacting luminous IR galaxies)
represented in the CFE - SFR density plot, the efficiency of Arp 147 to form bound clusters during the  $1 - 10$ Myr age interval
is relatively higher than those of quiescent dwarfs and gas-poor spirals. These galaxies are labelled, for instance, as grey triangle-right markers \citep{2012ApJ...751..100C} or a grey hexagon  \citep{2016MNRAS.460.2087H} in Fig. \ref{fig:cfe-sfr}. 
\textcolor{black}{\citet{1987ApJ...312..566A} reported that recent SF level in RiGs is more extreme and higher than that of normal galaxies. Their findings are based on the comparison of FIR luminosities ($L_{FIR}$) and infrared to optical luminosity ratios of a sample of 26 RiGs (including Arp 147) with those of normal disk galaxies. A high value of $L_{FIR}$ is due to the predominant IR excess from of a large population of young OB stars. Thereafter, the authors suggest that intense and coherent starbursts across the ring must have triggered the birth of these massive blue stars. This would then explain the high value of $\Gamma_{\Delta t_1}$ in this work.} 
In fact, it is consistent with similar estimates for targets with extreme SF activity such as galaxy mergers \citep[grey stars,][]{2020MNRAS.499.3267A} and blue compact galaxies \citep[grey diamond markers,][]{2011MNRAS.417.1904A}. 
According to \citetalias{2010ApJ...721.1348R}, the major epoch of starburst activity in Arp 147 occurred over the last 15 Myr, which is then likely to trigger more significant cluster formation.
Previous star cluster works have also reported that dense gas-rich environments with high gas pressures 
result in elevated SFEs and, subsequently, in a higher CFE \citep[e.g.][]{2010MNRAS.405..857G,2012MNRAS.426.3008K,2020MNRAS.499.3267A,2023MNRAS.526.2341R}. 
The primary mechanism that can generate such high gas densities is rapid gas accretion resulting from an encounter of an intruder with a larger galaxy's disk.
Although the high value of \textcolor{black}{$\Gamma_{\Delta t_1} \sim$ 36} per cent is in agreement with 
a recent episode of SF triggered by a head-on collision, and also considering our conservative approach of utilising a mass-limited subsample with a non-H$\alpha$ based SFR, it is still, nevertheless, plausible that blended unbound clusters \textcolor{black}{and old knots erroneously fitted with young ages} could partially contaminate the catalogue leading to an overestimation of the CFE. 

The value of $\Gamma_{\Delta t_5}$ (10 per cent) is almost 2 times higher than that of $\Gamma_{\Delta t_4}$ (\textcolor{black}{7} per cent).
This means that fewer bound clusters have been formed before the collision as opposed to the CFE of a second generation of massive blue knots post-collision.
This trend, which can be explained by the abundance of high gas densities as a result of the collision,
is in agreement with a recent numerical work by \citet{2020ApJ...891....2L}. The authors find the CFE to vary with the merging phase, increasing from 20 to 80 per cent between the early merger stage and at the peak of the starburst. The difference between $\Gamma_{\Delta t_4}$ and $\Gamma_{\Delta t_5}$ can also be explained by a cessation of SF happening in star clusters older than 50 Myr (and hence a relatively lower $\Gamma_{\Delta t_4}$) due to cluster dissolution over time via two-body relaxation and/or secular evolution \citep[e.g.][]{2017ApJ...849..128C,2023MNRAS.519.3749C}. 
The decrease in $\Gamma$ over the $1 - 100$ Myr timescale (\textcolor{black}{36} per cent in the first 10 Myr, then dropping to 4 per cent during the $10 - 100$ Myr interval) can also be attributed to the same mechanism. This has also been observed by, e.g., \citet{2017ApJ...849..128C} (from 27 down to 7 per cent) and \citet{2023MNRAS.519.3749C} (from 24 down to 5 per cent) using the same age intervals.
 Finally, the corresponding CFE of the age interval ${\Delta t_1}$ falls within 2$\sigma$ uncertainty of the \citet{2012MNRAS.426.3008K} model assuming the Kennicutt-Schmidt law. This can be interpreted by the cluster formation process in Arp 147 to be driven by the surface gas density. 
None of the estimated CFE values clearly match the \citet{2008AJ....136.2846B} curve 
nor within the same range as the constant CFE value of 24 per cent from \citet{2017ApJ...849..128C}. 

Based on these comparisons and the simulations predicting the formation of the ring and the major epoch of SF, we suggest that an outward-moving ringlike density enhancement  
has condensed to form massive gravitationally bound clusters 
in large numbers across the ring, at least during the last 10 to 50 Myr \citep{2005ApJ...628..231B}. 
Overall, our results are more consistent with $\Gamma$ being dependent of the galactic environment.  
We note, however, that the derived CFEs are upper limits because of uncertainties and possible bias from blending (at the distance of the CRG, not all sources with masses below $\rm 10^{6.5}\,M_{\odot}$ are a single cluster detection), \textcolor{black}{age-extinction degeneracy (an underestimate in the age of genuinely old knots leads to their inclusion when computing the CFR)},
and unbound cluster contamination.

\subsection{Knot properties on sub-galactic scales} \label{sec:sub-gal}

\begin{figure}
\centering
  \resizebox{.85\hsize}{!}
 {\includegraphics[trim= 1cm 0.5cm 1cm 0cm]{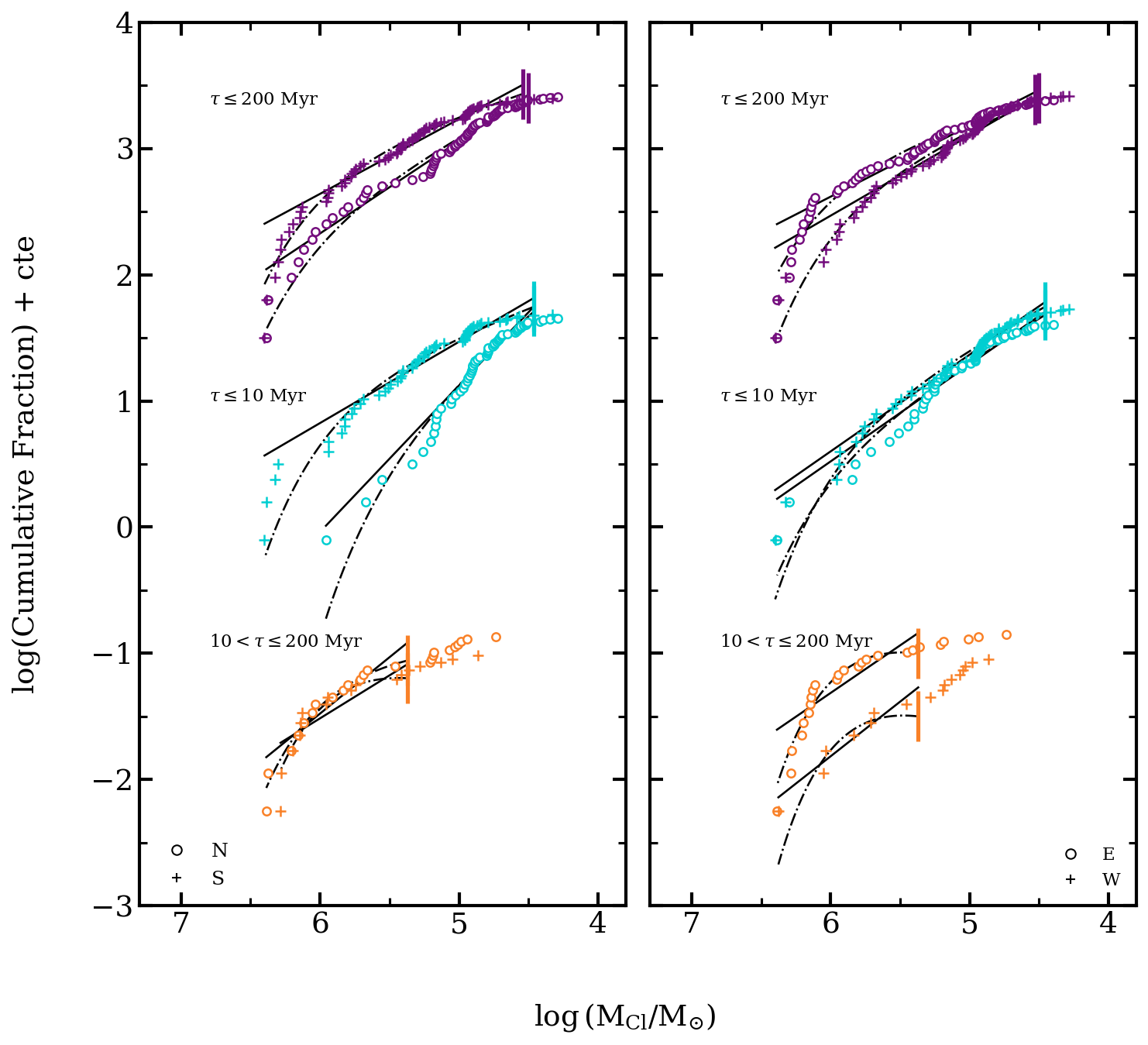}}  
\caption{Cumulative CMFs on sub-galactic scales: N/S (left panel) and E/W regions (right). The open circles represent data points associated with the northern or eastern region, and the crosses those of the southern or western region. Three age intervals of $1 - 200$ Myr (purple),  $ 1 - 10$ Myr (cyan) and $10 - 200$ Myr (orange) were considered for each sub-galactic sample. Other labels the same as in Fig. \ref{fig:CMF}. Knots  with masses $\rm > 10^{6.5} \ M_{\odot}$ are excluded for consistency. Table \ref{tab:beta-mf} reports the fit results.} 
\label{fig:subsample}
\end{figure}

We also built CMFs and estimated the CFE on sub-galactic scales to further investigate the cluster formation and disruption mechanisms of Arp 147. To this end, we split the cluster catalogue based on the source positions across the field: sources in the North (N) vs. South (S) and then East (E) vs. West (W). To minimise statistical bias, the regions (N/S and E/W) 
are delimited in a way that they have equal number of knots 
rather than conforming to the centre of the ring.

Purple data points in Fig. \ref{fig:subsample} show the cumulative CMFs of various regions of the CRG for all knots younger than 200 Myr. We also draw the corresponding CMFs for $\tau \leq$ 10 Myr (cyan) and $10 \leq \tau < 200$ Myr (orange). The derived distributions of the N and S regions (left panel) are similar in shape regardless of the age interval, though the fitted paramaters of the N region are generally larger than those of the S area (see Table \ref{tab:beta-mf}). A PL function is still a reasonable fit to the MFs of knots younger than 10 Myr while the distributions are best-represented by a Schechter function within the age range $10 - 200$ Myr.
There is a turnover of the CMF at $\rm \sim 1 \times 10^{6} \, M_{\odot}$ (N) and \textcolor{black}{$\rm \sim 6.6 \times 10^{5} \, M_{\odot}$} (S) for the latter age interval. The CMF shapes of the E and W regions (right panel) are different, especially for $\Delta t_2 = 10 - 200$ Myr where the CMF shape of the E region clearly deviates from a PL as opposed to that of the W region. The former is best described by a Schechter fit with a truncated characteristic mass of \textcolor{black}{$\rm M_{\it c}= 6.2 \times 10^{5} \, M_{\odot}$} (see Table \ref{tab:beta-mf}). 

 By adopting the same method as in \citet{2013ApJ...775L..38R}, we used {\it Spitzer}/IRAC 8\,$\mu$m data to estimate the flux from each region. The 8\,$\mu$m mid-IR luminosity of the ring was then redistributed with respect to the estimated ratio for each region. We get the following SFR values in units of $\rm M_{\odot}\,yr^{-1}$: 2.5 (N), 2.8 (S), 3.5 (E), and 1.8 (W), and hence SFR densities of 0.03, 0.03, 0.04, 0.02 ${\rm M_{\odot} \,yr^{-1}\,kpc^{-2}}$, respectively. Based on mass-limited sub-samples with $\Delta t_2 = 10 - 200$ Myr and the derived SFRs, the values of the CFE are respectively \textcolor{black}{$\Gamma^{\rm N} \sim 3.3$}, \textcolor{black}{$\Gamma^{\rm S} \sim 2.8$},  \textcolor{black}{$\Gamma^{\rm E} \sim 3.5$},  \textcolor{black}{$\Gamma^{\rm W} \sim 2.0$} per cent for N, S, E, and W regions. The CFE varies between 4.3 and \textcolor{black}{7.4} per cent had we used $\Delta t_3 = 10 - 100$ Myr, with the smallest value of $\Gamma$ associated with the western region. We also add to the CFE $- \Sigma_{\rm SFR}$ distribution in Fig. \ref{fig:cfe-sfr} the data points associated with the E/W regions (open circle and open diamond) given the discrepancy of their CMFs. Although the two data points do not fall within the $2\sigma$ uncertainty of both theoretical models, they seem to follow the trend of increasing $\Gamma$ for higher $\Sigma_{\rm SFR}$. 

It has been shown that measured truncation masses $\rm M_{\it c}$ increase systematically with SF level, not only on galaxy-wide scales, but within individual galaxies as well \citep[e.g.][]{2015MNRAS.452..246A}. 
In that case, the eastern and northern regions having larger $\rm M_{\it c}$ $\rm \sim 1.5 \times 10^{6} \, M_{\odot}$ (for $\tau < 200$ Myr) are expected to host more intense SF activity than the other parts of the CRG. This is in agreement with the distribution of ULXs \citepalias{2010ApJ...721.1348R} and previous numerical simulations \citep{1992ApJ...399L..51G} which indicate that the dominant star-forming sites are found in these regions. 
 It is thus not surprising to observe the age spatial distribution in Fig. \ref{fig:spatial} (where the youngest knots mostly inhabit these regions) and to derive relatively high CFE associated with the regions. 
The differences in the observed CMFs can also be interpreted such that the sub-galactic regions (especially the eastern area) undergo different levels of disruption. 
Based on the truncated CMF shape of the E region and assuming that age-extinction degeneracy does not significantly affect our mass estimation, knots in this region are then the most vulnerable to the mild cluster disruption reported in Section \ref{sec:caf}. In fact, this side of the ring is closest to the intruder,  $\sim 13$ kpc away, and hence the most affected by any residual disruption caused by the intruder exiting the potential well of the distorted spiral. 

The plume on the N-NE side of the ring hosts 25 knots out of the 40 outside the ring (i.e. the ones flagged as 1 in Table \ref{tab:knots-cat}). The sources in that region have a median age of $\sim 5$ Myr with the majority (80 per cent) younger than 50 Myr. Such a distribution implies that the plume is an ideal star-forming site in the CRG, \textcolor{black}{which makes sense given the substantial filamentary structures as traced by H$\alpha$ in that particular area \citep{2008AJ....136.1259R}}. It also suggests that this young and outflowing feature from the ring \citepalias{2011MNRAS.417..835F} likely formed as a result of the collision. \textcolor{black}{In that case, the few clusters with ages older than the dynamical age of Arp 147 could have migrated from their original birthplaces or they are merely misidentified as bound star clusters, if not severely affected by age-extinction degeneracy (see Section \ref{sec:CFE}).} 
The cluster masses are all smaller than $10^6 \ {\rm M_{\odot}}$ with a median value of  $\rm \sim 8.1 \times 10^4\,M_{\odot}$. The absence of the most massive objects can be expected if the gas density within the plume is  lower than that of regions with strong gas compressions, like the S-SE reddish area of the ring. 

\section{Comparison with other CRGs}\label{sec:compare}

\textcolor{black}{Here, we systematically compare the SF activity (as traced by the blue knots and/or other star-forming regions) of Arp 147 with other CRGs studied with HST or instruments of similar resolution. This is essential to ensure unbiased comparisons not affected by artificial differences from instrument limitations. We note that there are no prior CFE analyses of CRGs in the literature that could have been used for comparison.}
 
\textcolor{black}{{\it The Cartwheel.}
This prototypical CRG is well-known to host  a large population of HII regions, where ionising star clusters are actively forming \citep[e.g.][]{1995ApJ...455..524H,2022MNRAS.514.1689Z,2023MNRAS.519.5492M,2024MNRAS.527.2816M}. While \citet{1995ApJ...455..524H} only detected 29 complexes of star clusters across the outer ring, recent works by \citet{2022MNRAS.514.1689Z} resolved these star-forming regions into 221 H$\alpha$-emitting individual knots using HST and VLT/MUSE data. The presence of such a large population associated with ionising star clusters indicates that the CRG is heavily populated by sources younger than 10 Myr. 
The authors report that most of the HII regions are also H$\beta$-emitters, with H$\beta$ acting as a direct tracer of current SF. This is in agreement with the works done  by \citet{2024MNRAS.527.2816M} and \citet{2024A&A...688A..89D}, though these follow-up studies also emphasize that star-forming regions in the Cartwheel have a wide range of ages between $\sim 20 - 170$ Myr rather than a narrow range with an upper limit of $\sim 10$ Myr. Such distribution is similar to that of Arp 147, i.e. predominant young knots combined with a small fraction of post-collision sources younger than the CRG's dynamical age of $\sim 50$ Myr. \citet{2024MNRAS.527.2816M} suggest  that the wide age range in the Cartwheel arises from bursty SF triggered by the dragging effect of the collision-driven shock waves. Clusters that initially formed elsewhere were pushed by the wave to their current location. Stellar populations older than the Cartwheel's dynamical age of $300$ Myr are mostly found in the spokes and the inner ring. These areas are HI deficient \citep{1996ApJ...467..241H} with a low level of SF activity \citep[e.g.][]{2024A&A...688A..89D}. For Arp 147, knots that formed after the collision mostly reside in the northern part of the ring which is opposite the remnant nucleus where the most  massive and young star clusters recently formed (see Fig. \ref{fig:spatial}). The observed pattern could be a mere reflection of age-extinction degeneracy in our case, or otherwise explained by the same process as in the Cartwheel.}

\textcolor{black}{As already reported in Section 5.1, while we did not find any prominent knot age gradient along the ring of Arp 147 (see Fig. \ref{fig:azrad}), previous works have instead supported the presence of such trend and discrediting a self-propagating SF \citepalias[][]{2011MNRAS.417..835F}. The ages of the star-forming regions in the Cartwheel are also known to vary along its outer ring but also across its different structures. This is expected since the level and distribution of SF are non-homogeneous on sub-galactic scales. According to \citet{1996ApJ...467..241H}, the inner ring and the optical spokes are HI deficient, which implies that a weak or suppressed SF is occurring in these regions. In contrast, a peak of massive SF associated with a high concentration of radio continuum emission is registered along the outer ring, in the southern quadrant more precisely. The enhanced cold gas with a surface density well above the critical one, necessary for strong starbursts to easily kick in, in a particular region supplies the extensive birth of radio emitting OB stars. Follow-up observations of Arp 147 using high-resolution UV and/or H$\alpha$ sensitive instruments are definitely required to strengthen our cluster analysis.}  

\textcolor{black}{The cluster age functions of both Arp 147 and the Cartwheel respectively show an abrupt increase in their cluster formation rate over the last 50 Myr and 150 Myr; see Fig. \ref{fig:caf} for Arp 147 and refer to \citet{2024MNRAS.527.2816M} for the Cartwheel. In particular, the Cartwheel's SFR further increased to $\rm \sim 18\,M_{\odot}\,yr^{-1}$ in the recent $\sim 20$ Myr while it used to be at a steady rate of $\rm \sim 5\,M_{\odot}\,yr^{-1}$ over the past 150 Myr. Before the collision, i.e. more than 300 Myr ago, \citet{2018MNRAS.473..585R} reported a SFR of $\rm \sim 1\,M_{\odot}\,yr^{-1}$. Recent studies by \citet{2024A&A...688A..89D} revealed that a small fraction of stellar component that formed before the collision co-exists with a large population of young stars within the outer ring of the Cartwheel. \citet{2024MNRAS.527.2816M} and \citet{2024A&A...688A..89D} explain the fluctuations in the SFR as a result from bursty SF episodes triggered by the collision. For Arp 147, we find the cluster formation rate to have increased in the recent 50 Myr, i.e. after the collision. Based on the shape of its CAF and ULX work by \citetalias{2010ApJ...721.1348R}, Arp 147 has also experienced a major epoch of starburst activity in the last 15 Myr. For both CRGs, it is thus clear that the head-on collision played a strong influence in triggering a recently significant cluster formation arising from strong starburst events. Their SFHs also reveal the presence of two distinct stellar populations across their ring, i.e. older sources that migrated away from their birth sites and the younger ones that recently formed in-situ. Again, it is worth noting that the ages of Arp 147's "older" sources might be attributed to age-extinction degeneracy.} 

\textcolor{black}{{\it NGC 922.} \citet{2010AJ....139.1369P} recovered more than 2000 UBVI-detected massive star clusters across the C-shaped ring and off-center bar of NGC 922. The latter structure along with the eastern side of the ring are highly populated by star-forming regions associated with strong H$\alpha$ emission. They found that more than two-thirds of the sources have ages younger than 7 Myr and only 3 per cent are older than 100 Myr. Overall, the derived ages of our knots are in agreement with the age range of the majority of clusters in NGC 922. Based on numerical simulations \citep[e.g.][]{2006MNRAS.370.1607W,2012MNRAS.420.1158M}, both CRGs experienced a strong starburst event triggered by the collision (as reflected by their similar SFR levels of $\rm \sim 8\,M_{\odot}\,yr^{-1}$), and hence the overabundance of young clusters across their ring. Compared with the CLF slopes of Arp 147 ($\rm \alpha_{Bayes} \sim 2$), this CRG's CLF slopes of $\alpha \sim 2.2$ are slightly steeper. If the difference in the slopes is not caused by statistical bias, then one may consider the distinct periods over which the two CRGs entered in head-on collision with their respective companion ($\sim$ 50 Myr for Arp 147 and 330 Myr for NGC 922). The cluster population of NGC 922 consists of sources with masses smaller than $\rm 7 \times10^3\,M_{\odot}$ and massive ones ($\rm  > 1 \times 10^5\,M_{\odot}$) at older ages. This mass range is relatively smaller than ours which spans between $\approx 1.9 \times 10^4 - 10^7\,{\rm M}_{\odot}$ with a median value of $\rm \sim 2 \times 10^5\,M_{\odot}$. Age-extinction degeneracy and blending effects can overestimate our knot masses, especially for masses larger than $\rm 10^{6.5}\,M_{\odot}$. Alternatively, it could be that the environment of Arp 147 presents more favourable conditions for massive clusters to form than that of NGC 922. In fact, although the CAFs of the two CRGs show that environment-dependent disruption is at play, \citet{2010AJ....139.1369P} recovered a CAF slope of $\delta = 0.7$ for NGC 922, which is three times steeper than ours. It means that their clusters undergo stronger dissolution. Such hypothesis is in agreement with their dynamical N-body simulations, requiring the inclusion of cluster dissolution and disruption model to reconcile observations with the simulated results.}

\textcolor{black}{{\it AM 0644\m741 (The Lindsay–Shapley RG).} 
HII regions across the ring of AM 0644\m741 have recently been studied by \citet{2024MNRAS.529.4369G} to constrain the triggering mechanisms of SF in the rings of this CRG. They explored the ionising mechanisms and chemical properties in the VLT/MUSE optical spectra of 179 HII regions. With a dynamical age of $\sim 175$ Myr and a SFR of $\rm \sim 11\,M_{\odot}\,yr^{-1}$, AM 0644\m741 is best known for its double-ringed structures hosting abundant H$\alpha$-emitting HII regions and young blue knots. The authors paired each MUSE-identified HII region with a population of massive star clusters from high-resolution HST images. They found that clusters with ages of $\sim 2 - 20$ Myr are the ones to photoionise the HII regions of the CRG. Both AM 0644\m741 and Arp 147 thus host very young knots that likely formed from recent bursts of SF being triggered by the drop-through collision.} 

\section{Summary and Conclusions}\label{sec:conclusion} 
We have investigated the star-forming knots and clumpy regions in one of the few known CRGs with an empty ring, Arp 147, using HST \textcolor{black}{F450W, F606W, and F814W archival images}.
We derived the fundamental parameters (age, mass, and extinction) of the sources to explore the star cluster formation and evolution scenarios of the galaxy. The main findings from this work are summarised as follows. 

\begin{itemize} 
    \item We recover 211 clusters and 6 kpc-sized clumps ($\rm \gtrsim 0.2\,kpc^2$) with a clear overdensity of blue sources in the colour diagrams. By fitting their SEDs with {\tt Yggdrasil} SSP models, we find that the clumps are all younger than 10 Myr and the blue knots have a median age of $\sim$ 8 Myr with more than half of the population younger than 10 Myr. \textcolor{black}{Overall, the fitted ages are consistent with the global distribution of UV and H$\alpha$ emissions across the ring}. Their masses range between  $\rm 1.9 \times 10^4 - 2.5 \times 10^7\,M_{\odot}$, though clusters with $\rm M_{cl} > 10^{6.5}\,M_{\odot}$ should be interpreted with caution.

    \item The \textcolor{black}{overall} azimuthal age distribution of the blue knots indicates an absence of any age gradient. \textcolor{black}{This likely stems from possible bias smearing out the age sequence given that the CAF and knot properties on sub-galactic scales are consistent with collision-triggered starburst events happening across the ring.} 
    The mass distribution reveals a decreasing mass with an increasing radial distance from the dusty S-SE region of the ring. This could be due to a varying distribution of the gas density triggered by the collision. 
    
    \item  Power-law slopes of the CLFs \textcolor{black}{in the F450W, F606W, and F814W filters} are similar within their uncertainties.  We find that the power-law slopes of $\alpha \approx 2$ of the cumulative CLFs are consistent with the canonical slope of 2 when adopting a Bayesian fitting technique. Mass-limited samples were used to draw CMFs for different age intervals. A PL function with a slope of $\beta = 1.78$ is still a reasonable fit to the CMF of sources younger than 10 Myr. Sources of ages between $10 - 200$ Myr have a truncated CMF best described by a Schechter function with a characteristic mass of \textcolor{black}{${\rm M}_c = 6.2 \times 10^{5} \, M_{\odot}$}. The measured value of ${\rm M}_c$ denotes Arp 147 as a host galaxy with high SF intensity \textcolor{black}{assuming the knot masses are not highly overestimated.} 

    \item We find that the cluster age function is  steeper over the past 50 Myr compared to its shape within the age range of $50 - 200$ Myr, where it is rather flat. This suggests an increase in the cluster formation rate right after the collision. The dynamical interaction of the disk galaxy with its companion must have played a role behind the enhanced SF. The truncated CMF along with the declining shape of the CAF 
    between $10 - 200$ Myr, however, could be evidence for cluster evolutionary fading combined with mild disruption over time. These interpretations should not be taken at face value since we did not disentangle cluster formation from disruption as both mechanisms are known to affect the shape of the observed CAF.
    
    \item At the commonly used age intervals of $1 - 10$ Myr and $10 - 100$ Myr, we find a cluster formation efficiency of \textcolor{black}{$\Gamma_{\Delta t_1} \sim$ 36} per cent and $\Gamma_{\Delta t_3} \sim$ 4 per cent. 
    The values of the CFE are respectively \textcolor{black}{$\sim$\,7} and $\sim\,10$ per cent, 50 Myr before and after the collision. The fraction of SF happening in bound clusters for Arp 147 is thus not constant, but rather decreases over time, likely caused by cluster dissolution. These values also indicate that the collision is an external factor influencing the CFE by triggering new starburst episodes, and hence a larger fraction of recent SF occurring in clusters.

\item Cluster analyses on sub-galactic scales (N/S, E/W, and plume) reveal that the northern and eastern regions have the largest truncated masses of \textcolor{black}{${\rm M}_c \gtrsim   1.4 \times 10^6\,{\rm M}_{\odot}$} as well as the highest CFE of \textcolor{black}{$\Gamma \gtrsim 3.3$} per cent compared to other regions. One would associate these trends as supporting evidence for environment-dependent cluster formation and/or disruption, especially since the eastern region is likely affected by the gravitational force from the nearby intruder. However, the presence of a highly extinguished dusty region therein should be taken into account.

\end{itemize}

Although careful constraints were applied to both extinction and metallicity, age-extinction degeneracy is not entirely broken due to the lack of \textcolor{black}{e.g. HST/UVIS and VLT/MUSE high-resolution}  observations. Results from this work should thus be viewed as preliminary \textcolor{black}{especially the derived ages of highly extinguished young clusters and genuinely old ones with low intrinsic extinction. Nevertheless,} they already provide insight into the SFH of the CRG. 
With a high SFR of $\sim7\,{\rm M_{\odot} \,yr^{-1}}$, 
all these elements taken together suggest that Arp 147 is the birthplace of massive star clusters and kpc-sized clumps.
While the knot ages, which are mostly below the age of the ring, are consistent with an intense SF recently triggered by a coherent event all around the ring, spatial/radial clustering, and a varying CFE on sub-galactic scales, \textcolor{black}{further} hint towards a non-uniform environment-dependent SF. 

Finally, this work has shown that archival data products can be as important as the intended main science programs, though additional data \textcolor{black}{are required} to strengthen the derived results \textcolor{black}{in our case}. 
The comprehensive cluster analysis of Arp 147 is also meant to establish the baseline of a much larger ongoing project dubbed the \textit{Stellar Clusters in Collisional Ring Galaxies} (SC2RG, Randriamanakoto et al. in prep). SC2RG is a systematic study of a representative sample of nearby CRGs to better understand SF mechanisms in these intriguing galaxies. 

\section*{Acknowledgements}
We thank the referee for their helpful comments, which greatly improved this manuscript. Based on observations made with the NASA/ESA {\it Hubble Space Telescope}, obtained from the Space Telescope Science Institute, which is operated by the Association of Universities for Research in Astronomy, Inc., under NASA contract NAS 5–26555. This research was supported by the South African Astronomical Observatory, which is a facility of the National Research Foundation, an agency of the Department of Science, Technology and Innovation. ZR also acknowledges funding from the 2020 L'Or{\'e}al - UNESCO For Women In Science Sub-Saharan Africa regional Programme and the National Geographic Society, grant number: EC-99523R-23 awarded in 2023.

\section*{Data availability}
The main optical data underlying this article are publicly available on the Mikulski Archive for Space Telescopes. The full catalogue of the blue knots presented in Table \ref{tab:knots-cat} is published online as supplementary material.


\bibliographystyle{mnras}

\begin{thebibliography}{99}

\bibitem[Adamo \& Bastian(2018)]{2018ASSL..424...91A} Adamo, A. \& Bastian, N.\ 2018, The Birth of Star Clusters, The Lifecycle of Clusters in Galaxies, 424, 91

\bibitem[Adamo et al.(2020\natexlab{a})]{2020MNRAS.499.3267A} Adamo, A., Hollyhead, K., Messa, M., et al.\ 2020a, \mnras, 499, 3267

\bibitem[Adamo et al.(2015)]{2015MNRAS.452..246A} Adamo, A., Kruijssen, J.~M.~D., Bastian, N., et al.\ 2015, \mnras, 452, 246

\bibitem[Adamo et al.(2011)]{2011MNRAS.417.1904A} Adamo, A., {\"O}stlin, G., \& Zackrisson, E.\ 2011, \mnras, 417, 3, 1904

\bibitem[Adamo et al.(2020\natexlab{b})]{2020SSRv..216...69A} Adamo, A., Zeidler, P., Kruijssen, J.~M.~D., et al.\ 2020b, \ssr, 216, 69


\bibitem[Aromal et al.(2025)]{2025ApJ...994...90A} Aromal, P., Gallagher, S.~C., Fedotov, K., et al.\ 2025, \apj, 994, 1, 90

\bibitem[Athanassoula \& Bosma(1985)]{1985ARA&A..23..147A} Athanassoula, E. \& Bosma, A.\ 1985, \araa, 23, 147

\bibitem[Anders et al.(2004)]{2004MNRAS.347..196A} Anders, P., Bissantz, N., Fritze-v. Alvensleben, U., et al.\ 2004, \mnras, 347, 196

\bibitem[Appleton \& Marston(1997)]{1997AJ....113..201A} Appleton, P.~N. \& Marston, A.~P.\ 1997, \aj, 113, 201

\bibitem[Appleton \& Struck-Marcell(1987)]{1987ApJ...312..566A} Appleton, P.~N. \& Struck-Marcell, C.\ 1987, \apj, 312, 566

\bibitem[Appleton \& Struck-Marcell(1996)]{1996FCPh...16..111A} Appleton, P.~N. \& Struck-Marcell, C.\ 1996, \fcp, 16, 111

\bibitem[Bastian(2008)]{2008MNRAS.390..759B} Bastian, N.\ 2008, \mnras, 390, 759

\bibitem[Bastian et al.(2012)]{2012MNRAS.419.2606B} Bastian, N., Adamo, A., Gieles, M., et al.\ 2012, \mnras, 419, 2606

\bibitem[Bastian et al.(2005)]{2005A&A...431..905B} Bastian, N., Gieles, M., Lamers, H.~J.~G.~L.~M., et al.\ 2005, \aap, 431, 905

\bibitem[Beir{\~a}o et al.(2009)]{2009ApJ...693.1650B} Beir{\~a}o, P., Appleton, P.~N., Brandl, B.~R., et al.\ 2009, \apj, 693, 1650

\bibitem[Bertin \& Arnouts(1996)]{1996A&AS..117..393B} Bertin, E. \& Arnouts, S.\ 1996, \aaps, 117, 393

\bibitem[Bigiel et al.(2008)]{2008AJ....136.2846B} Bigiel, F., Leroy, A., Walter, F., et al.\ 2008, \aj, 136, 2846

\bibitem[Bik et al.(2003)]{2003A&A...397..473B} Bik, A., Lamers, H.~J.~G.~L.~M., Bastian, N., et al.\ 2003, \aap, 397, 473

\bibitem[Bransford et al.(1998)]{1998AJ....116.2757B} Bransford, M.~A., Appleton, P.~N., Marston, A.~P., et al.\ 1998, \aj, 116, 2757

\bibitem[Boutloukos \& Lamers(2003)]{2003MNRAS.338..717B} Boutloukos, S.~G. \& Lamers, H.~J.~G.~L.~M.\ 2003, \mnras, 338, 717

\bibitem[Burkert et al.(2005)]{2005ApJ...628..231B} Burkert, A., Brodie, J., \& Larsen, S.\ 2005, \apj, 628, 231 

\bibitem[Buta \& Crocker(1993)]{1993AJ....105.1344B} Buta, R. \& Crocker, D.~A.\ 1993, \aj, 105, 1344

\bibitem[Calzetti et al.(2000)]{2000ApJ...533..682C} Calzetti, D., Armus, L., Bohlin, R.~C., et al.\ 2000, \apj, 533, 682

\bibitem[Calzetti et al.(2015)]{2015AJ....149...51C} Calzetti, D., Lee, J.~C., Sabbi, E., et al.\ 2015, \aj, 149, 51

\bibitem[Caputo et al.(2024)]{2024AJ....168..259C} Caputo, M., Chandar, R., Mok, A., et al.\ 2024, \aj, 168, 6, 259

\bibitem[Chandar et al.(2017)]{2017ApJ...849..128C} Chandar, R., Fall, S.~M., Whitmore, B.~C., et al.\ 2017, \apj, 849, 128

\bibitem[Chandar et al.(2023)]{2023ApJ...949..116C} Chandar, R., Caputo, M., Mok, A., et al.\ 2023, \apj, 949, 116

\bibitem[Cook et al.(2023)]{2023MNRAS.519.3749C} Cook, D.~O., Lee, J.~C., Adamo, A., et al.\ 2023, \mnras, 519, 3749

\bibitem[Cook et al.(2012)]{2012ApJ...751..100C} Cook, D.~O., Seth, A.~C., Dale, D.~A., et al.\ 2012, \apj, 751, 100

\bibitem[Davies et al.(2017)]{2017ApJ...847..112D} Davies, B., Kudritzki, R.-P., Lardo, C., et al.\ 2017, \apj, 847, 2, 112

\bibitem[de Grijs et al.(2003)]{2003MNRAS.343.1285D} de Grijs, R., Anders, P., Bastian, N., et al.\ 2003, \mnras, 343, 1285

\bibitem[de Grijs et al.(2013)]{2013MNRAS.431.2917D} de Grijs, R., Anders, P., Zackrisson, E., et al.\ 2013, \mnras, 431, 2917

\bibitem[de Vaucouleurs et al.(1991)]{1991S&T....82Q.621D} de Vaucouleurs, G., de Vaucouleurs, A., Corwin, H.~G., et al.\ 1991, Sky Telesc., 82, 621 

\bibitem[Ditrani et al.(2024)]{2024A&A...688A..89D} Ditrani, F.~R., Longhetti, M., Fossati, M., et al.\ 2024, \aap, 688, A89

\bibitem[D'Onghia et al.(2008)]{2008MNRAS.389.1275D} D'Onghia, E., Mapelli, M., \& Moore, B.\ 2008, \mnras, 389, 1275

\bibitem[Egorov et al.(2023)]{2023ApJ...944L..16E} Egorov, O.~V., Kreckel, K., Sandstrom, K.~M., et al.\ 2023, \apjl, 944, 2, L16

\bibitem[Elagali et al.(2018)]{2018MNRAS.481.2951E} Elagali, A., Lagos, C.~D.~P., Wong, O.~I., et al.\ 2018, \mnras, 481, 2951

\bibitem[Elmegreen(2008)]{2008ApJ...672.1006E} Elmegreen, B.~G.\ 2008, \apj, 672, 1006

\bibitem[Elmegreen \& Efremov(1997)]{1997ApJ...480..235E} Elmegreen, B.~G. \& Efremov, Y.~N.\ 1997, \apj, 480, 235

\bibitem[Elmegreen \& Elmegreen(2005\natexlab{a})]{2005ApJ...627..632E} Elmegreen, B.~G. \& Elmegreen, D.~M.\ 2005a, \apj, 627, 632

\bibitem[Elmegreen \& Elmegreen(2006)]{2006ApJ...651..676E} Elmegreen, D.~M. \& Elmegreen, B.~G.\ 2006, \apj, 651, 676

\bibitem[Elmegreen et al.(1994)]{1994ApJ...425...57E} Elmegreen, D.~M., Elmegreen, B.~G., Lang, C., et al.\ 1994, \apj, 425, 57

\bibitem[Elmegreen et al.(2005\natexlab{b})]{2005ApJ...631...85E} Elmegreen, D.~M., Elmegreen, B.~G., Rubin, D.~S., et al.\ 2005b, \apj, 631, 85

\bibitem[Fabbiano(1989)]{1989ARA&A..27...87F} Fabbiano, G.\ 1989, \araa, 27, 87

\bibitem[Fall et al.(2005)]{2005ApJ...631L.133F} Fall, S.~M., Chandar, R., \& Whitmore, B.~C.\ 2005, \apjl, 631, L133

\bibitem[Fa{\'u}ndez-Abans et al.(2013)]{2013A&A...559A...8F} Fa{\'u}ndez-Abans, M., da Rocha-Poppe, P.~C., Fernandes-Martin, V.~A., et al.\ 2013, \aap, 559, A8

\bibitem[Fogarty et al.(2011)]{2011MNRAS.417..835F} Fogarty, L., Thatte, N., Tecza, M., et al.\ 2011, \mnras, 417, 835 (F11)

\bibitem[Fouesneau et al.(2014)]{2014ApJ...786..117F} Fouesneau, M., Johnson, L.~C., Weisz, D.~R., et al.\ 2014, \apj, 786, 117

\bibitem[Fouesneau \& Lan{\c{c}}on(2010)]{2010A&A...521A..22F} Fouesneau, M. \& Lan{\c{c}}on, A.\ 2010, \aap, 521, A22

\bibitem[Gallagher et al.(2010)]{2010AJ....139..545G} Gallagher, S.~C., Durrell, P.~R., Elmegreen, D.~M., et al.\ 2010, \aj, 139, 545

\bibitem[Gao et al.(2003)]{2003ApJ...596L.171G} Gao, Y., Wang, Q.~D., Appleton, P.~N., et al.\ 2003, \apjl, 596, L171

\bibitem[Gazak et al.(2014)]{2014ApJ...787..142G} Gazak, J.~Z., Davies, B., Bastian, N., et al.\ 2014, \apj, 787, 2, 142

\bibitem[Georgakakis et al.(2004)]{2004MNRAS.349..135G} Georgakakis, A., Georgantopoulos, I., Vallb{\'e}, M., et al.\ 2004, \mnras, 349, 1, 135

\bibitem[Gerber et al.(1992)]{1992ApJ...399L..51G} Gerber, R.~A., Lamb, S.~A., \& Balsara, D.~S.\ 1992, \apjl, 399, L51

\bibitem[Gerola \& Seiden(1978)]{1978ApJ...223..129G} Gerola, H. \& Seiden, P.~E.\ 1978, \apj, 223, 129

\bibitem[Gieles et al.(2006)]{2006A&A...450..129G} Gieles, M., Larsen, S.~S., Bastian, N., et al.\ 2006, \aap, 450, 129

\bibitem[Ghosh \& Mapelli(2008)]{2008MNRAS.386L..38G} Ghosh, K.~K. \& Mapelli, M.\ 2008, \mnras, 386, L38

\bibitem[Goddard et al.(2010)]{2010MNRAS.405..857G} Goddard, Q.~E., Bastian, N., \& Kennicutt, R.~C.\ 2010, \mnras, 405, 857

\bibitem[G{\'o}mez-Gonz{\'a}lez et al.(2024)]{2024MNRAS.529.4369G} G{\'o}mez-Gonz{\'a}lez, V.~M.~A., Mayya, Y.~D., Zaragoza-Cardiel, J., et al.\ 2024, \mnras, 529, 4, 4369

\bibitem[Higdon(1995)]{1995ApJ...455..524H} Higdon, J.~L.\ 1995, \apj, 455, 524

\bibitem[Higdon(1996)]{1996ApJ...467..241H} Higdon, J.~L.\ 1996, \apj, 467, 241

\bibitem[Higdon et al.(2011)]{2011ApJ...739...97H} Higdon, J.~L., Higdon, S.~J.~U., \& Rand, R.~J.\ 2011, \apj, 739, 97

\bibitem[Hollyhead et al.(2016)]{2016MNRAS.460.2087H} Hollyhead, K., Adamo, A., Bastian, N., et al.\ 2016, \mnras, 460, 2087

\bibitem[Inoue et al.(2021)]{2021MNRAS.507.6140I} Inoue, S., Yoshida, N., \& Hernquist, L.\ 2021, \mnras, 507, 6140

\bibitem[Johnson et al.(2016)]{2016ApJ...827...33J} Johnson, L.~C., Seth, A.~C., Dalcanton, J.~J., et al.\ 2016, \apj, 827, 33

\bibitem[Kaaret et al.(2004)]{2004MNRAS.348L..28K} Kaaret, P., Alonso-Herrero, A., Gallagher, J.~S., et al.\ 2004, \mnras, 348, L28 

\bibitem[Kaaret et al.(2017)]{2017ARA&A..55..303K} Kaaret, P., Feng, H., \& Roberts, T.~P.\ 2017, \araa, 55, 303

\bibitem[Kennicutt(1998)]{1998ARA&A..36..189K} Kennicutt, R.~C.\ 1998, \araa, 36, 189

\bibitem[Kruijssen(2012)]{2012MNRAS.426.3008K} Kruijssen, J.~M.~D.\ 2012, \mnras, 426, 3008

\bibitem[Lada \& Lada(2003)]{2003ARA&A..41...57L} Lada, C.~J. \& Lada, E.~A.\ 2003, \araa, 41, 57 

\bibitem[Lah{\'e}n et al.(2020)]{2020ApJ...891....2L} Lah{\'e}n, N., Naab, T., Johansson, P.~H., et al.\ 2020, \apj, 891, 2

\bibitem[Larsen(2009)]{2009A&A...494..539L} Larsen, S.~S.\ 2009, \aap, 494, 2, 539

\bibitem[Larsen \& Richtler(2000)]{2000A&A...354..836L} Larsen, S.~S. \& Richtler, T.\ 2000, \aap, 354, 836

\bibitem[Lavery et al.(2004)]{2004ApJ...612..679L} Lavery, R.~J., Remijan, A., Charmandaris, V., et al.\ 2004, \apj, 612, 679

\bibitem[Leitherer et al.(2014)]{2014ApJS..212...14L} Leitherer, C., Ekstr{\"o}m, S., Meynet, G., et al.\ 2014, \apjs, 212, 14

\bibitem[Leitherer et al.(1999)]{1999ApJS..123....3L} Leitherer, C., Schaerer, D., Goldader, J.~D., et al.\ 1999, \apjs, 123, 3

\bibitem[Li et al.(2015)]{2015ApJS..216....6L} Li, S., de Grijs, R., Anders, P., et al.\ 2015, \apjs, 216, 1, 6

\bibitem[Linden et al.(2021)]{2021ApJ...923..278L} Linden, S.~T., Evans, A.~S., Larson, K., et al.\ 2021, \apj, 923, 278

\bibitem[Linden et al.(2017)]{2017ApJ...843...91L} Linden, S.~T., Evans, A.~S., Rich, J., et al.\ 2017, \apj, 843, 91 

\bibitem[Lynds \& Toomre(1976)]{1976ApJ...209..382L} Lynds, R. \& Toomre, A.\ 1976, \apj, 209, 382

\bibitem[Madore et al.(2009)]{2009ApJS..181..572M} Madore, B.~F., Nelson, E., \& Petrillo, K.\ 2009, \apjs, 181, 572

\bibitem[Mapelli \& Mayer(2012)]{2012MNRAS.420.1158M} Mapelli, M. \& Mayer, L.\ 2012, \mnras, 420, 1158

\bibitem[Mapelli et al.(2010)]{2010MNRAS.408..234M} Mapelli, M., Ripamonti, E., Zampieri, L., et al.\ 2010, \mnras, 408, 234

\bibitem[Mayya et al.(2024)]{2024MNRAS.527.2816M} Mayya, Y.~D., Barway, S., G{\'o}mez-Gonz{\'a}lez, V.~M.~A., et al.\ 2024, \mnras, 527, 2, 2816

\bibitem[Mayya et al.(2023)]{2023MNRAS.519.5492M} Mayya, Y.~D., Plat, A., G{\'o}mez-Gonz{\'a}lez, V.~M.~A., et al.\ 2023, \mnras, 519, 4, 5492

\bibitem[Messa et al.(2018\natexlab{a})]{2018MNRAS.477.1683M} Messa, M., Adamo, A., Calzetti, D., et al.\ 2018a, \mnras, 477, 1683

\bibitem[Messa et al.(2018\natexlab{b})]{2018MNRAS.473..996M} Messa, M., Adamo, A., {\"O}stlin, G., et al.\ 2018b, \mnras, 473, 996

\bibitem[Mora et al.(2015)]{2015AJ....150...93M} Mora, M.~D., Chanam{\'e}, J., \& Puzia, T.~H.\ 2015, \aj, 150, 93

\bibitem[Pellerin et al.(2010)]{2010AJ....139.1369P} Pellerin, A., Meurer, G.~R., Bekki, K., et al.\ 2010, \aj, 139, 1369

\bibitem[Portegies Zwart et al.(2010)]{2010ARA&A..48..431P} Portegies Zwart, S.~F., McMillan, S.~L.~W., \& Gieles, M.\ 2010, \araa, 48, 431

\bibitem[Poutanen et al.(2013)]{2013MNRAS.432..506P} Poutanen, J., Fabrika, S., Valeev, A.~F., et al.\ 2013, \mnras, 432, 506

\bibitem[Renaud et al.(2018)]{2018MNRAS.473..585R} Renaud, F., Athanassoula, E., Amram, P., et al.\ 2018, \mnras, 473, 585

\bibitem[Randriamanakoto(2023)]{2023afas.confE..83R} Randriamanakoto, Z.\ 2023, African Astronomical Society, 83

\bibitem[Randriamanakoto(2025)]{2025afas.confE.136R} 
Randriamanakoto, Z.\ 2025, African Astronomical Society, 136

\bibitem[Randriamanakoto et al.(2013\natexlab{a})]{2013ApJ...775L..38R} Randriamanakoto, Z., Escala, A., V{\"a}is{\"a}nen, P., et al.\ 2013a, \apjl, 775, L38

\bibitem[Randriamanakoto et al.(2022)]{2022MNRAS.513.4232R} Randriamanakoto, Z., V{\"a}is{\"a}nen, P., Ranaivomanana, P., et al.\ 2022, \mnras, 513, 4232

\bibitem[Randriamanakoto et al.(2013\natexlab{b})]{2013MNRAS.431..554R} Randriamanakoto, Z., V{\"a}is{\"a}nen, P., Ryder, S., et al.\ 2013b, \mnras, 431, 554

\bibitem[Randriamanakoto et al.(2019)]{2019MNRAS.482.2530R} Randriamanakoto, Z., V{\"a}is{\"a}nen, P., Ryder, S.~D., et al.\ 2019, \mnras, 482, 2530

\bibitem[Rappaport et al.(2010)]{2010ApJ...721.1348R} Rappaport, S., Levine, A., Pooley, D., et al.\ 2010, \apj, 721, 1348 (R10)

\bibitem[Rodruck et al.(2023)]{2023MNRAS.526.2341R} Rodruck, M., Charlton, J., Borthakur, S., et al.\ 2023, \mnras, 526, 2341

\bibitem[Romano et al.(2008)]{2008AJ....136.1259R} Romano, R., Mayya, Y.~D., \& Vorobyov, E.~I.\ 2008, \aj, 136, 1259

\bibitem[Salvaggio et al.(2023)]{2023MNRAS.522.1377S} Salvaggio, C., Wolter, A., Belfiore, A., et al.\ 2023, \mnras, 522, 1, 1377

\bibitem[Schweizer(1976)]{1976ApJS...31..313S} Schweizer, F.\ 1976, \apjs, 31, 313

\bibitem[Shabani et al.(2018)]{2018MNRAS.478.3590S} Shabani, F., Grebel, E.~K., Pasquali, A., et al.\ 2018, \mnras, 478, 3590

\bibitem[Silva-Villa \& Larsen(2011)]{2011A&A...529A..25S} Silva-Villa, E. \& Larsen, S.~S.\ 2011, \aap, 529, A25

\bibitem[Sun et al.(2016)]{2016ApJ...816....9S} Sun, W., de Grijs, R., Fan, Z., et al.\ 2016, \apj, 816, 9

\bibitem[Spergel et al.(2007)]{2007ApJS..170..377S} Spergel, D.~N., Bean, R., Dor{\'e}, O., et al.\ 2007, \apjs, 170, 2, 377

\bibitem[Swartz et al.(2011)]{2011ApJ...741...49S} Swartz, D.~A., Soria, R., Tennant, A.~F., et al.\ 2011, \apj, 741, 49

\bibitem[Swartz et al.(2009)]{2009ApJ...703..159S} Swartz, D.~A., Tennant, A.~F., \& Soria, R.\ 2009, \apj, 703, 159

\bibitem[Theys \& Spiegel(1976)]{1976ApJ...208..650T} Theys, J.~C. \& Spiegel, E.~A.\ 1976, \apj, 208, 650

\bibitem[Theys \& Spiegel(1977)]{1977ApJ...212..616T} Theys, J.~C. \& Spiegel, E.~A.\ 1977, \apj, 212, 616

\bibitem[Thompson \& Theys(1978)]{1978ApJ...224..796T} Thompson, L.~A. \& Theys, J.~C.\ 1978, \apj, 224, 796

\bibitem[V{\"a}is{\"a}nen et al.(2014)]{2014ApJ...797L..16V} V{\"a}is{\"a}nen, P., Barway, S., \& Randriamanakoto, Z.\ 2014, \apjl, 797, L16

\bibitem[Waldron et al.(2023)]{2023MNRAS.522..173W} Waldron, W., Sun, M., Luo, R., et al.\ 2023, \mnras, 522, 173

\bibitem[Whitmore(2003)]{2003dhst.symp..153W} Whitmore, B.~C.\ 2003, A Decade of Hubble Space Telescope Science, 14, 153

\bibitem[Whitmore et al.(2007)]{2007AJ....133.1067W} Whitmore, B.~C., Chandar, R., \& Fall, S.~M.\ 2007, \aj, 133, 1067

\bibitem[Whitmore et al.(2023)]{2023MNRAS.520...63W} Whitmore, B.~C., Chandar, R., Lee, J.~C., et al.\ 2023, \mnras, 520, 63

\bibitem[Whitmore et al.(2020)]{2020ApJ...889..154W} Whitmore, B.~C., Chandar, R., Lee, J.~C., et al.\ 2020, \apj, 889, 2, 154

\bibitem[Whitmore et al.(1999)]{1999AJ....118.1551W} Whitmore, B.~C., Zhang, Q., Leitherer, C., et al.\ 1999, \aj, 118, 1551

\bibitem[Wolter et al.(2018)]{2018ApJ...863...43W} Wolter, A., Fruscione, A., \& Mapelli, M.\ 2018, \apj, 863, 43

\bibitem[Wong et al.(2006)]{2006MNRAS.370.1607W} Wong, O.~I., Meurer, G.~R., Bekki, K., et al.\ 2006, \mnras, 370, 1607

\bibitem[Wong et al.(2017)]{2017MNRAS.466..574W} Wong, O.~I., Vega, O., S{\'a}nchez-Arg{\"u}elles, D., et al.\ 2017, \mnras, 466, 574

\bibitem[Wuyts et al.(2012)]{2012ApJ...753..114W} Wuyts, S., F{\"o}rster Schreiber, N.~M., Genzel, R., et al.\ 2012, \apj, 753, 114

\bibitem[Yuan et al.(2020)]{2020NatAs...4..957Y} Yuan, T., Elagali, A., Labb{\'e}, I., et al.\ 2020, Nature Astronomy, 4, 957

\bibitem[Zackrisson et al.(2001)]{2001A&A...375..814Z} Zackrisson, E., Bergvall, N., Olofsson, K., et al.\ 2001, \aap, 375, 814

\bibitem[Zackrisson et al.(2011)]{2011ApJ...740...13Z} Zackrisson, E., Rydberg, C.-E., Schaerer, D., et al.\ 2011, \apj, 740, 13

\bibitem[Zaragoza-Cardiel et al.(2022)]{2022MNRAS.514.1689Z} Zaragoza-Cardiel, J., G{\'o}mez-Gonz{\'a}lez, V.~M.~A., Mayya, D., et al.\ 2022, \mnras, 514, 1689

\bibitem[Zhang \& Fall(1999)]{1999ApJ...527L..81Z} Zhang, Q. \& Fall, S.~M.\ 1999, \apjl, 527, L81

\end{thebibliography}



\appendix

\section{Extinction map and ULX properties} 

\begin{figure}
\begin{center}
  \resizebox{0.8\hsize}{!}{\includegraphics[trim= 0cm 0.7cm 0cm 0cm]{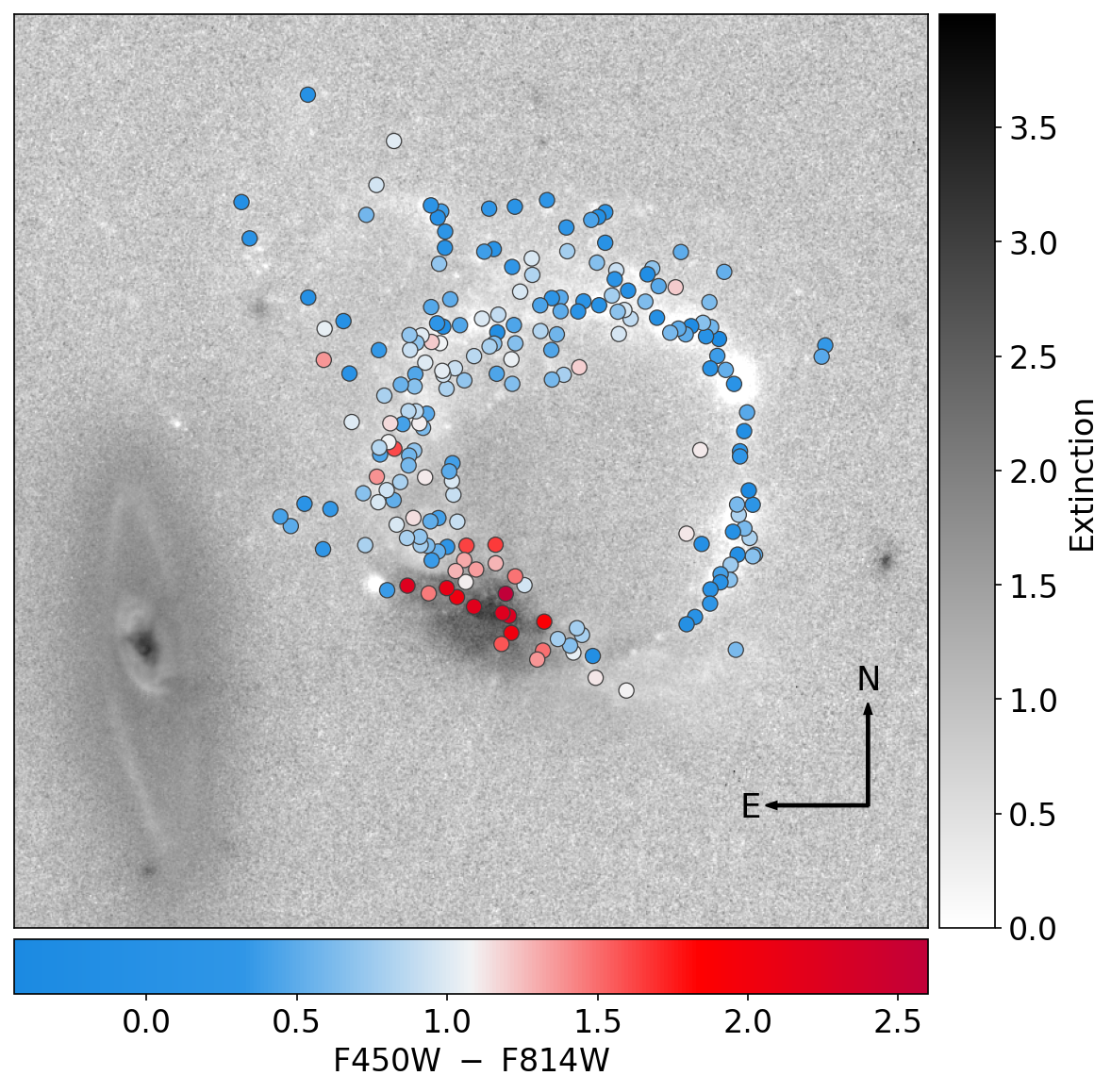}}  
\end{center}
\caption{Spatial distribution of the knots as a function of their \textcolor{black}{F450W - F814W} colour overplotted on the extinction map of the CRG.  The horizontal and vertical colour bars denote a linear scale of \textcolor{black}{F450W - F814W} and extinction values, respectively.  
The most extinguished regions such as in the S-SE side of the ring are represented by the darker shades for the latter colour gradient.}
\label{fig:Av-map}
\end{figure}

\section{PDF distributions of the CMF fitted parameters} 

\begin{figure}
\centering
\begin{tabular}{cc}
  \resizebox{0.35\hsize}{!}{\includegraphics[trim= 2.5cm 0.7cm 0cm 0cm]{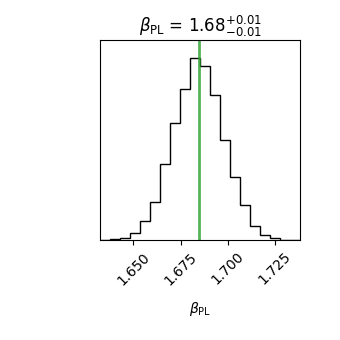}}  
  \resizebox{0.65\hsize}{!}{\includegraphics[trim= 1.cm 0.7cm 0cm 0cm]{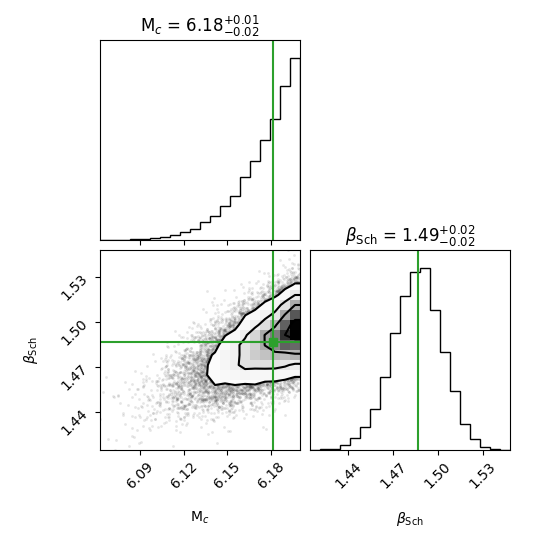}}\\ 
\end{tabular}
\caption{{\it Left panel:} one-dimensional posterior probability distribution of the power-law slope $\beta_{\rm PL}$ for the CMF of knots younger than 200 Myr shown in Fig. \ref{fig:CMF}. {\it Right panel:} Two-dimensional posterior probability distribution of the Schechter slope $\beta_{\rm Sch}$ and the truncation mass $\rm M_{\it c}$ of the same CMF. Contours represent the 1$\sigma$, 2$\sigma$, and 3$\sigma$ confidence intervals. The histograms show the marginalised distributions of the fitted parameters.}
\label{fig:PDF-lt200}
\end{figure}

\begin{figure}
\centering
\begin{tabular}{cc}
  \resizebox{0.36\hsize}{!}{\includegraphics[trim= 2.cm 0.7cm 0cm 0cm]{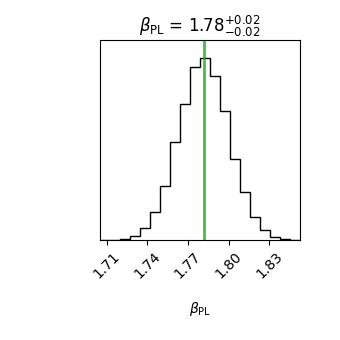}}  
  \resizebox{0.65\hsize}{!}{\includegraphics[trim= 1cm 0.7cm 0cm 0cm]{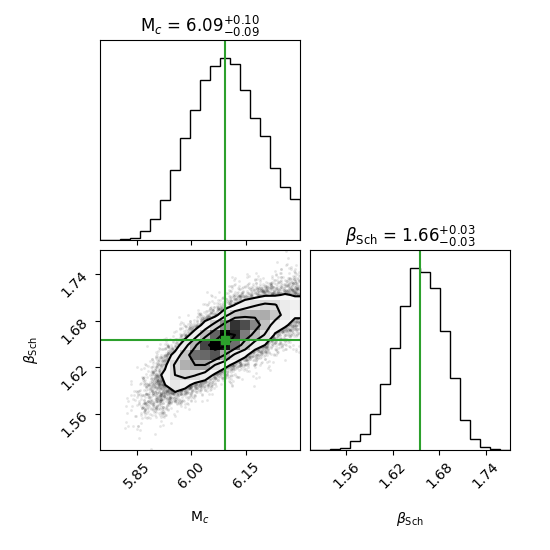}}\\ 
  
  \resizebox{0.36\hsize}{!}{\includegraphics[trim= 2.cm 0.7cm 0cm 0cm]{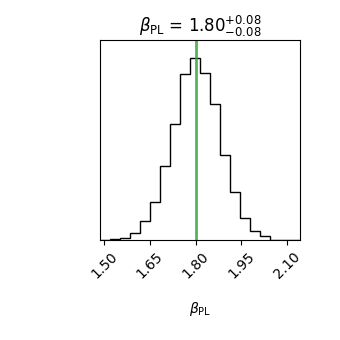}}  
  \resizebox{0.65\hsize}{!}{\includegraphics[trim= 1cm 0.7cm 0cm 0cm]{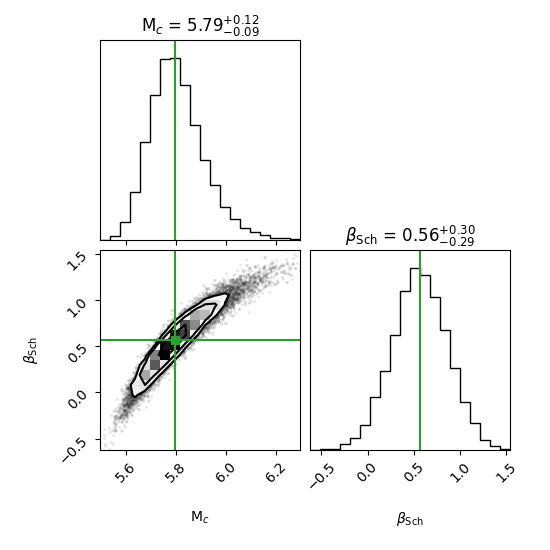}}
\end{tabular}
\caption{Posterior probability distributions similar to the ones in Fig. \ref{fig:PDF-lt200} but for the CMFs of $1 - 10$ Myr (top panels) and  $10 - 200$ Myr (bottom) age intervals shown in Fig. \ref{fig:CMF}.}
\label{fig:res-Bayes-CMFs}
\end{figure}

\begin{table}
\caption{X-ray luminosities of the ULXs and the properties of their closest optical knots within 0.5 arcsec.}
\begin{center}
\setlength{\tabcolsep}{3.5pt}
\scalebox{.87}{
\begin{tabular}{l c c c c c c}
 \hline \hline
  ID & RA, DEC  & log $L_X$ & \# YMCs & $\overline{\rm \textcolor{black}{F450W - F814W}}$ & log\,($\overline{\tau}$) & log\,(${\rm \overline{M}_{cl}}$) \\ 
  
  & (deg, deg) & (${\rm erg\,s^{-1}}$) & ($d \leq r$) & (mag) & (yr) & (M$_{\odot}$) \\ 
  \hline 

    ULX-1 & 47.825167, 1.317306 & 39.88 & 3 & 0.50 & 6.84 & 5.55 \\
    ULX-2 & 47.826417, 1.317333 & 39.53 & 2 &0.32 & 6.74 & 4.93 \\
    ULX-3 & 47.827375, 1.317222 & 39.17 & 1 & 0.89 & 8.30 & 5.55 \\
    ULX-4 & 47.827875, 1.316889 &39.31 & 1 & 1.07 & 8.30 & 6.00 \\
    ULX-5 & 47.828333, 1.316028 &39.56 & 4 & 0.74 & 8.11 & 6.59 \\
    ULX-6 & 47.827750, 1.315111 &39.38 & 1 & 0.91 & 6.84 & 5.40 \\
    ULX-7 & 47.827875, 1.314306 &39.75 & 2 & 1.70 & 6.87 & 6.53 \\
    ULX-8 & 47.826958, 1.313389 &39.79 & 1 & 1.37 & 6.84 & 5.81 \\
    ULX-9 & 47.829750, 1.314889 &39.74 & 2 & 0.45 & 6.90 & 5.12 \\

  \hline
\end{tabular}
}
\end{center}
\label{tab:ULXs}
\end{table}

\bsp	
\label{lastpage}
\end{document}